\documentstyle[psfig]{mn2e}
\newcommand{\gsim}{\raisebox{-3.8pt}{$\;\stackrel{\textstyle >}{\sim}\;$}}

\newcommand{\Msol}{$M_{\odot}$}

\newcommand{\etal}{\mbox{{\rm et~al.\ }}}

\def\ozs{\Omega_Z^{SFH}}
\def\ozo{\Omega_Z^{OBS}}

\def\ie{{\frenchspacing\it i.e., }}

\title{Where are the Cosmic Metals at $z \sim 3$\,?}

\author[Sommer--Larsen \& Fynbo]{Jesper Sommer--Larsen$^{1,2,3}$ and Johan P. U.
       Fynbo$^2$\\ 
       $^1$ Excellence Cluster Universe, Technische Universit\"at 
       M\"unchen, Boltzmanstr. 2, D-85748 Garching, Germany\\ 
       $^2$ Dark Cosmology Centre, Niels Bohr Institute, Juliane Maries Vej 
             30, DK-2100 Copenhagen, Denmark\\
       $^3$ Institute of Astronomy, University of Tokyo, Osawa 2-21-1,
Mitaka, Tokyo, 181-0015, Japan\\
E-mail: {\tt jslarsen@astro.ku.dk,jfynbo@dark-cosmology.dk}
}

\date{\tt accepted October 18, 2007}

\pubyear{2007}
\begin{document}
\maketitle
\title{Where are the Cosmic Metals?}
%
%
\begin{abstract}
The global temperature distribution of the cosmic gas-phase oxygen at
$z$$\sim$3 is determined by combining high resolution cosmological  
simulations
of individual proto-galactic, as well as larger, regions with the  
observed,
extinction-corrected, rest-frame V-band galaxy luminosity function. The
simulations have been performed with three different stellar initial  
mass
functions (IMFs), a Kroupa (K98), a Salpeter (S) and an Arimoto- 
Yoshii (AY),
spanning a range of a factor of five in chemical yield and specific  
SNII energy
feedback. Gas-phase oxygen is binned according to $T$ as log$(T)\la4.0$
(``cold''), log$(T)\sim4.5$ (``warm''), and log$(T)\sim$5.0, 5.5,  
6.0, 6.5, 7.0
(``hot'' phases). Oxygen is found to be distributed over all $T$  
phases, in
particular for the top-heavy AY IMF. But, at variance with previous  
works, it
is found that for the K98 and S IMFs the cold phase is the most
important. For these IMFs
it contains 47 and 37\%, respectively, of all gas-phase oxygen,  
mainly at
fairly high density, $n_{\rm{H}}$$\ga$0.1 cm$^{-3}$. The implications  
of this
in relation to observational damped Lyman-$\alpha$ absorber (DLA)  
studies are
discussed. In relation to ``missing metals'' it is found that a  
significant
fraction of the oxygen is located in a warm/hot phase that may be very
difficult to detect. Moreover, it is found that less than about 20-25\% 
of the cosmic oxygen
is associated with galaxies brighter than $M_{\rm{V}} 
\sim-22$, \ie
the faintest galaxy luminosities probed by current metallicity  
determinations
for Lyman Break Galaxies (LBGs). Hence, 75-80\% of the oxygen is also  
in this
sense ``missing''. From the LBG based, $\lambda$$\sim$1500 {\AA} UV  
luminosity
density history at $z$$\ge$3, we obtain an essentially IMF independent
constraint on the mean oxygen density at $z$=3. We compare this to  
what is
obtained from our models, for the three different IMFs. We find that  
the K98
IMF is strongly excluded, as the chemical yield is simply too small, the
Salpeter is marginally excluded, and the AY matches the constraint  
well. The K98 IMF can only match the data if the $\lambda$$\sim$1500 {\AA} 
extinction corrections have been overestimated by factor of $\sim$4,
which seems highly unlikely. The
yields for K98 are also far too small to match the observational data  
for
\hbox{C\,{\sc iv}}. The optimal IMF should have a yield intermediate between the S  
and AY.
\end{abstract}


\section{Introduction}
\label{sect:introduction}
A number of ``problems'' have been discussed in relation to the cosmic
mean metallicity and/or mean metal density. Pagel (1999) discussed the
low redshift ($z$$\sim$0) ``excess metals problem'' showing that
a comparison of the average metal to stellar densities in the low-$z$
Universe indicates that the average cosmic stellar initial mass function 
(IMF) has a higher yield, than what is obtained for the ``standard''
Salpeter (1955) IMF. This is a well known result for ellipticals in
clusters (e.g., Romeo \etal 2006 and references therein), 
but Pagel's analysis indicates that it could be more universal. 

Conversely, Pettini (1999) formulated the high-$z$ ``missing metals problem'' 
as follows: Studies of the
co-moving rest-frame UV luminosity density of high-$z$ Lyman break galaxies 
(LBGs) allow us to trace the cosmic star formation density (or history, SFH), 
$\dot\rho_\star(z)$, up to redshifts $z_{max} \approx 7-8$. 
Assuming an IMF of such 
stars, one can compute the specific fraction of heavy elements 
(``metals'') they produce, $y$, and derive 
the metal production rate $\dot\rho_Z(z) = y \dot\rho_\star(z)$. The 
integral from $z_{max}$ gives, at any given $z$, the density of cosmic 
metals $\rho_Z^{SFH}(z)$, or, expressed in units of the critical
density, $\ozs$.     
Moreover, if one restricts the analysis to elements, such as the 
$\alpha$-elements, produced almost exclusively in massive stars undergoing 
core collapse, the apparent 
dependence of $\rho_Z^{SFH}(z)$ on the IMF is essentially removed, 
since both the UV light and the oxygen production
originate from massive stars. This will be discussed further
in Section~\ref{sect:UV}, and was already noticed by, e.g., Songaila \etal
(1990), Madau \etal (1996), Pagel (1999) and Pettini (1999). 

Early searches in cosmic structures for which the metal/baryon mass ratio
(metallicity,
$Z=\Omega_Z/\Omega_b$) can be derived either via intergalactic gas quasar 
absorption line experiments (DLAs
or the Ly$\alpha$ ``forest'') or through direct spectroscopic studies of
Lyman break galaxies have found that only $\ozo \la 0.20\ozs$ is 
stored in these components, \ie the large majority of the metals are 
``missing''.
Similar missing metal problems have been formulated by Wolfe \etal (2003)
and Prochaska \etal (2003, 2006) on the basis of star formation rates and 
metallicities of DLAs alone. 

Ferrara \etal (2005, FSB05) attempted to quantify more 
precisely the extent of the missing metals deficit, and also to suggest
where the ``missing metals'' might be found. They found from considering
typical stellar masses and metallicities of Lyman break galaxies, and an
estimate of the co-moving density of LBGs a contribution of 
$\Omega_Z^{LBG} = 3.4 \times 10^{-6}$ from Lyman break galaxies. They also
estimated the contribution from metals in DLAs
to about $\Omega_Z^{DLA} = 3.8 \times 10^{-7}$. Using the SFH estimate of
Bouwens \etal (2004b), assuming a Salpeter IMF, and applying a dust
correction factor of 4.5 (Reddy \& Steidel 2004) they estimated that the
Universe should be characterised by a total metallicity density of
$\ozs = 1.84 \pm 0.34\times 10^{-5}$ at $z$=2.3. From the above numbers
they concluded that about 80\% of the metals in the Universe are missing,
in the sense that they are not directly associated with the above two
galactic components.

FSB05 suggested that the ``missing metals'' are associated with
galaxies, mainly residing in their ``hot'' gas halos, having 
been deposited there by star-burst driven super-winds (see also
Pettini 2004). This hypothesis
seems reasonable, given that outflow velocities of 300-400 km/s are 
routinely inferred from the spectra of LBGs (e.g., Pettini \etal 2001,
Shapley \etal 2003).
Moreover, semi-analytical as well as fully hydrodynamical models of
galaxy formation, based on Cold Dark Matter (CDM), require such ``feedback'' 
in order to enable the formation of realistic galaxies, solving 
the ``over-cooling'', ``angular momentum'', ``missing satellites'' and 
other problems (e.g., Sommer-Larsen \etal 1999, Cole \etal 2001, 
Thacker \& Couchman 2001, Sommer-Larsen \etal 2003, SGP03).

Considering the widely used probes of Ly$\alpha$ forest metal absorption
in QSO spectra, 
\hbox{C\,{\sc iv}}  and \hbox{O\,{\sc vi}}  (e.g., Songaila 2001, Bergeron \etal 2002, Carswell \etal 2002, Boksenberg, Sargent \& Rauch 2003,
Schaye \etal 2003, Simcoe \etal 2004, Aracil \etal 2004, Scannapieco \etal 2005; and Bergeron \& Herbert-Fort 2005, Tripp \etal 2006), 
FSB05 inferred observed integrated \hbox{C\,{\sc iv}}  and \hbox{O\,{\sc vi}}  column densities 
over the
redshift range 1.7$<$$z$$<$3.8, and the corresponding cosmic average 
densities of these. Assuming a two-phase intergalactic medium
(IGM), consisting of a ``cold'' part ($T$$\sim$10$^4$K), responsible for
the hydrogen Ly$\alpha$ forest absorption, and a ``hot'' component
($T$$\sim$10$^6$K), they showed, using a number of simplifying
assumptions, that more than 90\% of the metals can reside in the hot
component, without violating the \hbox{C\,{\sc iv}}  and \hbox{O\,{\sc vi}}  Ly$\alpha$ forest 
metallicity constraints (and with the cold phase as the major contributor
to the \hbox{C\,{\sc iv}}  and \hbox{O\,{\sc vi}}  column densities). This result follows provided that 
$T_{hot}\sim$10$^{5.5}$K, and the density of the hot metal-enriched gas
is only factors of a few times the mean cosmic baryonic density, which, e.g.,
at $z$=3, corresponds to a hydrogen number density of 
$n_{\rm{H}}$$\sim$1.2$\times$10$^{-5}$cm$^{-3}$. FSB05 suggest this metal
containing gas to be identified as wind-blown galaxy halo gas.

The low luminosity galaxies at high redshift are traced  by the DLAs (e.g.
Fynbo \etal  1999;  Haehnelt \etal  2000, Schaye 2001a; M\o ller \etal 2002).
From very early on it was found that DLAs contain different regions with very
different temperatures and ionisation states (Turnshek et  al. 1989).  The DLA
metal content discussed by FSB05 only accounts for the cold,
mainly neutral component.  The \hbox{C\,{\sc iv}} and \hbox{Si\,{\sc iv}}
lines in DLAs corresponds to a warmer  phase with more turbulent kinematics.
The cross-section for strong \hbox{C\,{\sc iv}}  and \hbox{Si\,{\sc iv}}
absorption is much larger than for DLAs consistent with the picture  that this
gas is located in an extended wind-blown halo around the stellar  components
(Petitjean \& Bergeron 1994, see also Adelberger \etal 2003, 2005, Porciani \&
Madau 2005, Scannapieco 2005).  An  even
hotter component, traced by \hbox{O\,{\sc vi}}  and \hbox{N\,{\sc v}}
(temperatures of order $3\times10^5$  K), has recently been identified in DLAs
(Fox \etal 2007).  Fox \etal  find that if the temperature of the
\hbox{O\,{\sc vi}}  bearing gas is 10$^6$ K or higher,  then this hot phase can
contribute significantly to the metal mass budget --- qualitatively consistent
with the proposal of FSB05.

Bouch\'e \etal (2006a,b) discussed the missing metals problem at
$z$$\sim$2.2-2.5 and found that the problem is not quite as severe
as at $z$$\sim$3. Counting metal contributions from stars in BX
galaxies (the equivalent of LBGs at $z$=2.2; e.g., Adelberger \etal 2004)
and DRGs (``distant red galaxies''; e.g., Franx \etal 2003, van Dokkum \etal
2003), as well as from gas in SMGs (sub-millimeter galaxies; e.g.,
Blain \etal 2004, Greve \etal 2005) and DLAs (e.g., Pettini \etal 2003)
they could account for about 1/3 of the metals expected. The BX
galaxies contain the largest fraction of the identified metals,
about 55\%. Moreover, Bouch\'e \etal (2007) found that another about 1/3
of the metals expected can be identified in the IGM at such redshifts.

Recently,
Dav\'e \& Oppenheimer (2006, DO06) modeled the enrichment history of the 
Universe and address the missing metals problem using a fully numerical 
approach. They perform a moderate resolution cosmological hydro/gravity 
simulation of a cubic region of the Universe of 32 $h^{-1}$Mpc box size. 
Their simulation
invokes metallicity and UV background dependent radiative cooling, 
star-formation and chemical evolution, in the instantaneous recycling
approximation, and based on the Salpeter IMF. The simulation resolves
galaxies of stellar masses down to $\sim$10$^9$ \Msol, corresponding
to a V-band absolute magnitude $M_{\rm{V}}$$\sim-20$ at $z$$\sim$3.
Due to the limited resolution, the authors adopt a parameterized description
of star-burst driven outflows in the form of a ``momentum-driven'' wind,
found by Oppenheimer \& Dav\'e (2006, OD06) to yield the best match of the
results of their simulations to various observational data.

DO06 quantify their results in terms of cosmic metal fractions in five phases:
a) stars, b) star-forming gas, c) halo gas (gas inside of the virial radius
of galaxy halos, which is not star-forming), d) shocked IGM (gas outside
of galaxy halos of $T$$\ge$3$\times$10$^4$ K), and e) diffuse IGM (gas outside
of galaxy halos of $T$$<$3$\times$10$^4$ K). At $z$$\sim$3, they find that
the dominant
metal phase is the diffuse IGM, which contains about 40\% of the cosmic
metals, with the other four phases containing approximately equal fractions 
of about 15\%. Hence, DO06 find that indeed only about 20\% of the metals
in the $z$$\sim$3 Universe reside in stars, while the ``missing'' about
80\% are located in the gas phase. Moreover, they find that the hot
halo gas is not the dominant, metal containing phase. Instead, a large
fraction of the missing metals are ``hidden'' in the diffuse IGM.
DO06 suggest that this is possible, in relation to observations of, e.g.,
low density IGM \hbox{C\,{\sc iv}} abundances, because the diffuse gas is somewhat hotter 
than
assumed in previous estimates, which at gas over-densities of
$\delta=\rho_{\rm{gas}}/\bar{\rho}_{\rm{gas}}$$\sim1-100$ implies larger
\hbox{C\,{\sc iv}} to C ionisation corrections, and hence that larger amounts of
C can be ``hidden'' in this phase.

In this paper we take another approach from that of DO06
(see also Calura \& Matteucci 2004, 2006). 
As we will show in the paper, due to resolution limitations
DO06 likely fail to account for about half of the
metals in the Universe produced by $z$$\sim$3, on top of which adds
cosmic variance effects, given the relatively small computational box
of DO06. To undertake simulations of larger cosmological volumes, at the
same time probing 8-10 magnitudes deeper (Sec.~\ref{sect:cosmic}) is
at present, as well as in any foreseeable future, computationally
prohibitive. Using K-band observations of a sample of LBGs, 
Shapley \etal (2001) determined the $z$$\sim$3 rest-frame V-band galaxy 
luminosity function. In Sommer-Larsen \& Fynbo (2007, paper~I) we correct 
the Shapley \etal luminosity function for extinction and other effects, to obtain a ``true''
$z$$\sim$3 Lyman break galaxy LF. Subsequently, cosmological 
high-resolution hydro/gravity simulations of the formation and evolution
of individual galaxies are combined with the corrected galaxy luminosity 
function
to obtain estimates of the average cosmic density of metals (in particular
oxygen) residing in stars at $z$=3. We consider models based on three different 
stellar IMFs, spanning almost a factor of five in chemical yield as well
as thermal/kinetic energy feedback from supernova type II (SNII) explosions. 
These are the Kroupa (1998, K98) IMF, a typical IMF suited for chemical evolution 
models of the Solar Neighbourhoohd, e.g., Boissier \& Prantzos
(1999), the ``standard'' Salpeter (1955, S) IMF and
the Arimoto-Yoshii (1987, AY) IMF, which is well suited for describing the
chemical evolution of elliptical galaxies. The simulations have sufficiently
high resolution to allow a two-phase modeling of the star-forming
interstellar medium (ISM), consisting of a ``cold'' $T$$\sim$10$^4$ K
star-forming phase and a ``hot'' $T$$\sim$10$^5$- 10$^6$ K phase, intermixed
with the cold gas. Star-bursts drive galactic winds by depositing
thermal energy from multiple SNII explosions in the gas, part of which is 
subsequently converted into kinetic energy self-consistently by the 
hydro-code.

In paper~I we conclude that for none of 
the IMFs is it possible to reconcile the amount of oxygen locked in stars with
the amount predicted from the (observed) cosmic UV luminosity density history,
hence confirming the ``missing metals problem'' as stated by Pettini (1999)
and FSB05. 
 
In this paper we combine the high-resolution galaxy formation models
with the V-band luminosity function to determine the cosmologically
averaged amount and properties of $z$=3 gas-phase metals, mostly
focusing on oxygen, but also on carbon, in particular in relation
to QSO absorption line determinations of the cosmic \hbox{C\,{\sc iv}} 
density, as well as iron, in relation to DLA abundances. 
Combining the results obtained with those of paper I
the total cosmic metal distribution is obtained for the three IMFs
considered. Comparing in turn these results to inferences from
the observed cosmic UV luminosity density history allows us to
significantly constrain the ``true'' 
cosmic stellar IMF.
Finally, relations to the ``missing metals problem'', as well
as results obtained by other authors, are discussed.

The paper is organised as follows: In section~\ref{sect:UV} we derive
constraints on cosmic metal production from the cosmic UV luminosity 
density history, in 
section~\ref{sect:simulations} we briefly describe the hydro/gravity
galaxy formation simulations, and in sections~\ref{sect:approach} and
\ref{sect:cosmic} we present
the approach used in this paper to determine the temperature distribution
of the cosmic gas-phase metals. Section~\ref{sect:CIV} presents
results on gas-phase \hbox{C\,{\sc iv}} abundances, and, relating these to observations
of \hbox{C\,{\sc iv}} absorption lines in QSO spectra, we derive a constraint on the
cosmic IMF. In section~\ref{sect:comb} we combine the results of paper I
and those obtained here to derive the total cosmic metal distribution.
Comparing this to what is obtained from the cosmic UV luminosity density 
history we obtain an additional constraint on the cosmic stellar IMF.
In section~\ref{sect:mm} we relate our results to the ``missing metals
problem'' and compare them to those of other workers in the field,
in section~\ref{sect:numres} we demonstrate that our results are
robust to changes of the numerical resolution and, finally,
section~\ref{sect:conclusions} summarises our conclusions.

In the paper we assume the flat $\Lambda$ cosmology, with $\Omega_M$ = 0.3
and  
$\Omega_{\Lambda}$=0.7, and $H_0=100h$ km/s/Mpc, with $h$=0.7, unless it 
is explicitly stated otherwise.

\section{Constraints on metal production from the cosmic UV
luminosity density history}
\label{sect:UV}
%
\begin{figure}
\psfig{file=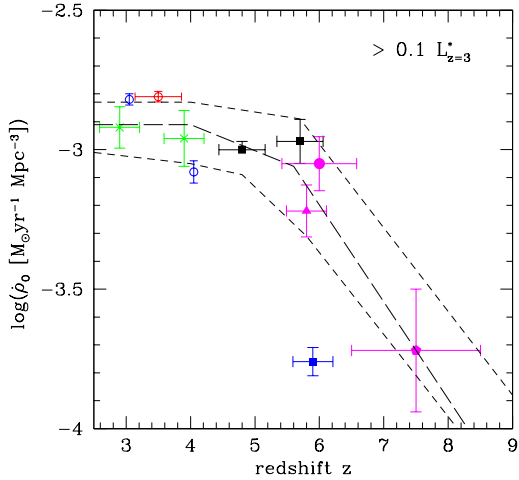,width=0.43\textwidth}
\caption{Cosmic oxygen enrichment history based on observations of
high-$z$ Lyman break galaxies. The data points shown are
based on Steidel \etal (1999; {\it green crosses}),
Schiminovich \etal (2005; {\it open red dot}),
Sawicki \& Thompson (2006; {\it open blue dots}),
Giavalisco \etal (2004; {\it black squares}),
Bunker \etal (2004; {\it blue square}),
Bouwens \etal (2004a; {\it magenta triangle}),
Bouwens \etal (2006; {\it magenta circle}),
Bouwens \etal (2004b; {\it magenta pentagon} --- note that the
latter only represents $L\ge$0.3$\times L^*_{UV,z=3}$). The
``maximum'' model is shown by the upper short-dashed curve, the
``median'' model by the long-dashed, and the ``minimum'' model
by the lower short-dashed.}
\label{fig:OFR}
\end{figure}
Constraints on metal production from the cosmic UV luminosity density 
history are discussed in paper~I, but for coherence we present the main
discussion in the following as well.

The luminosity of young galaxies at rest-frame $\lambda\sim$1500 {\AA} is a 
measure of the rate of formation of massive stars (mainly O and B type)
as shown by, e.g., Madau \etal (1998). Moreover, $\alpha$-elements like
oxygen are almost exclusively produced in such massive stars, and hence
there is a direct link between the cosmic average oxygen production rate
density and the average UV luminosity density, which is essentially
independent of the stellar initial mass function --- see, e.g., Pettini
(1999). To estimate the cosmic oxygen production rate density as a function
of redshift we use various recent estimates of the average 
cosmic star formation rate. The estimates have been obtained from
Steidel \etal (1999), Bouwens \etal (2004a,b), Bunker \etal (2004),
Giavalisco \etal (2004), Schimonovich \etal (2005), Bouwens \etal
(2006) and Sawicki \& Thompson (2006).  
The estimates used are all based on assuming a ``standard''
Salpeter IMF in converting from UV luminosity to star formation
rate. The oxygen (mass) yield of a Salpeter IMF is 0.01 (e.g., Lia \etal
2002a,b) --- hence in Fig.~\ref{fig:OFR} we have multiplied the published
SFRs by 0.01 to obtain the (essentially) IMF independent oxygen
production rate densities. The estimates have moreover been corrected to
correspond to (at any redshift) the UV luminosity density of galaxies 
brighter than 0.1$\times L^*_{UV,z=3}$ following an approach similar 
to that of Bouwens \etal (2006; note though that the 
$z$$\sim$7.5 data point of Bouwens \etal 2004b only 
represents $L\ge$0.3$\times L^*_{UV,z=3}$).
Finally, the values shown in the plot
result from multiplying the observed values by a dust-correction factor 
of 5.5. This value is intermediate between the $z\sim$3 values of 4.5
and 6.5 suggested by Reddy \& Steidel (2004) and Dahlen \etal (2007),
respectively (but see below).

The cosmic average oxygen density at $z$=3 can now be obtained by
integrating the oxygen production rate density, $\dot{\rho}_O(z)$, 
from $z$=3 and back in time. We shall consider three models for 
$\dot{\rho}_O(z)$: the ``median'' model, the ``maximum'' model, and
the ``minimum'' model. The median model is obtained by calculating
the median values of $\dot{\rho}_O$ in the bins $z$=3 to 4 and $z$=
4 to 6 (assigning equal weight to each data point, but excluding the 
$z$=5.9 value of Bunker \etal 2004; see below),
and connecting these values with the $z\simeq$7.5 value of Bouwens \etal
(2004b). This model is shown in Fig.~\ref{fig:OFR} by the long-dashed
curve. The maximum model is obtained by connecting the $z$=3 to 4
median value with the Giavalisco \etal (2004) $z\simeq$5.7 value, and the
Bouwens \etal (2004b) $z\simeq$6.0 value, and 
multiplying the resulting ``upper envelope'' by a factor 6.5/5.5
to maximize also the dust correction. This model is shown by the upper
short-dashed curve in Fig.~\ref{fig:OFR}. Finally, the minimum model
is obtained by connecting the $z$=3 to 4 median value with the Giavalisco 
\etal (2004) $z\simeq$4.8 value, the Bouwens \etal (2004a) value and the
Bouwens \etal (2006) $z\simeq$7.5 value, and multiplying this 
``lower envelope''
by a factor 4.5/5.5 to minimize the dust correction (but see below).
This model is shown by the lower short-dashed curve in Fig.~\ref{fig:OFR}.

Assuming a flat space world model, the $z$=3 cosmic average oxygen
density can now be evaluated as
\[
\rho_O(z\mathrm{=3}) = -\int_{t(z=3)}^0 \dot{\rho}_O(z(t))~dt~~=
\]
\begin{equation}
\frac{1}{H_0}
\int_{3}^{\infty} \frac{\dot{\rho}_O(z)}{(1+z) \sqrt{(1+z)^2 (1+z\Omega_M) -
z (2+z) \Omega_{\Lambda}}}~dz,     
\end{equation}
where values of $H_0=100h$ km/s/Mpc, with $h$=0.7, $\Omega_M$ = 0.3 and
$\Omega_{\Lambda}$=0.7 are assumed in this paper. Using eq.\,1 we obtain
$\rho_O(z$=3) = 0.32, 0.42 and 0.56 $\times 10^7$ \Msol/($h^{-1}$Mpc)$^3$
for the ``minimum'', ``medium'' and ``maximum'' models, respectively.
These numbers represent an integral constraint that any successful model
of galaxy formation must meet. As will be shown in section~\ref{sect:cosmic}
this enables us to strongly constrain the ``cosmic'' stellar IMF at $z\ge$3.
\begin{figure}
\psfig{file=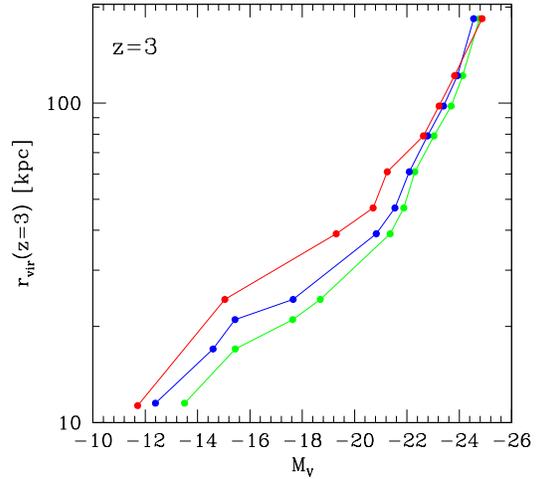,width=0.43\textwidth}
\caption{Relation between virial radius and absolute V-band (rest-frame)
magnitude at
$z$=3 for the K98 (green), Salpeter (blue) and AY (red) IMFs.}
\label{fig:rvir}
\end{figure}

In determining the ``median'' and ``minimum'' models we excluded the
$z$=5.9 value of Bunker \etal (2004). This is simply done on the basis
of the large discrepancy between this value, and all other $z$=4 to 6
values. To indicate the effect of this, we determined an alternative
median model in which the Bunker \etal (2004) value is included in the
$z$=4 to 6 bin. This changed the median model estimate of $\rho_O(z$=3)
from 0.42 to 0.40 $\times 10^7$ \Msol/($h^{-1}$Mpc)$^3$, i.e. a $\sim$2 \%
change, and hence quite small effect, which, given all other uncertainties,
we shall ignore in the following.
 
Bouwens \etal (2006) propose that the $\lambda\sim$1350-1500 {\AA} 
extinction correction decreases with increasing $z$ from $z\sim$3.
Converting their proposed, extinction-corrected SFR($z$) to a
corresponding $\dot{\rho}_O(z)$ (including changing the UV luminosity
density from their limit of 0.04$\times L^*_{UV,z=3}$ to the limit
of 0.1$\times L^*_{UV,z=3}$ used here) results in a value of $\rho_O(z$=3) = 
0.33 $\times 10^7$ \Msol/($h^{-1}$Mpc)$^3$, hence within the bounds derived 
above. 
\begin{figure*}
\leavevmode
\psfig{file=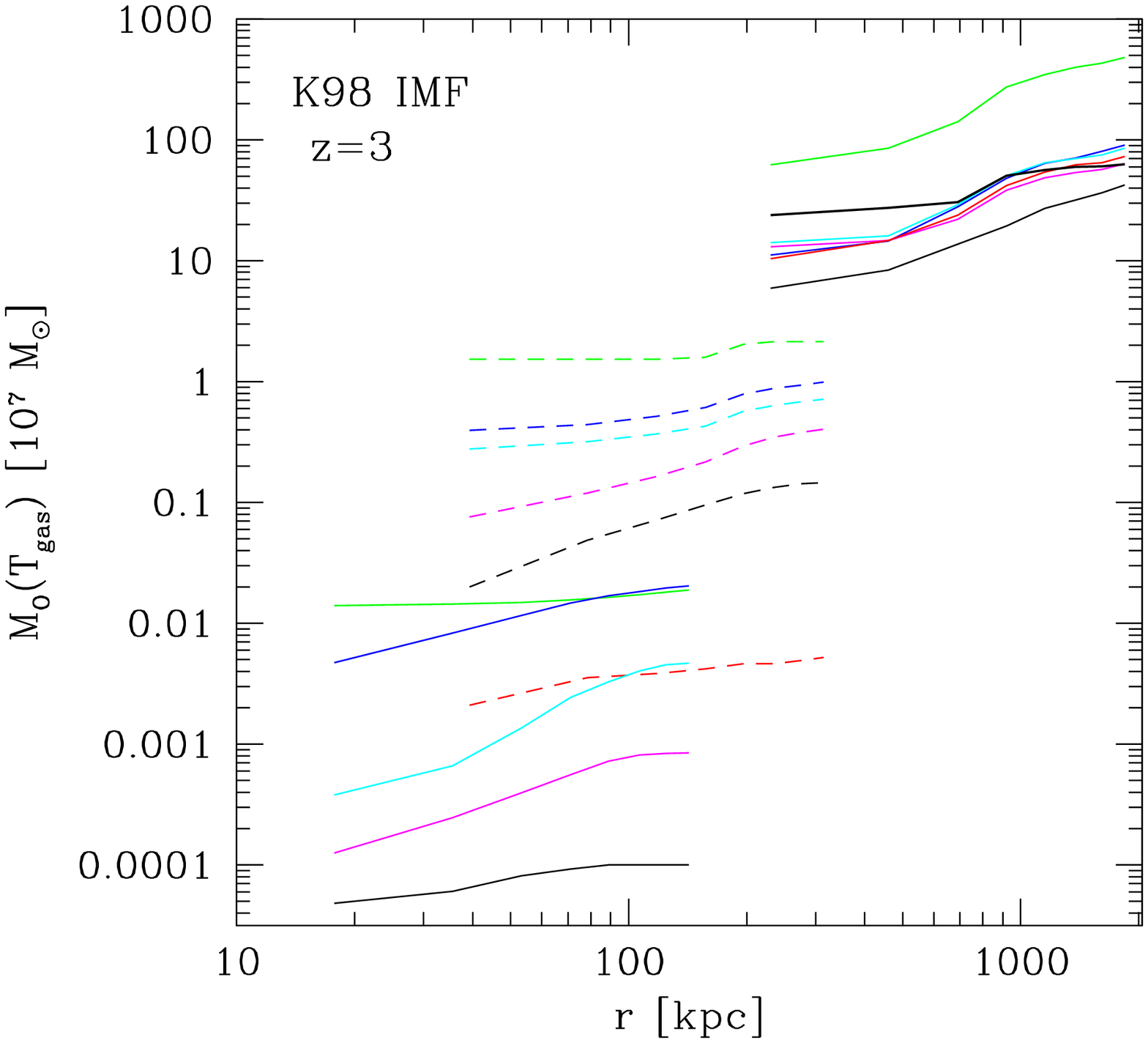,width=0.41\textwidth}
\psfig{file=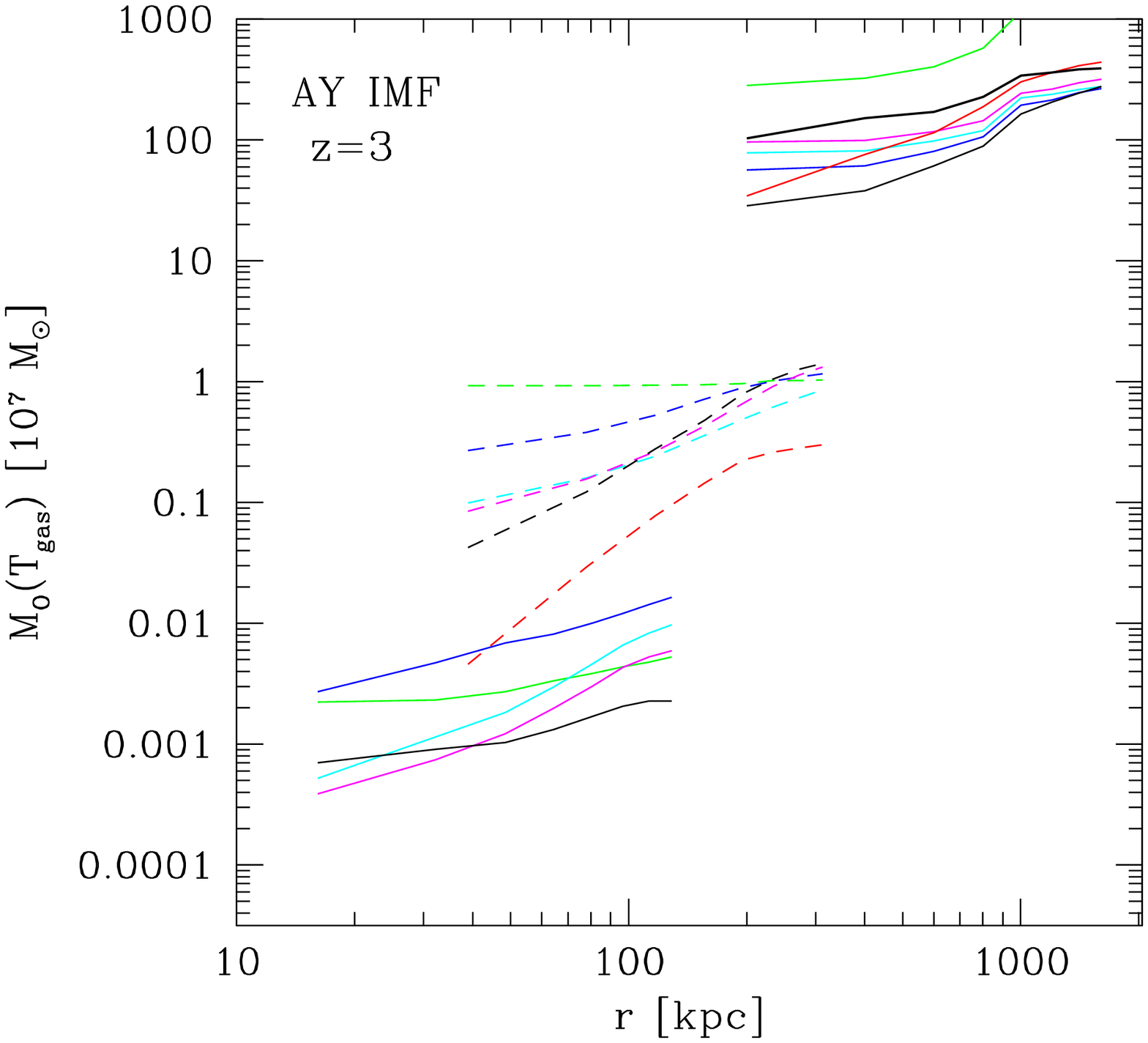,width=0.41\textwidth}

\caption{Cumulative mass of gas-phase oxygen with galacto-centric 
distance for the {\it same} three galaxies at $z$=3, simulated with the 
K98 IMF
(left) and the AY IMF (right). Gas temperature colour coding is
as follows: $\log(T)$=4: green, 4.5: blue, 5: cyan, 5.5: magenta, 6: black,
6.5: red, 7 black (thick line). For all galaxies the region from $r_{vir}$ to 8~$r_{vir}$
is shown. For the K98 IMF, the galaxies have V-band absolute magnitudes
of $-16.1$, $-21.4$ and $-25.3$. For the AY IMF, the corresponding numbers are 
$-13.8$, $-19.3$ and $-25.4$. The corresponding stellar masses are
5.8$\times 10^7$~\Msol, 7.3$\times 10^9$~\Msol~and 
3.4$\times 10^{11}$~\Msol~for the K98, and
6.4$\times 10^6$~\Msol, 5.2$\times 10^8$~\Msol~and 
2.5$\times 10^{11}$~\Msol~for the AY IMF.}
\label{fig:Omassr}
\end{figure*}

The integral constraint obtained in this section can also be expressed
in units of the critical density: we obtain 
$\Omega_O$($z$=3) = $\rho_O(z$=3)/$\rho_{\rm{crit}}$ = 0.81, 1.05 and 
1.41$\times 10^{-5}$ for the minimum, median and maximum models,
respectively. 

FSB05 obtained for the total density of heavy elements
in units of the critical $\Omega_Z$ = (1.84$\pm$0.34)$\times 10^{-5}$,
by integrating the observed cosmic star formation history (SFH, based on
a Salpeter IMF), as presented by Bouwens \etal (2004b), to $z$=2.3 and
applying a dust correction factor of 4.5. The above value translates into 
$\Omega_O$($z$=2.3) = (0.77$\pm$0.14)$\times 10^{-5}$. If the oxygen density
production rate models shown in Fig.~\ref{fig:OFR} are continued 
$z$=2.3, we would obtain 1.22, 1.55 and 2.01 $\times 10^{-5}$. The reason
for the apparent discrepancy is two-fold: a) we use a dust correction of
a factor of 5.5, and b) more importantly, the Bouwens \etal (2004b) SFH
was based on a UV luminosity to a limit of 0.3$\times L^*_{UV,z=3}$,
rather than the limit of 0.1$\times L^*_{UV,z=3}$ adopted in this work.
Using the $z\sim3$ UV luminosity function of Adelberger \& Steidel (2000)
we estimate the latter correction to be a factor of 1.5. Multiplying
this by a factor of 5.5/4.5 the findings of FSB05 would 
translate into about (1.4$\pm$0.3)$\times 10^{-5}$, 
in good agreement with our values above. 

Extremely dust-obscured galaxies such as high-$z$ SCUBA sources (e.g.,
Smail \etal 1997; Barger \etal 1998; Eales \etal 1999; Chapman
\etal 2005) are not included in the dust-corrected star formation
rate densities shown in Fig.~\ref{fig:OFR}.  Most such
galaxies have UV luminosities that heavily under-represent their star
formation rates and thus such galaxies are not properly included 
in the estimate of the oxygen production rate density.
Their numbers may be sufficiently large that their contribution is
significant, however we neglect this contribution for two reasons:
a) the contribution from such galaxies would be largest at $z$$\sim$2,
where the redshift distribution of SCUBA sources appears to peak 
(Chapman \etal 2005, Reddy \etal 2005), whereas we are concerned with the 
range $z$$\ga$3,
and b) more importantly, we are concerned with the oxygen production rate 
density history of Lyman Break Galaxies {\it only}, as we wish to compare
the integral constraint thus obtained to results of combining the observed
LBG $z$$\sim$3 optical rest-frame luminosity function with detailed,
high-resolution models of galaxy formation.

\begin{figure*}
\leavevmode
\psfig{file=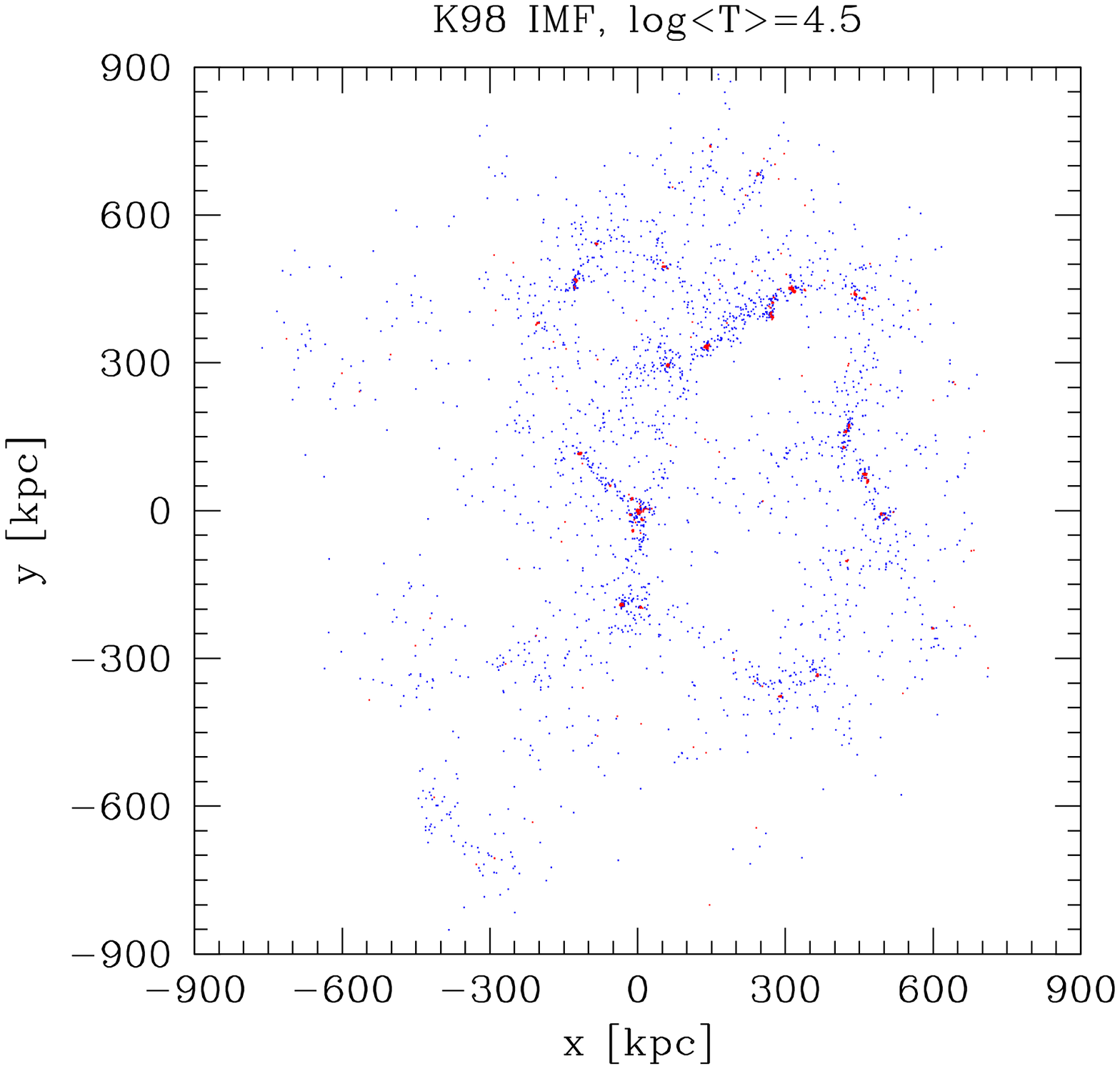,width=0.41\textwidth}
\psfig{file=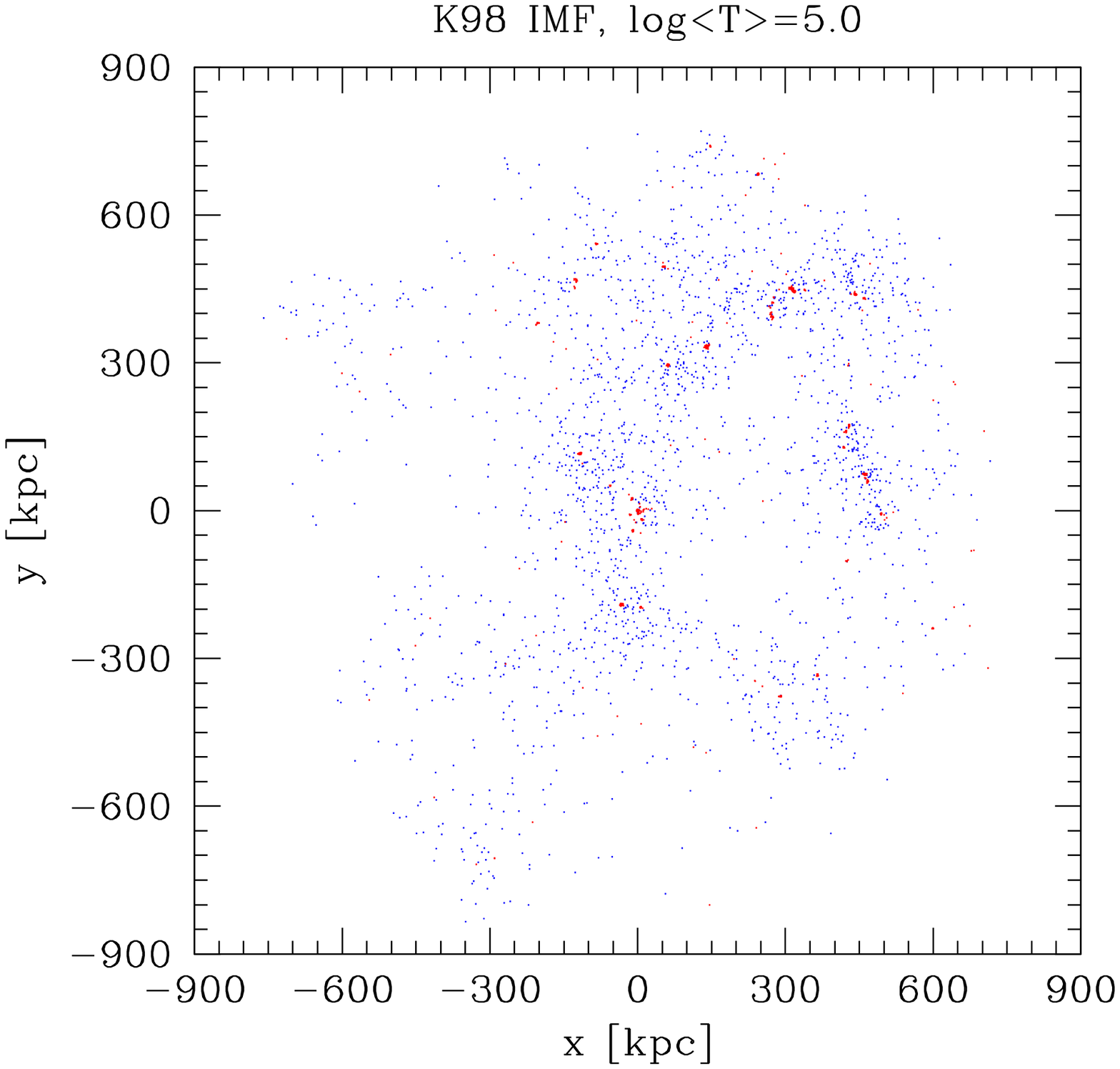,width=0.41\textwidth}

\leavevmode
\psfig{file=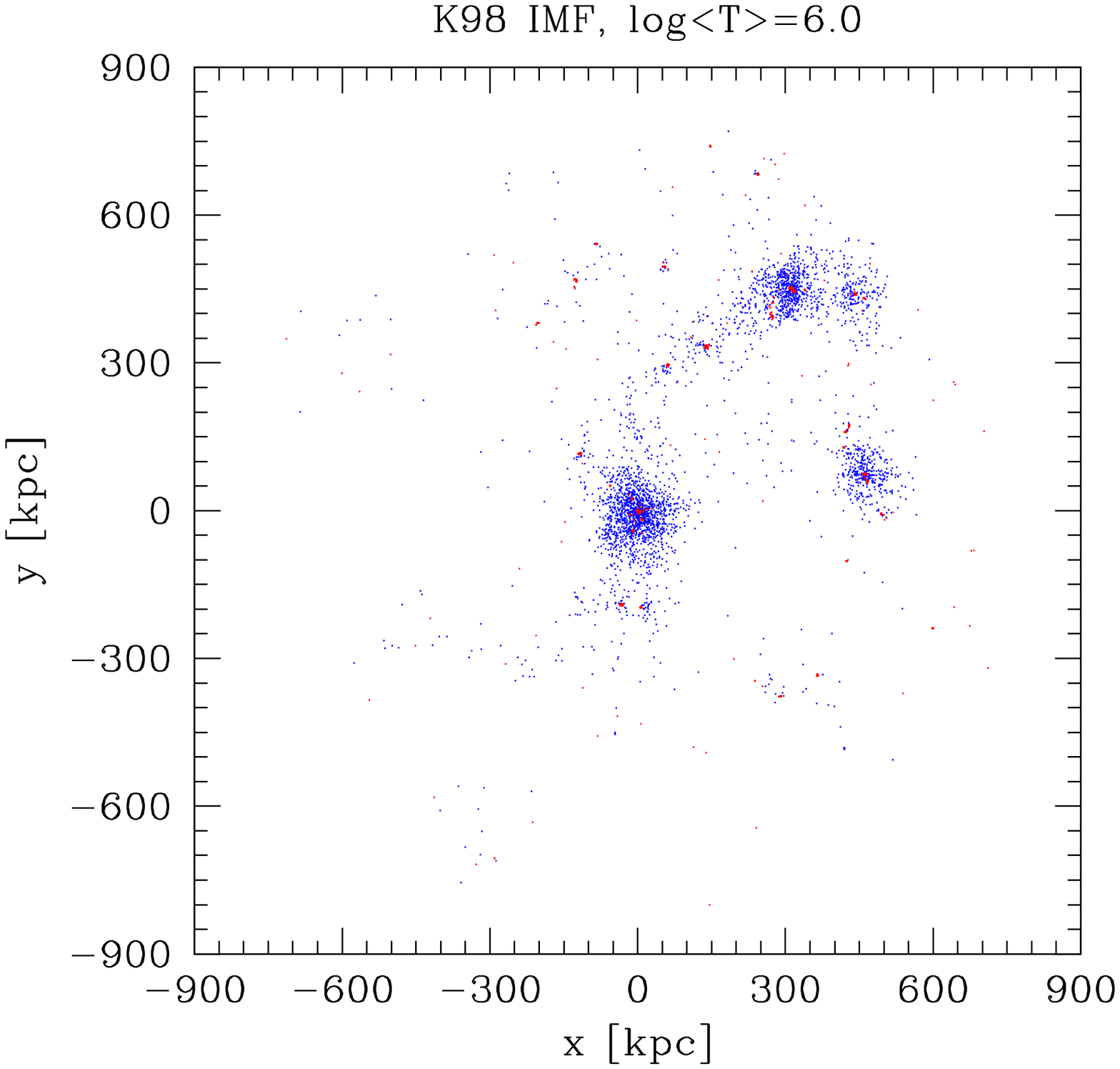,width=0.41\textwidth}
\psfig{file=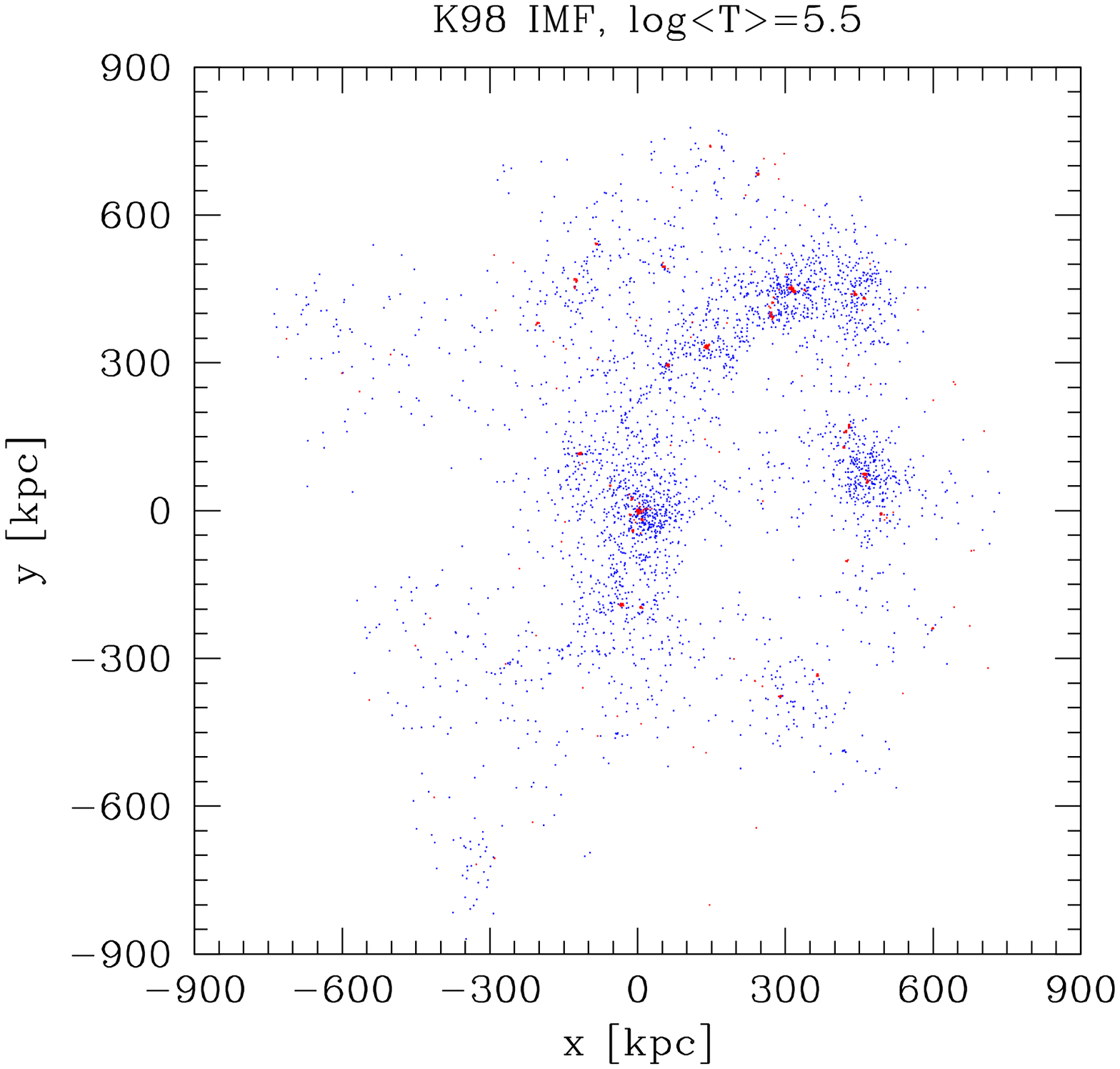,width=0.41\textwidth}
\caption{Spatial (projected) distributions of gas-phase oxygen in the 
``medium-density''
field galaxy region, at $z$=3, for the K98 simulation. 
Results for 5 different temperature bins,
corresponding to $\log(T)$=4, 4.5, .., 6, are shown. All temperature bins
are represented using about 5000 ``dots'', which for each temperature
bin correspond to equal amounts of oxygen mass. 
Hence, each figure illustrates directly
the spatial distribution of the gas-phase oxygen mass, {\it not} gas
mass. The cold-phase ($\log(T)$=4) is shown by red
dots in all 4 figures. In addition the $\log(T)$=4.5, 5, 5.5, and 6
phases are shown by blue dots in the corresponding panels.}
\label{fig:T}
\end{figure*}

\begin{figure*}
\leavevmode
\psfig{file=K6.0small.ps,width=0.41\textwidth}
\psfig{file=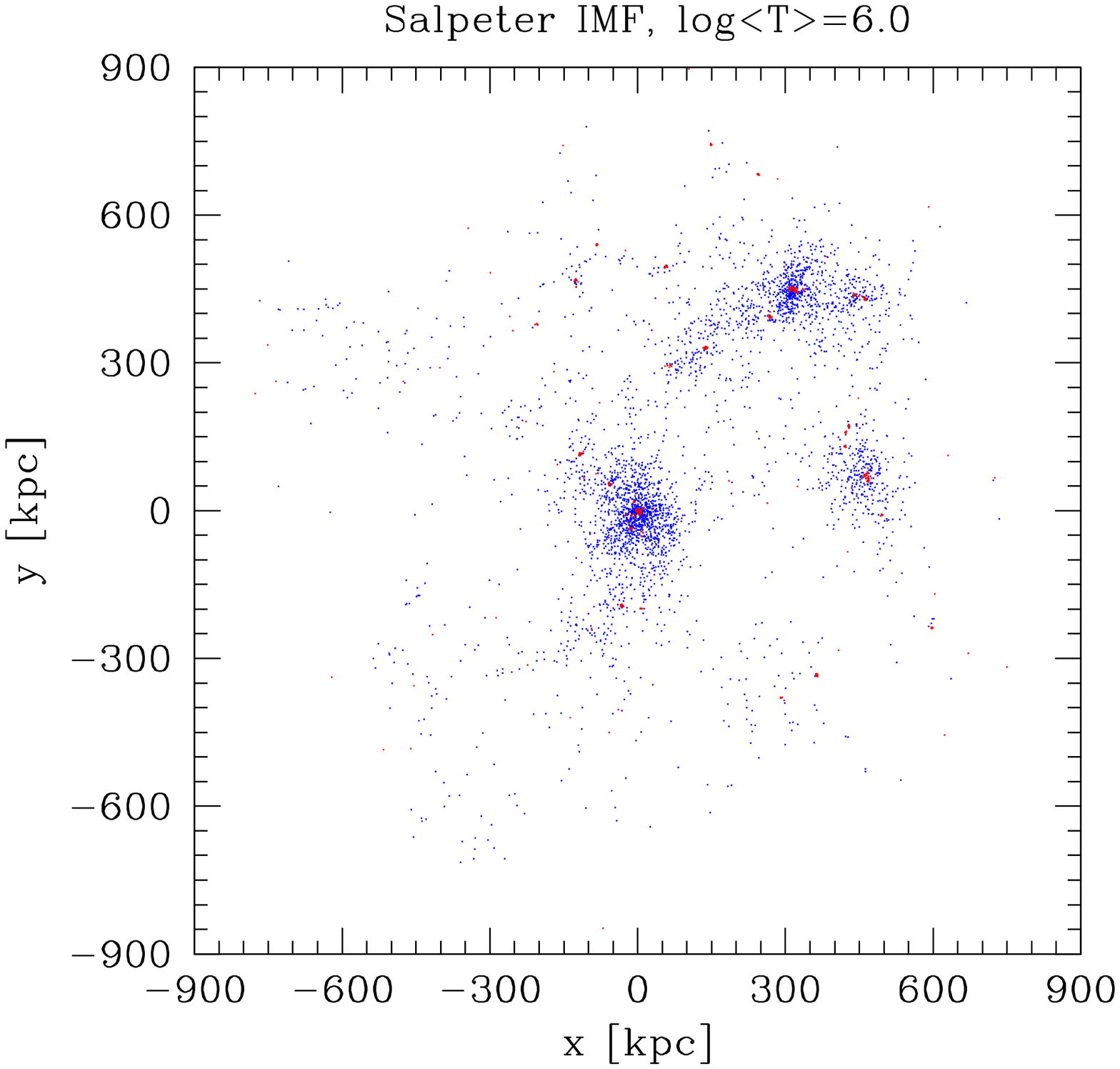,width=0.41\textwidth}

\leavevmode
\psfig{file=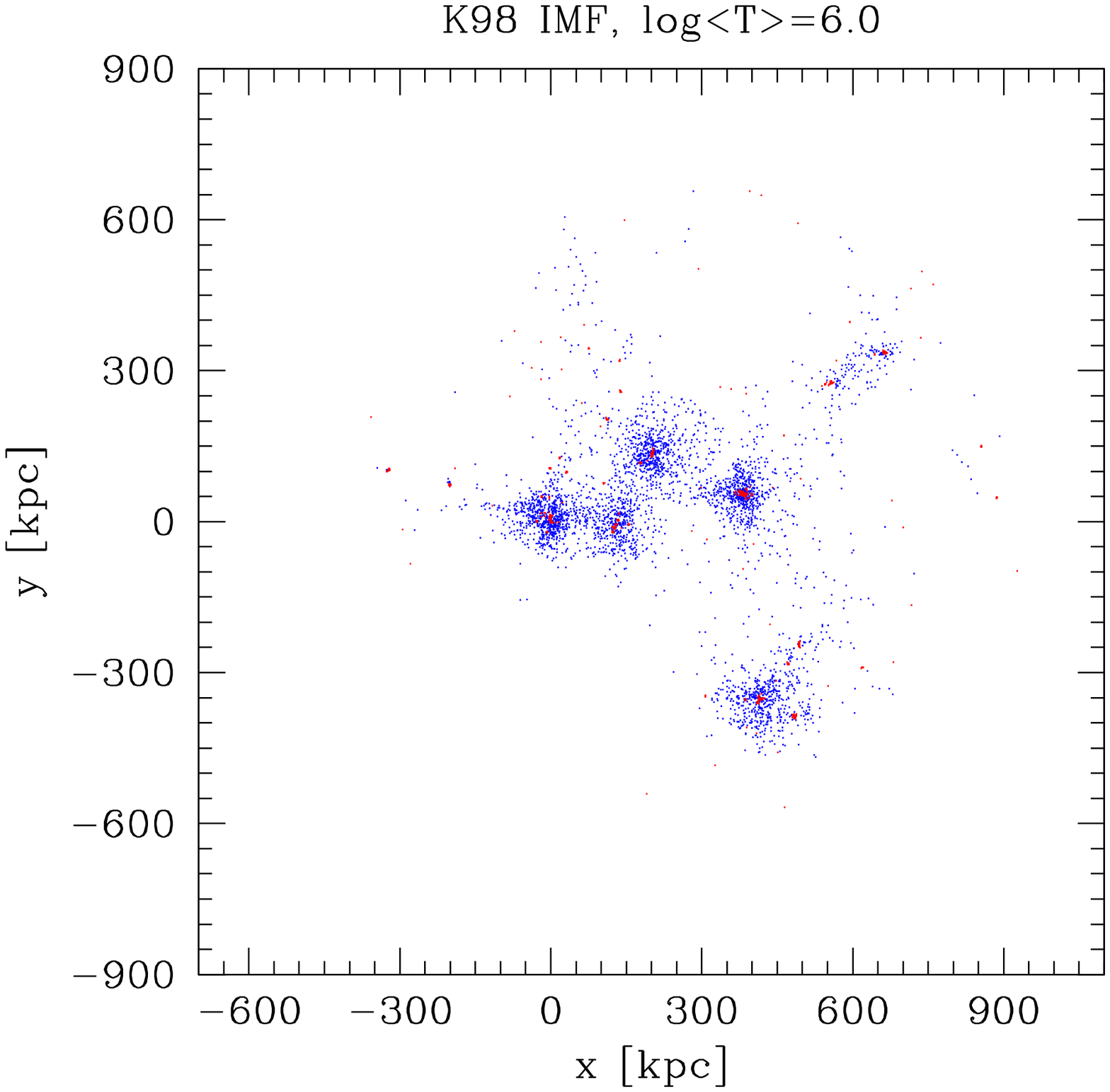,width=0.41\textwidth}
\psfig{file=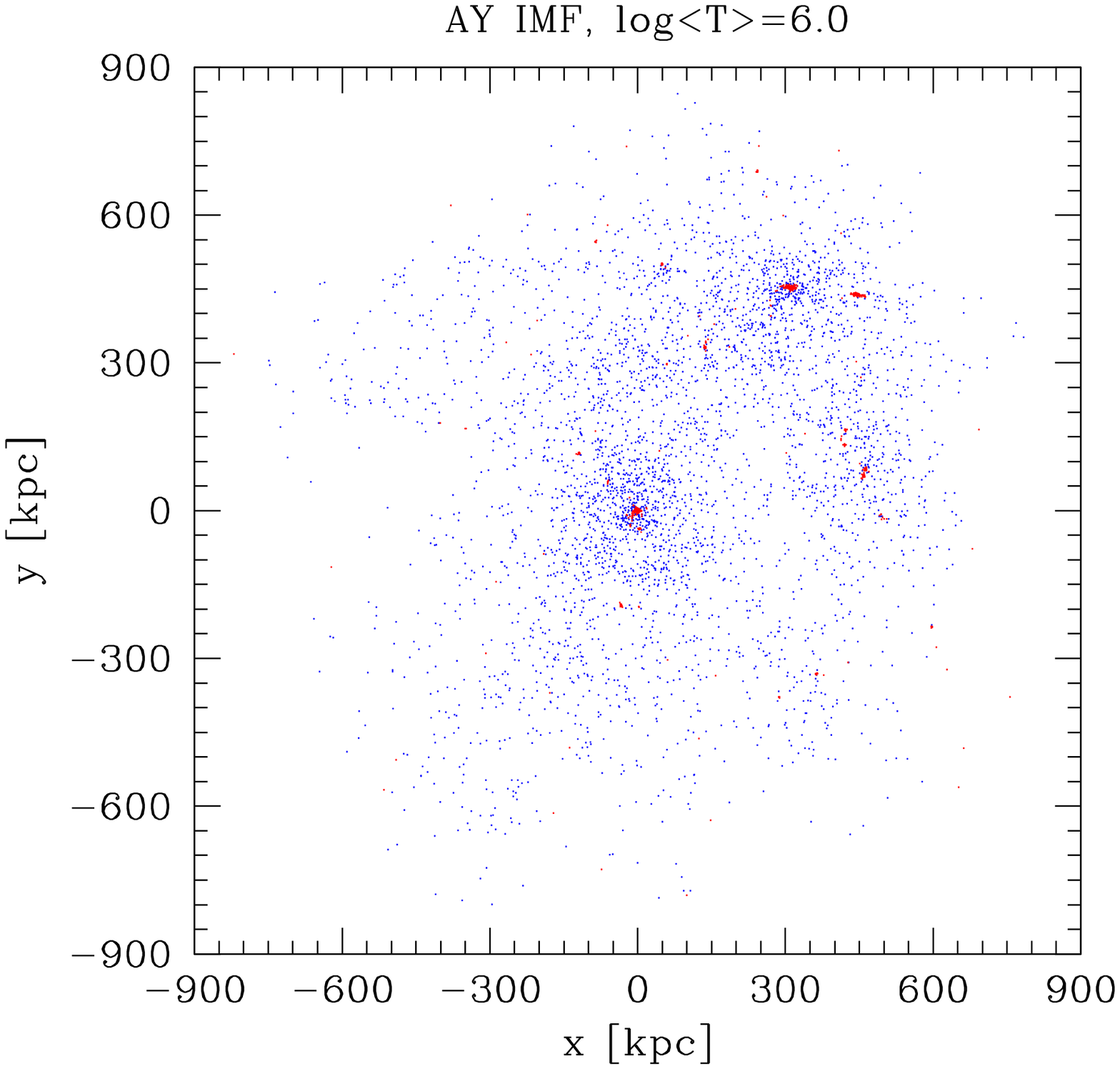,width=0.41\textwidth}
\caption{Spatial (projected) distributions of $\log(T)$=6 gas-phase 
oxygen in the ``medium-density''
field galaxy region, at $z$=3, for the K98 (top left),
Salpeter (top right), AY (bottom right) simulations. Also shown is
the result for a K98 simulation of a proto-elliptical region (bottom
left) --- see Fig.~\ref{fig:T} for more detail}
\label{fig:imfs}
\end{figure*}

\begin{figure*}
\leavevmode
\psfig{file=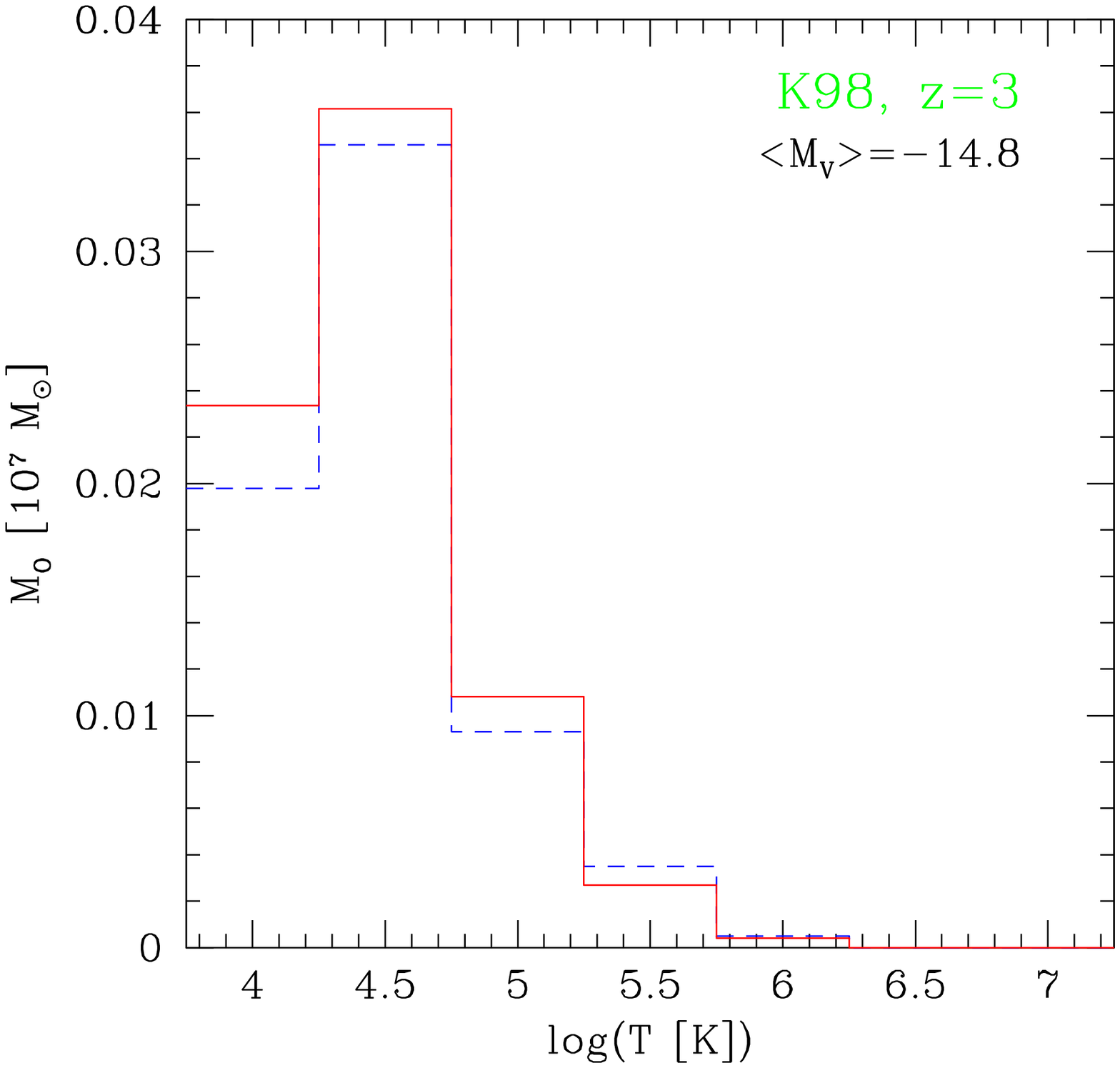,width=0.41\textwidth}
\psfig{file=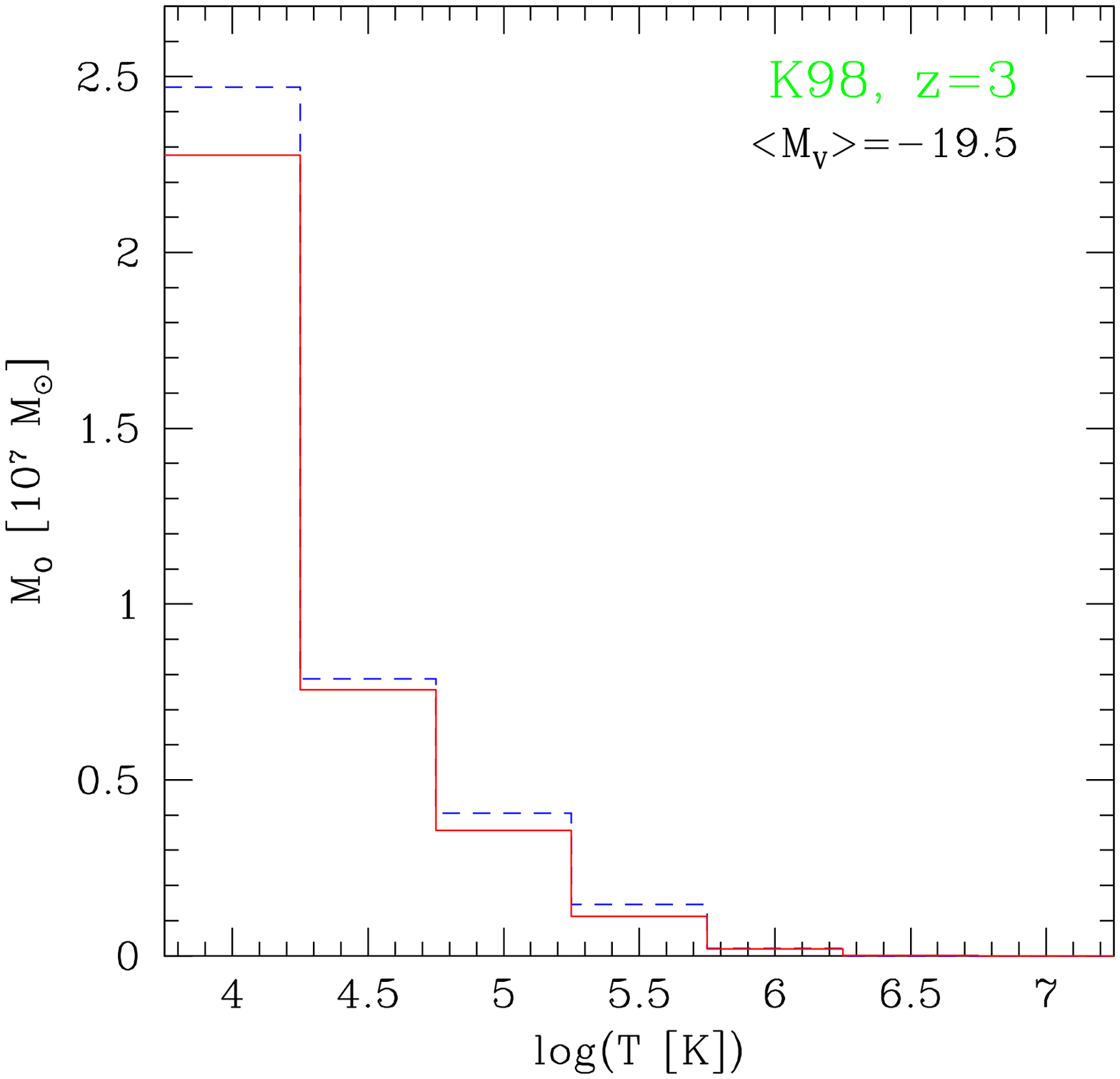,width=0.41\textwidth}

\leavevmode
\psfig{file=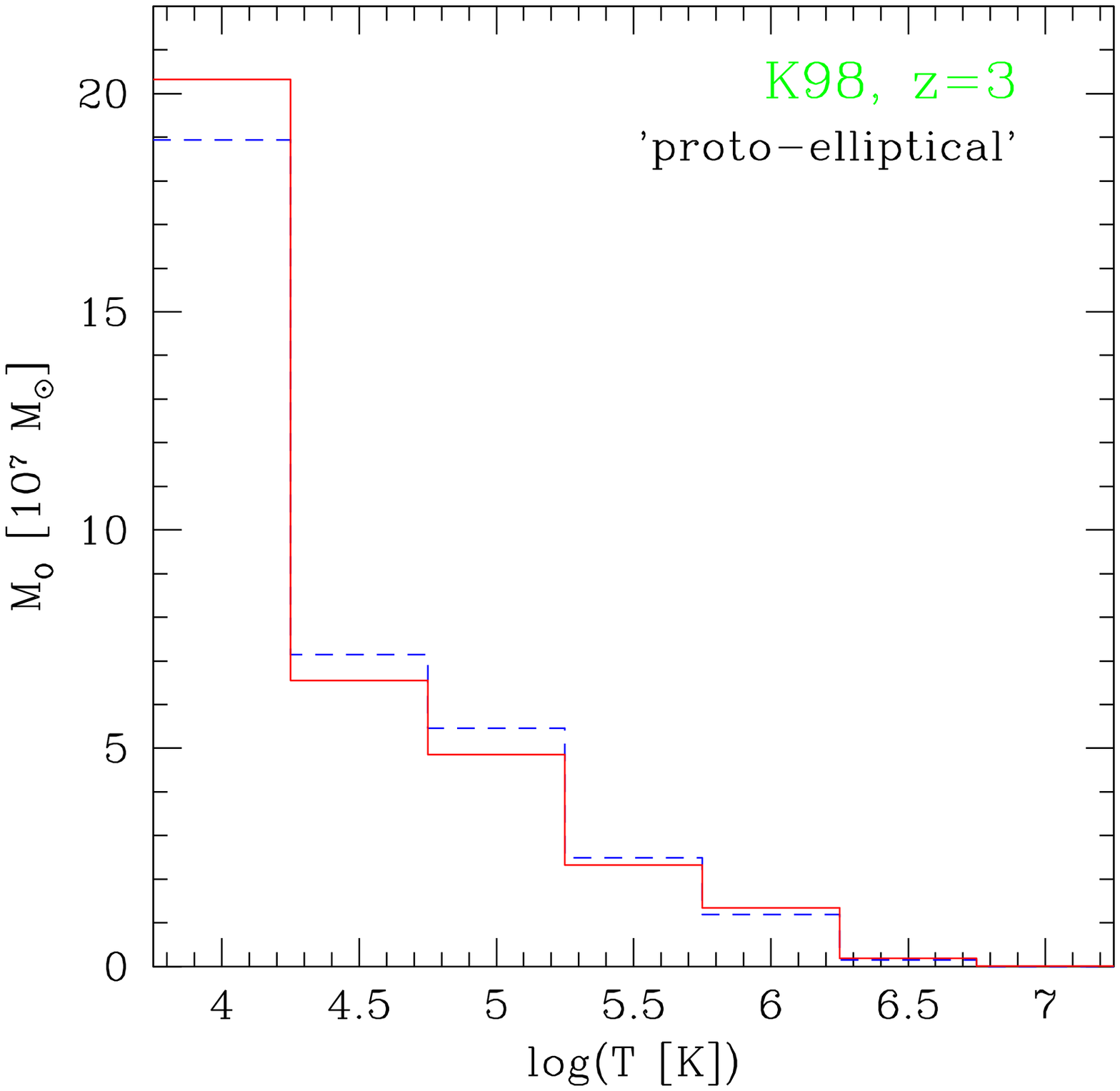,width=0.41\textwidth}
\psfig{file=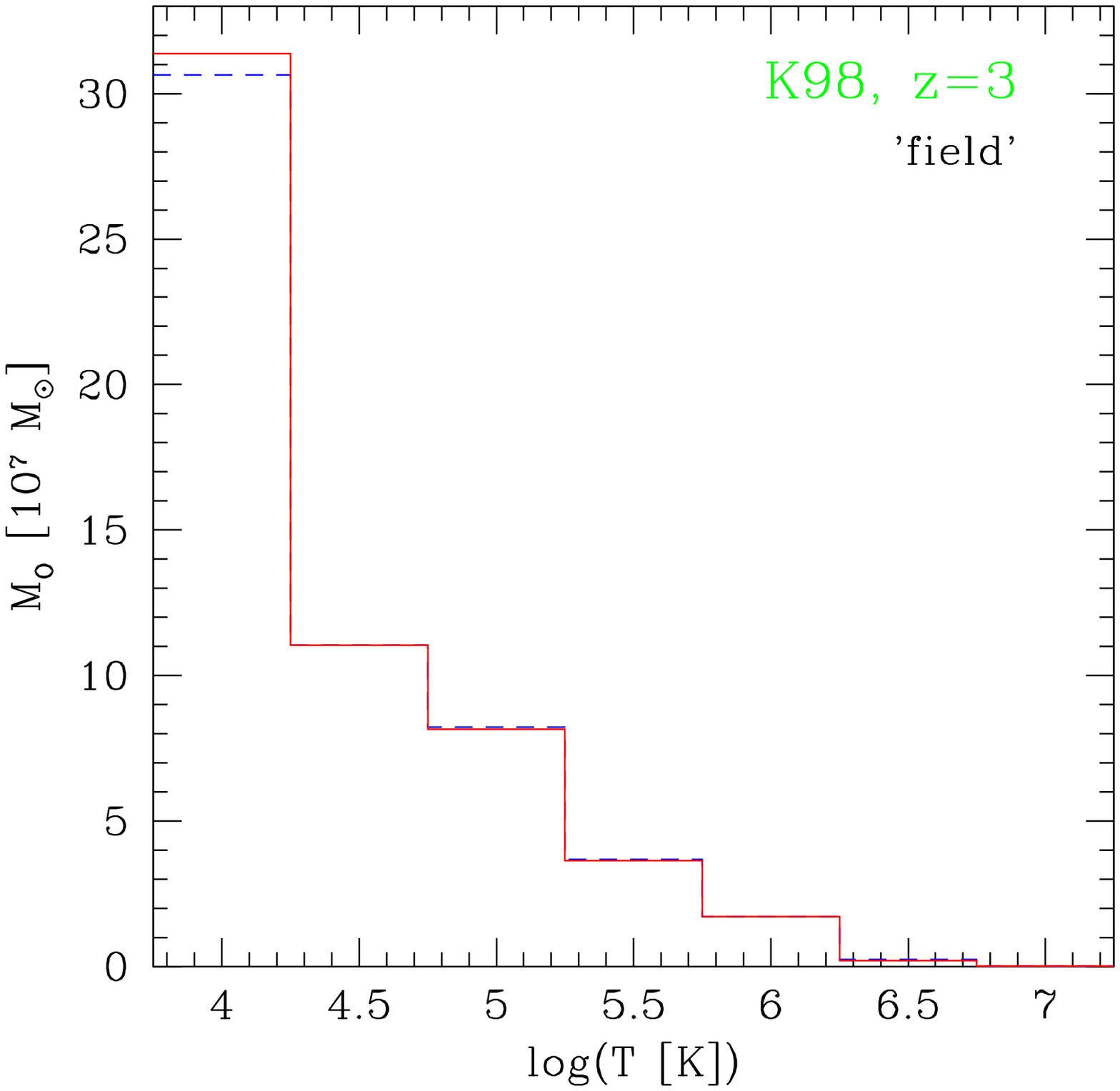,width=0.41\textwidth}
\caption{Comparison of predicted and actual gas-phase oxygen mass 
distributions for combinations of various regions simulated using the
K98 IMF at high resolution. The top left plot shows the result for
the combination of three regions, each containing a galaxy simulated
using the ``zoom-in'' technique plus a number of companions. The three
main galaxies have a mean $M_{\rm{V}}$ of $-14.8$. The red solid histogram
shows the actual oxygen mass distribution of the three regions
combined. The blue dashed histogram shows what the method outlined
in the text predicts on the basis of the combined galaxy luminosity
function of the three regions. The top right plot shows a similar comparison
for a combination of 5 high-resolution regions containing main galaxies
of an average $M_{\rm{V}}$ of $-19.5$. The bottom right plot is for the combination
of the low and medium density ``field'' regions described in the text,
containing a total of 103 resolved galaxies of $M_{\rm{V}}$ down to about 
$-22.5$.
Finally, the bottom left plot shows the results for a higher density
``proto-elliptical'' region, containing 60 galaxies of $M_{\rm{V}}$ down to 
about -22.5.}
\label{fig:gmpK}
\end{figure*}

\begin{figure*}
\leavevmode
\psfig{file=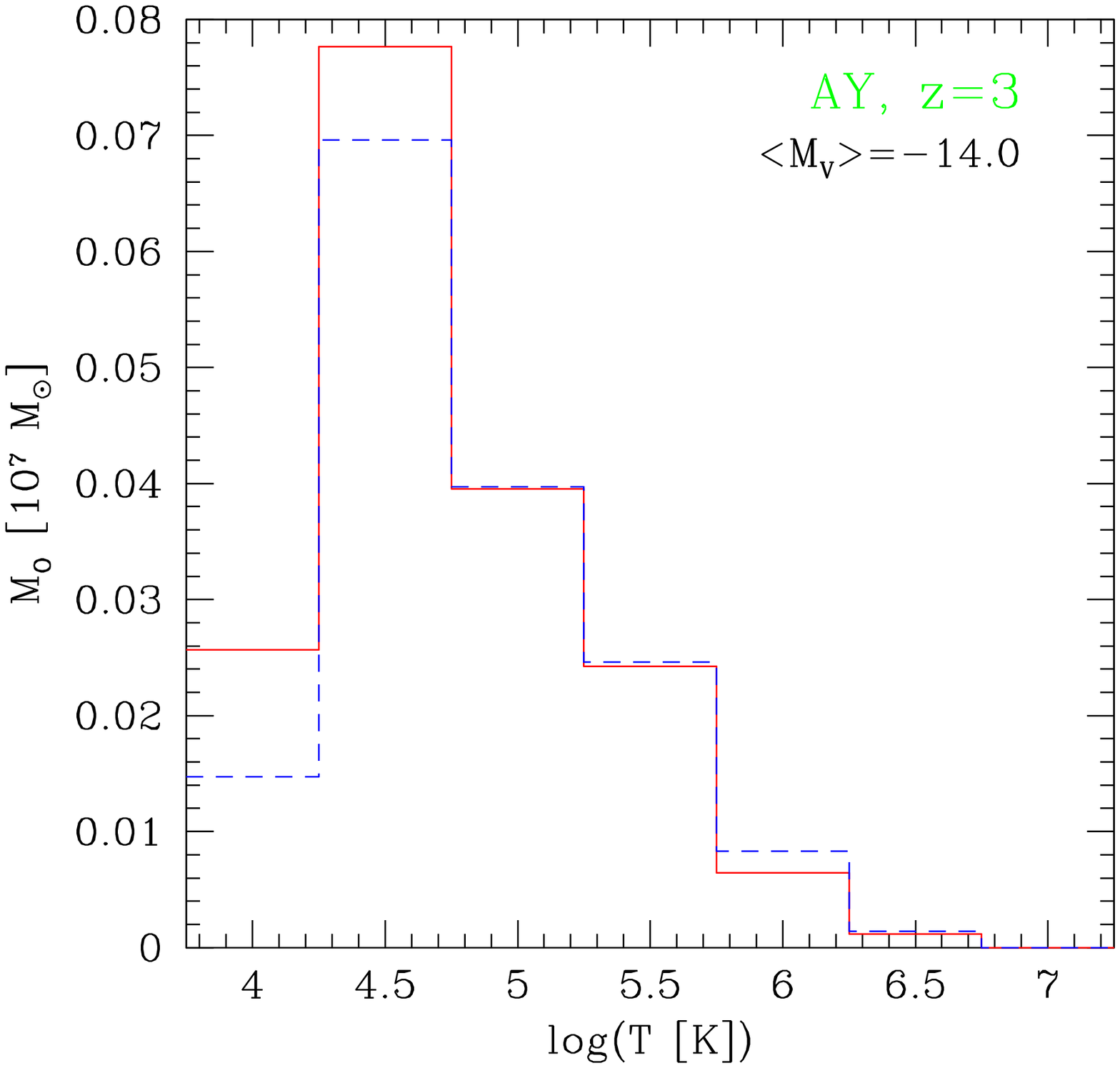,width=0.41\textwidth}
\psfig{file=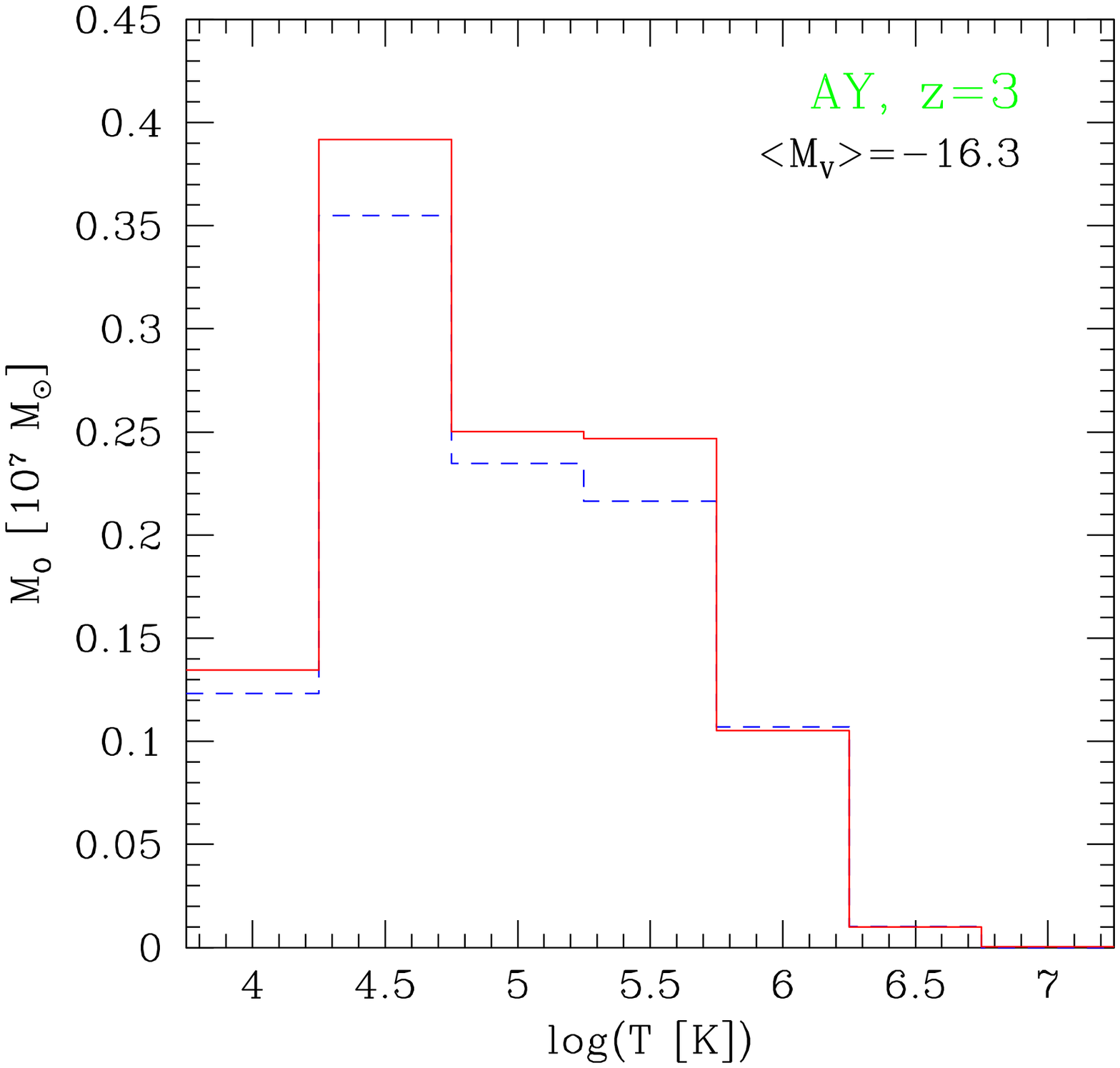,width=0.41\textwidth}

\leavevmode
\psfig{file=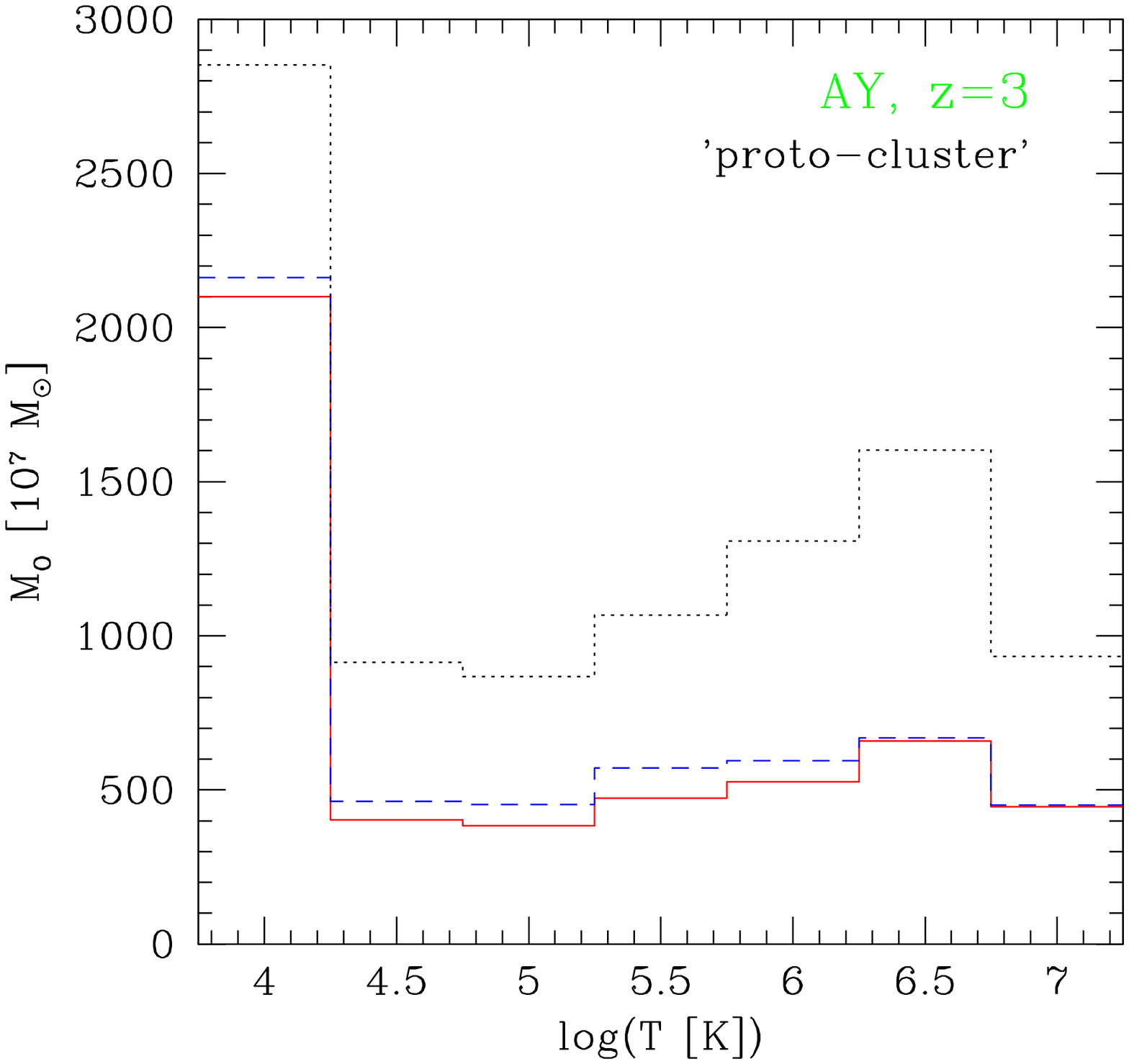,width=0.41\textwidth}
\psfig{file=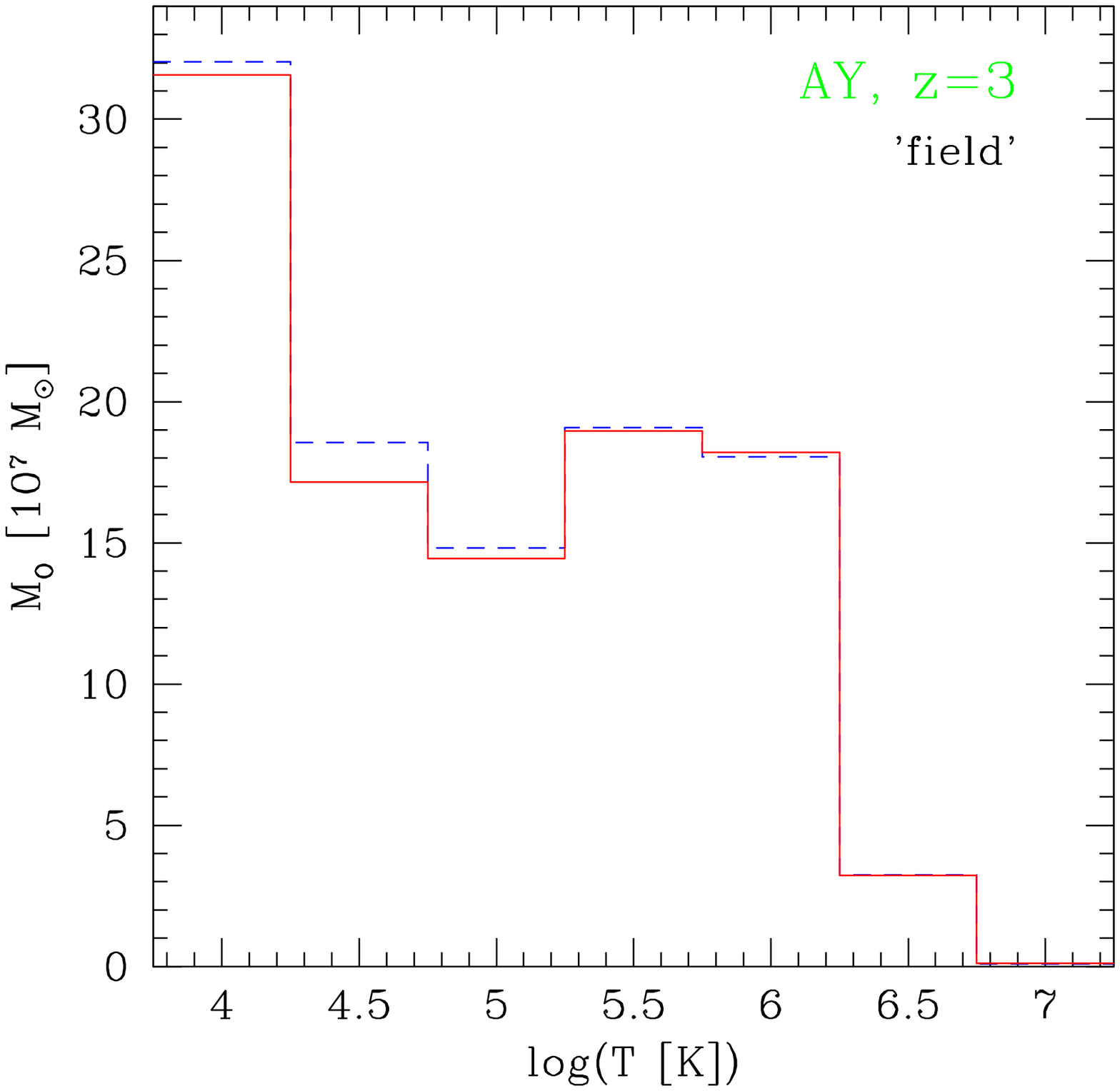,width=0.41\textwidth}
\caption{Similar to Fig.~\ref{fig:gmpK}, but for simulations using the
AY IMF. The top left plot shows the result for
the combination of three regions, each containing a galaxy simulated
using the ``zoom-in'' technique plus a number of companions. The three
main galaxies have a mean $M_{\rm{V}}$ of $-14.0$. The top right plot shows results
for a combination of 2 high-resolution regions containing main galaxies
of an average $M_{\rm{V}}$ of $-16.3$. The bottom right plot is for the combination
of the low and medium density ``field'' regions described in the text,
containing a total of 40 resolved galaxies of $M_{\rm{V}}$ down to about 
$-21.5$.
Finally, the bottom left plot shows the results for a very high density
``proto-cluster'' region, containing 86 galaxies of $M_{\rm{V}}$ down to 
about $-25.5$. The black dotted histogram shows the prediction using
a constant $\tilde{r}$ approach (see text); the blue dashed histogram
shows the prediction resulting from making the $M_{\rm{V}}<-22$ $\tilde{r}$
modification described in the text}
\label{fig:gmpAY}
\end{figure*}
\section{The simulations}
\label{sect:simulations}
The code used for the simulations is a significantly improved version of
the TreeSPH code we used for our previous work on galaxy formation 
(SGP03). A similar version of the code has been used recently
to simulate clusters of galaxies, and a detailed description can be found 
in Romeo \etal (2006). Here we briefly mention its main features and the 
upgrades over the previous version of SGP03 --- see also Sommer-Larsen (2006).
\begin{enumerate}
\item
The basic equations are integrated by incorporating the ``conservative'' 
entropy equation solving scheme of Springel \& Hernquist (2002), which
improves the numerical accuracy in lower resolution regions.
\item
Cold high-density gas is turned into stars in a probabilistic way as
described in SGP03. In a star-formation event 
an SPH particle is converted fully into a star particle. Non-instantaneous 
recycling of gas and heavy elements is described through probabilistic 
``decay'' of star particles back to SPH particles as discussed by 
Lia \etal (2002a). In a decay event a star particle is converted fully 
into a SPH particle, so that the number of baryonic particles in the simulation
is conserved.
\item
Non-instantaneous chemical evolution tracing
10 elements (H, He, C, N, O, Mg, Si, S, Ca and Fe) has been incorporated
in the code following Lia \etal (2002a,b); the algorithm includes 
supernov\ae\ of type II and type Ia, and mass loss and chemical enrichment
from stars of all masses.
Most of the simulations presented in this paper have been undertaken using
three different Initial Mass Functions: the Kroupa (1998) IMF (denoted
K98 in the following), derived for field stars in the Solar 
Neighborhood, the standard Salpeter IMF (S), and the more top-heavy
Arimoto \& Yoshii (1987) IMF (AY), which is well suited for the modeling
of elliptical galaxies, as well as galaxy clusters. More detail is
given in Lia \etal (2002a,b).
\item 
Atomic radiative cooling is implemented, depending both on the 
metallicity of the gas (Sutherland \& Dopita 1993) and on the meta--galactic 
UV field, modeled after Haardt \& Madau (1996). Moreover, a simplified 
treatment of radiative transfer, switching off the
UV field where the gas becomes optically thick to Lyman limit photons on
scales of $\sim$ 1~kpc, is invoked.
 \item
Star-burst driven winds are incorporated in the simulations
at early epochs ($z$$\ga$5-6), as strong early feedback is crucial to 
largely overcome the angular momentum problem (SGP03). A burst of star 
formation is modeled in the same way as in SGP03: when a star particle 
is formed, further self-propagating star formation is triggered in the 
surroundings; the energy from the resulting, correlated
SNII explosions is released initially into the interstellar medium 
as thermal energy, and gas cooling is locally halted to reproduce the 
adiabatic super--shell expansion phase; a fraction of the supplied energy 
is subsequently converted (by the hydro code itself) into kinetic energy 
of the resulting expanding super--winds and/or shells.
The super--shell expansion also drives the dispersion of the metals produced
by type~II supernov\ae\ (while metals produced on longer timescales are 
restituted to the gaseous phase by the ``decay'' of the corresponding 
star particles, see point 2 above). 

At later epochs, only a fraction (typically, 20\%) of the stars induce 
efficient feedback, and star formation is no longer self--propagating
so that no strong star-bursts are triggered by correlated SN explosions.
This allows the smooth settling of the disc (see SGP03 for all details).
Star formation efficiencies have been chosen such that realistic disc
galaxy gas fractions result at $z$=0 (SGP03), and can also be shown to
match the Kennicutt (1998) star formation relation quite well.

AGN driven feedback has not been
invoked in the simulations, as it is unlikely to play a major role
in the formation of, at least, disc galaxies 
(e.g., Sommer-Larsen 2006) --- the most common galaxy type in the
Universe. The higher stellar UV escape fractions as well as LyC luminosities 
found 
by Razoumov \& Sommer-Larsen (2007) at $z=3.6$, compared to $z\la3$ ,
also support the notion that massive stars in galaxies become 
progressively more important
sources of ionizing photons as one goes back in time, as the comoving
number density of quasars declines rapidly at $z\gsim3$ (e.g., Richards
\etal 2006). This provides additional circumstantial evidence that AGN
feedback is of minor importance, at least at $z\ga3$, the relevant
redshift range in this paper.
\end{enumerate}
The galaxies were drawn and re-simulated from a $10 h^{-1}$~Mpc 
box-length dark matter (DM)-only 
cosmological simulation, based on the ``standard'' flat $\Lambda$ 
Cold Dark Matter cosmological model ($h=0.65$, $\Omega_0=0.3$, 
$\sigma_8=1.0$); our choice of $h$ and $\sigma_8$ is slightly
different from presently more popular values (0.7 and 0.9 respectively),
but this has little impact on the resulting galaxy properties
(e.g., Portinari \& Sommer-Larsen 2007).
When re-simulating with the hydro-code, baryonic 
particles were ``added" to the original DM ones, which were
split according to an adopted baryon fraction $f_b=0.15$. 
The gravity softening lengths were fixed in physical coordinates from $z$=6
to $z$=0 and in co-moving coordinates at earlier times.

The simulations are run with resolutions of $m_{SPH}=m_*=3.6 \times 
10^3$-4.5$\times 10^7$~\Msol, $m_{DM}=2.0 \times 10^4$-3.3$\times 10^7$ 
~\Msol\ and $\epsilon_{SPH}=\epsilon_*~=~0.08-3.8$~kpc going from the smallest individual
galaxy simulations to the largest proto-cluster simulation. Each region
containing an individual proto-galaxy was simulated using 0.2-2.2
million particles in total; in addition two ``low'' and ``medium''
galaxy density regions were simulated using 1.4 and 1.6 million particles,
a proto-elliptical region using 1.4 million particles, and a 
proto-cluster region using 2.3 million particles (the high resolution 
``Virgo'' simulation of Romeo \etal 2006).   

Images of some of the simulated galaxies are available at 
{\sf http://www.tac.dk/$\sim$jslarsen}.

\section{The approach}
\label{sect:approach}
The approach adopted is, based on the simulations, to characterize
a given galaxy at $z$=3 and of absolute 
magnitude $M_{\rm{V}}$ by a distribution of ISM/IGM oxygen mass 
$m_O(M_{\rm{V}},\log(T))$, 
where $\log(T)$=3.5, 4, 4.5 , ..,7, and a given value of $\log(T)$ 
corresponds to gas temperatures in the range
$T=[10^{\log(T)-0.25},10^{\log(T)+0.25}]$~K. 
Almost no oxygen is found at temperatures
lower than 10$^{3.5}$ K, mainly because the radiative cooling function is
effectively truncated below about 8000 K. If molecule formation and 
molecular cooling had been invoked in the hydro/gravity simulations,
part of this gas would be in a much colder, high-density, predominantly
H$_2$ phase (see also Sec.~\ref{sect:DLA}). Moreover, it is
found that the $\log(T)$=3.5 as well as 4 ISM/IGM oxygen is spatially 
associated
with the stellar galaxies (similar spatial extents), so 
the $\log(T)$=3.5 and 4 
bins have been merged, and are denoted by $\log(T)$=4. Hence, all gas colder 
than 10$^4$ K is included in this phase. Moreover, for field galaxies at
$z$$\sim$3 almost no oxygen is found in the $\log(T)$$>$7 phases, which will
therefore be ignored in the following.  

Once the functions $m_O(M_{\rm{V}},\log(T))$ have been determined for a given IMF, the 
cosmologically averaged (gaseous) oxygen temperature distribution can be 
determined by folding these functions with the (dust corrected) 
rest-frame V-band luminosity function at $z\sim3$. We stress that
$m_O(M_{\rm{V}},\log(T))$ pertains only to the galaxy itself, not including any
of its satellite galaxies --- see further below.

To determine $m_O(M_{\rm{V}},\log(T))$ we proceed as follows: in Fig.~\ref{fig:Omassr}
(left) is shown the cumulative mass of ISM/IGM oxygen, $M_O(r;M_{\rm{V}};\log(T))$, 
around three (mostly) isolated
galaxies of $M_{\rm{V}}$ = $-16.1$, $-21.4$ and $-25.3$, respectively, simulated using the
K98 IMF. Fig.~\ref{fig:Omassr} (right) is similar, but for the {\it same}
three galaxies, of $M_{\rm{V}}$ = $-13.8$, $-19.3$ and $-25.4$, simulated with the AY IMF  
For all three galaxies the region from $r_{vir}$ to 8~$r_{vir}$
is shown\footnote{In this paper the virial radius,
$r_{\rm{vir}}$, is the radius corresponding to, at $z$=3, an over-density 
181 times the mean matter density of the Universe at this redshift, 
as appropriate for a top--hat collapse
in the adopted $\Lambda$CDM cosmology (e.g., Bryan \& Norman 1998).}, 
and the different curves correspond to $\log(T)$=4, 4.5 , ..,7.
As can be seen from the figure, for a
given galaxy and $\log(T)$, the cumulative gas oxygen mass gently increases 
with $r$ by factors $\sim$1.5-4, over the radial range shown. 
The one exception is the $\log(T)$=4 case, where the gas metals are mainly 
spatially correlated with the central galaxies, and the curves are quite 
flat (with the exception of the very bright galaxy, which has been
selected from the proto-cluster simulations). We stress that the mass
of gas-phase oxygen around a given (isolated) galaxy has been contributed 
not only by
the stars in the galaxy itself, but also by stars in all its satellite 
galaxies (and in principle also by ``inter-galactic'' stars, as will be 
discussed below).

We select a sample of ``base'' galaxies consisting of (mainly) isolated
galaxies ranging in absolute V-band magnitude from about $-12$ to $-25$.
It is assumed that in order to account for the $m_O(M_{\rm{V}},\log(T))$ of a given
base galaxy one should include all gas oxygen mass inside of a radius
$\tilde{r}(M_{\rm{V}},\log(T)) = r(M_{\rm{V}},\log(T))/r_{vir}(M_{\rm{V}})$, to be determined
in the following, such that 
$m_O(M_{\rm{V}},\log(T))$=$M_O(\tilde{r}\times r_{vir};M_{\rm{V}};\log(T))$.
``Isolated'' galaxies are in this paper defined as galaxies not containing
companions of stellar mass larger than 1/3 of the mass of the galaxy
itself within 8~$r_{vir}$. The very brightest galaxies, however, of
$M_{\rm{V}}$$\la-23$, were selected from the higher density proto-cluster
simulations, and the criterion was relaxed to not containing
companions of stellar mass larger than 1/2 of the mass of the galaxy
itself within 4~$r_{vir}$. 
The relation between $r_{vir}$ and absolute V-band (rest-frame) 
magnitude $M_{\rm{V}}$ is shown in Fig.~\ref{fig:rvir} for the three IMFs
under consideration. 

\begin{figure}
\psfig{file=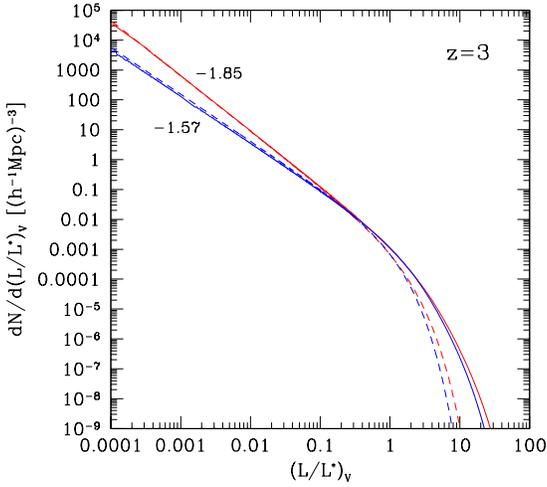,width=0.43\textwidth}
\caption{Observational and extinction corrected, modified V-band
restframe galaxy luminosity functions, based on K-band observations by
Shapley \etal (2001). Red dashed curve shows the original
Shapley \etal LF of faint end slope $\alpha=-1.85$, 
red solid curve the corresponding extinction corrected, modified LF.
Blue dashed curve shows the result of fitting a LF of faint end
slope $\alpha=-1.57$ to the Shapley \etal data, blue solid curve
the corresponding extinction corrected, modified LF.}
\label{fig:LF}
\end{figure}
\begin{figure}
\psfig{file=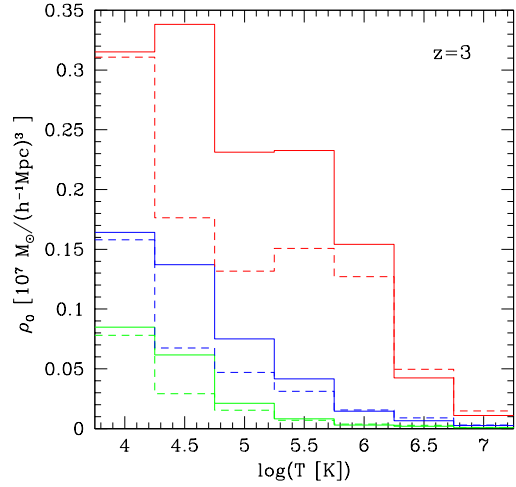,width=0.43\textwidth}
\caption{Cosmologically averaged gas-phase oxygen density
temperature distributions for the three IMFs considered.
Solid and dashed, green histograms correspond to the K98 IMF and 
the extinction corrected, modified galaxy luminosity functions of faint end
slopes -1.85 and -1.57, respectively. The blue and red histograms
show the corresponding quantities for the Salpeter and AY IMFs,
respectively. For comparison, a density of 10$^7$ \Msol/$(h^{-1}\rm{Mpc})^{-3}$
corresponds to a density parameter of 
$\Omega_O = \rho_O/\rho_{\rm{crit}} = 2.5 \times 10^{-5}$, for $h$=0.7} 
\label{fig:cosmic}
\end{figure}
For a given galaxy, $\tilde{r}$ will be finite for two reasons: a) no galaxy
is perfectly isolated, such that, even for isolated galaxies, with 
increasing $\tilde{r}$ increasing amounts of gas-phase oxygen associated
with other (smaller) companion galaxies are included, and b) for certain
values of $\log(T)$ (depending on the IMF) galaxies are found to ``share''
a common pool of gas-phase oxygen. In practice it is found that for none of 
the
IMFs considered, and any $\log(T)$, does $\tilde{r}$ exceed 8. For the brightest
galaxies, $M_{\rm{V}}$$\la-23$, $\tilde{r}$ does not exceed 4.    

Fig.~\ref{fig:T} shows, for $\log(T)$=4.5, 5, 5.5 and 6, the spatial distribution 
of the gas-phase oxygen for a region containing 52 resolved galaxies with
$M_{\rm{V}}$ in the range $-15$ to $-22.5$, simulated using the K98 IMF. Each dot
represents, for a given $\log(T)$, an equal amount of gas-phase oxygen mass,
so the figure displays the actual distribution of gas-phase oxygen mass,
{\it not} gas mass.
As can be seen, the spatial oxygen mass distributions depend strongly on the 
gas temperature. 

In Fig.~\ref{fig:imfs} is shown, for the three IMFs considered, the
distribution of $\log(T)$=6 oxygen mass. As can be seen, the oxygen mass
distribution also depends strongly on the IMF adopted. For comparison is also
shown the corresponding plot for a K98 simulation of a proto-elliptical,
high-density region, which is biased towards many large galaxies
(bottom left) It is
seen how the $\log(T)$=6 oxygen mass distribution in this case is concentrated
near the large galaxies, more so than for the more representative ``field''
galaxy case.

We now approximate $\tilde{r}(M_{\rm{V}},\log(T))$ by $\tilde{r}(\log(T))$, \ie independent of
$M_{\rm{V}}$.  We show further below that this is a good approximation, except for the
very brightest galaxies, for which a correction has to be made.  For each IMF,
$\tilde{r}(\log(T))$ is determined as follows: i) 
for a number of regions containing
up to about 200 resolved galaxies we a) determine the absolute V-band
magnitudes of the $N$ galaxies in the region, $M_{V,1}, ..., M_{V,N}$,
and b) the total gas-phase oxygen mass temperature distribution of the
region
\begin{equation}
f_{O,\rm{region}}^{\rm{actual}}(\log(T)) = 
\int_{region} \rho_O(\bar{x};\log(T))~d\bar{x}~~. 
\end{equation}
ii) for a given set of base galaxies, and
given values of the $\tilde{r}(\log(T))$'s, one can predict the total mass of
oxygen as a function of $\log(T)$ as
\begin{equation}
f_{O,\rm{region}}^{\rm{predicted}}(\log(T)) = 
\Sigma_{i=1}^{N} M_O^{pred}(\tilde{r} \times r_{vir}(M_{V,i});M_{V,i};\log(T)) ~~, 
\end{equation}
where $M_O^{pred}(\tilde{r} \times r_{vir};M_{V,i};\log(T))$ is determined by 
linear
interpolation in $M_{\rm{V}}$ as well as $\tilde{r}(\log(T))$ ($\tilde{r}$=0,1,...8)
using the corresponding values for the base galaxies. The values of
$\tilde{r}(\log(T))$ are then obtained such as to provide the best
simultaneous match of $f_{O,\rm{region}}^{\rm{predicted}}(\log(T))$ to 
$f_{O,\rm{region}}^{\rm{actual}}(\log(T))$ for a 
number of simulated regions, using standard least squares fitting.

For
all three IMFs this is done for the {\it same} two regions, representing 
``low-density'' and ``medium-density'' ``field'' environments, with
galaxies of $M_{\rm{V}}\ga-22.5$. The 
two regions are simulated using a total of 1.4 and 1.6 million particles,
respectively. For the AY IMF, two additional regions centered on smaller 
galaxies are also used (contrary to the K98 and S IMF cases, for the
AY simulations such small galaxies have substantial amounts of not only
$\log(T)$=4-6, but also $\log(T)$=6.5 and 7 gas-phase oxygen associated) --- these 
regions are each simulated using about 300,000 particles in total. 
Finally, to test the approach also in a proto-cluster environment,
for all three IMFs the same such region was simulated using a total
of 2.3 million particles (the high-resolution ``Virgo'' region
of Romeo \etal 2006).     
 
As will be shown in the next section, in particular for $\log(T)$=4.5, 5 and 5.5,
a significant contribution to the cosmic gas-phase oxygen budget comes from
small galaxies with $M_{\rm{V}}\ga-17$ to $-19$. To check that the  $\tilde{r}(\log(T))$
fits also work in this range, comparison was made to combined data-sets
of regions with $M_{\rm{V}}$ of the central galaxy ranging from 
$-14$ to $-19$, all
simulated at high resolution, with typically 200-300\,000 particles per
region.
The results for the K98 IMF are shown in Fig.~\ref{fig:gmpK}, including
also the result for a proto-elliptical region. In general, the predicted
distributions match the actual ones very well, except for the very
smallest galaxies, where the predicted distribution falls somewhat short
of the actual one in the $\log(T)$=4, 4.5 and 5 temperature bins. 
This implies that the results obtained in the following section for the
faintest galaxies are actually lower limits. Given the small magnitude
of the effect we shall, however, neglect this in the following.

In Fig.~\ref{fig:gmpAY} a similar comparison is shown for AY IMF
simulations. Again, the match between predicted and actual distributions
is, in general, good. For the smallest galaxies the distributions are
somewhat different (see above). For the proto-cluster region, the 
predicted distribution exceeds the actual one, with $f_{O,\rm{region}}^{\rm{predicted}}(\log(T))$
being up to about twice as large as $f_{O,\rm{region}}^{\rm{actual}}(\log(T))$. This is due to
the very large galaxy density in such environments. Using this region
to solve for an alternative set of $\tilde{r}(\log(T))$ results in
$\tilde{r}(\log(T))$ values of about half the ones obtained as described
above. As the main contribution to $f_{O,\rm{region}}^{\rm{actual}}(\log(T))$ is associated
with the larger galaxies in the region, $M_{\rm{V}} \sim-24$ to $-25.5$, 
$f_{O,\rm{region}}^{\rm{actual}}(\log(T))$ in all regions can be matched by a model, where
$\tilde{r}(\log(T))$ changes linearly from the ``field'' values to
the ``proto-cluster'' values between $M_{\rm{V}}$ = $-22$ to $-24$, as
shown in Fig.~\ref{fig:gmpK}. However, in any case this correction for
the very large galaxies is of no consequence to the main results 
presented in the following section. This is because the contribution from such
galaxies to the overall cosmic metal budget is very minor, as will be
detailed in the following section. 

Finally we note, that even in typical ``field'' environments, some fraction 
of the stellar systems originally formed have subsequently been 
disrupted through tidal stripping and other dynamical processes. The 
gas-phase metals produced by these ``inter-galactic'' stars are 
``automatically'' accounted for by the above approach, but in any case
the fraction of such stars is quite small, $\la$5-10\% (paper I).
   
\begin{figure*}
\leavevmode
\psfig{file=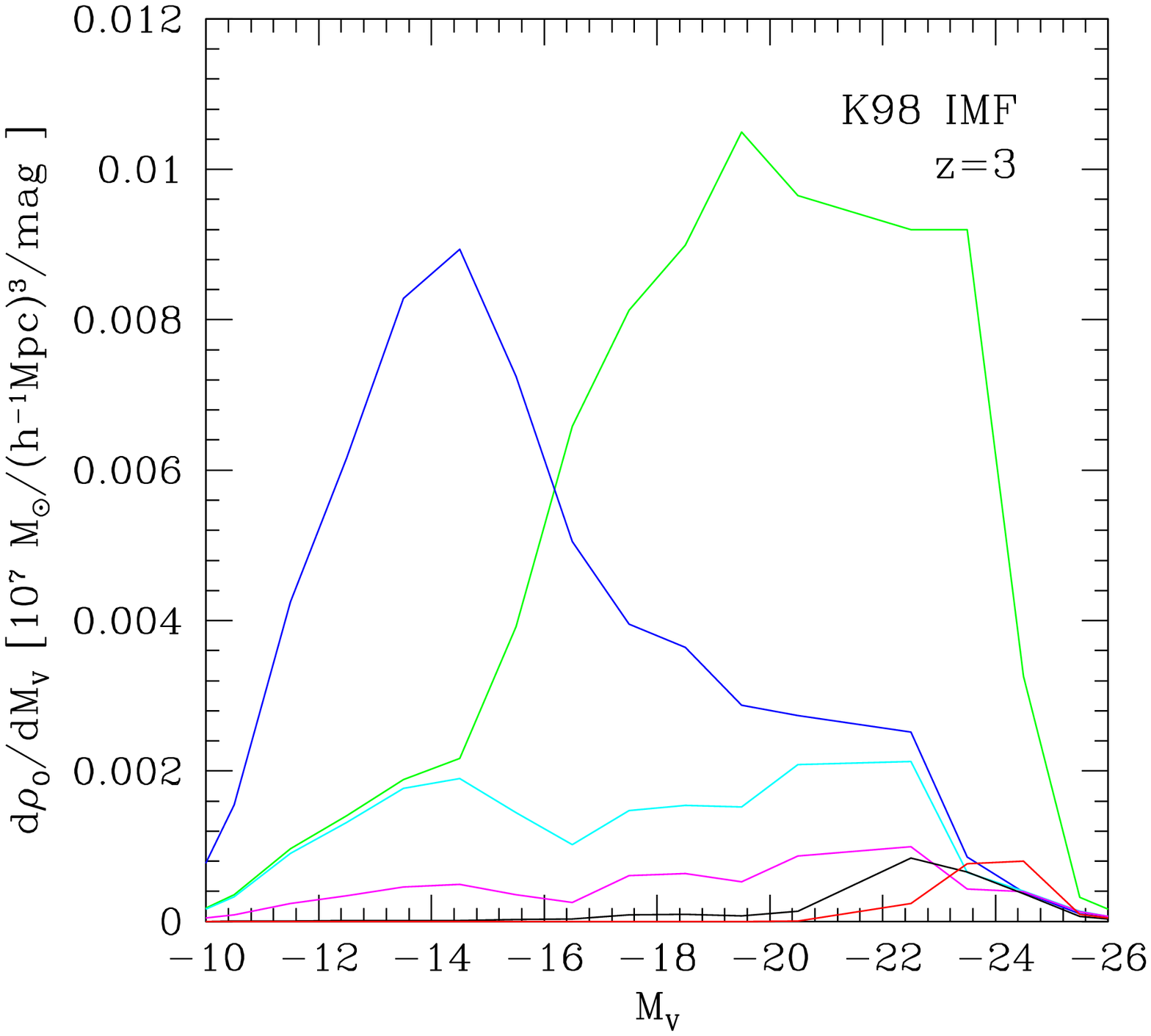,width=0.41\textwidth}
\psfig{file=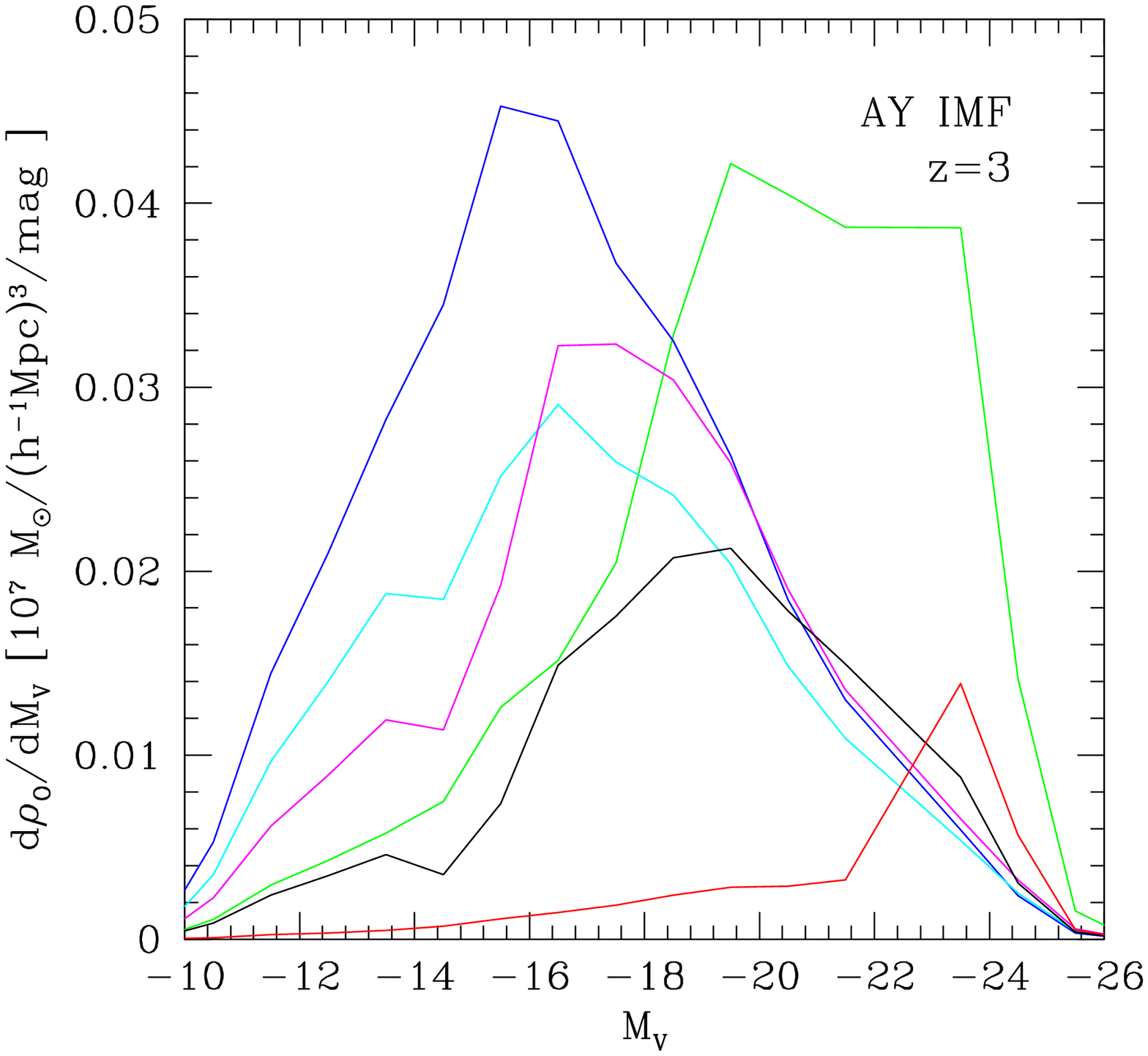,width=0.41\textwidth}

\caption{Contribution to cosmologically averaged gas-phase oxygen 
density as a function of galaxy V-band absolute magnitude for
the K98 (left) and AY (right) IMFs. Curves
corresponding to different gas temperatures are colour coded as in
Fig.~\ref{fig:Omassr} ($\log(T)$=4: green, 4.5: blue, 5: cyan, 5.5: magenta, 6: black, 6.5: red). 
The results shown are for the extinction corrected
modified Shapley \etal ($\alpha=-1.85$) galaxy luminosity function.
For comparison with the cosmological density parameter, see caption of 
Fig.~\ref{fig:cosmic}.}
\label{fig:cosmic_abunT_M_V}
\end{figure*}

\section{The temperature distribution of the cosmic gas-phase oxygen}
\label{sect:cosmic}
The average cosmic density of gas-phase oxygen at $z$=3, as a function of 
gas temperature, can now be obtained as 
\begin{equation}
\langle\rho_{O}(\log(T))\rangle = 
\int_{0}^{\infty} \frac{dN(L_V)}{dL_V}~m_O(M_{\rm{V}}(L_V),\log(T))~dL_V~~, 
\end{equation}
where $dN(L_V)/dL_V$ is the rest-frame V-band, extinction
corrected galaxy luminosity function at $z$=3. 

\begin{figure*}
\leavevmode
\psfig{file=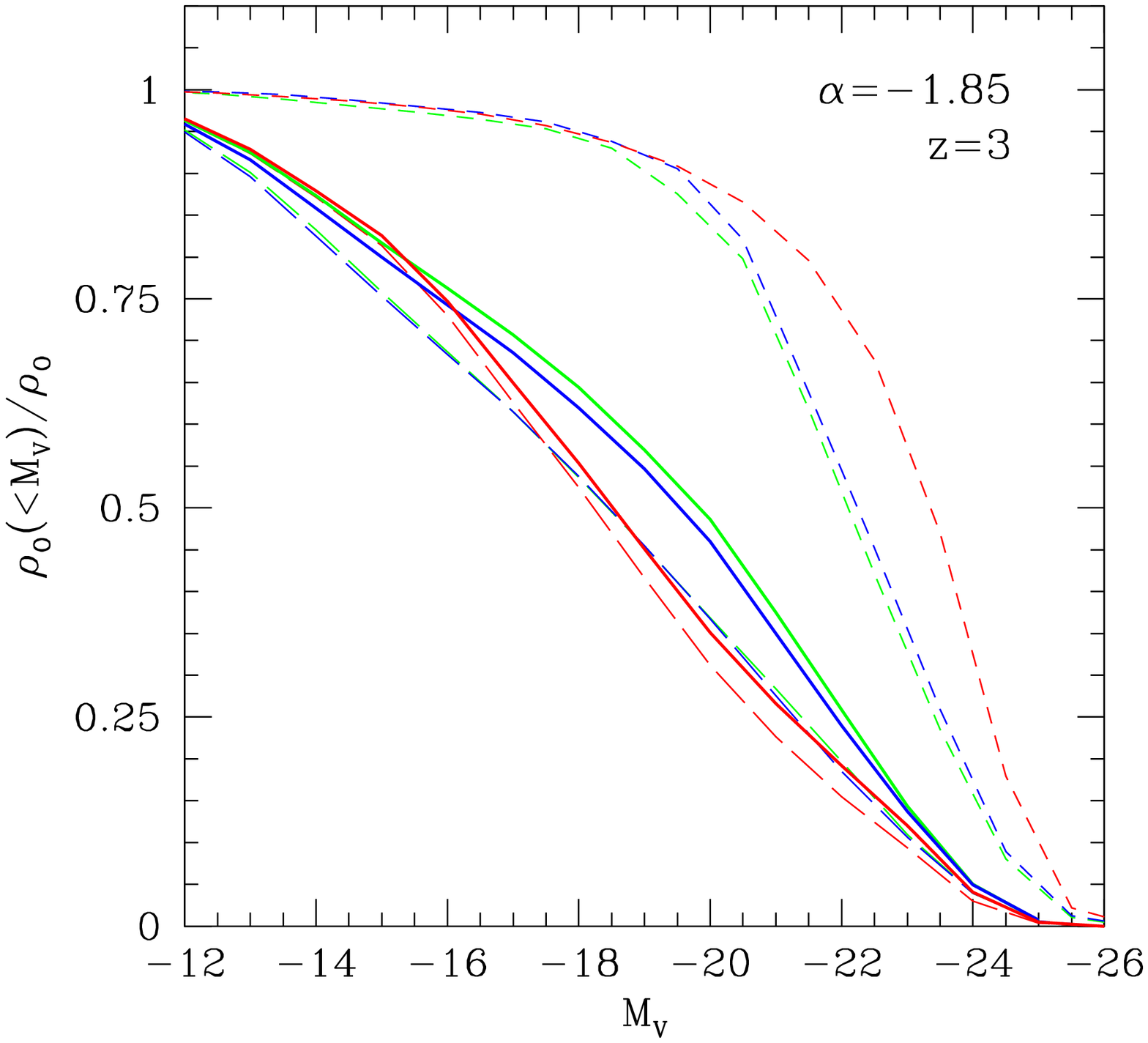,width=0.41\textwidth}
\psfig{file=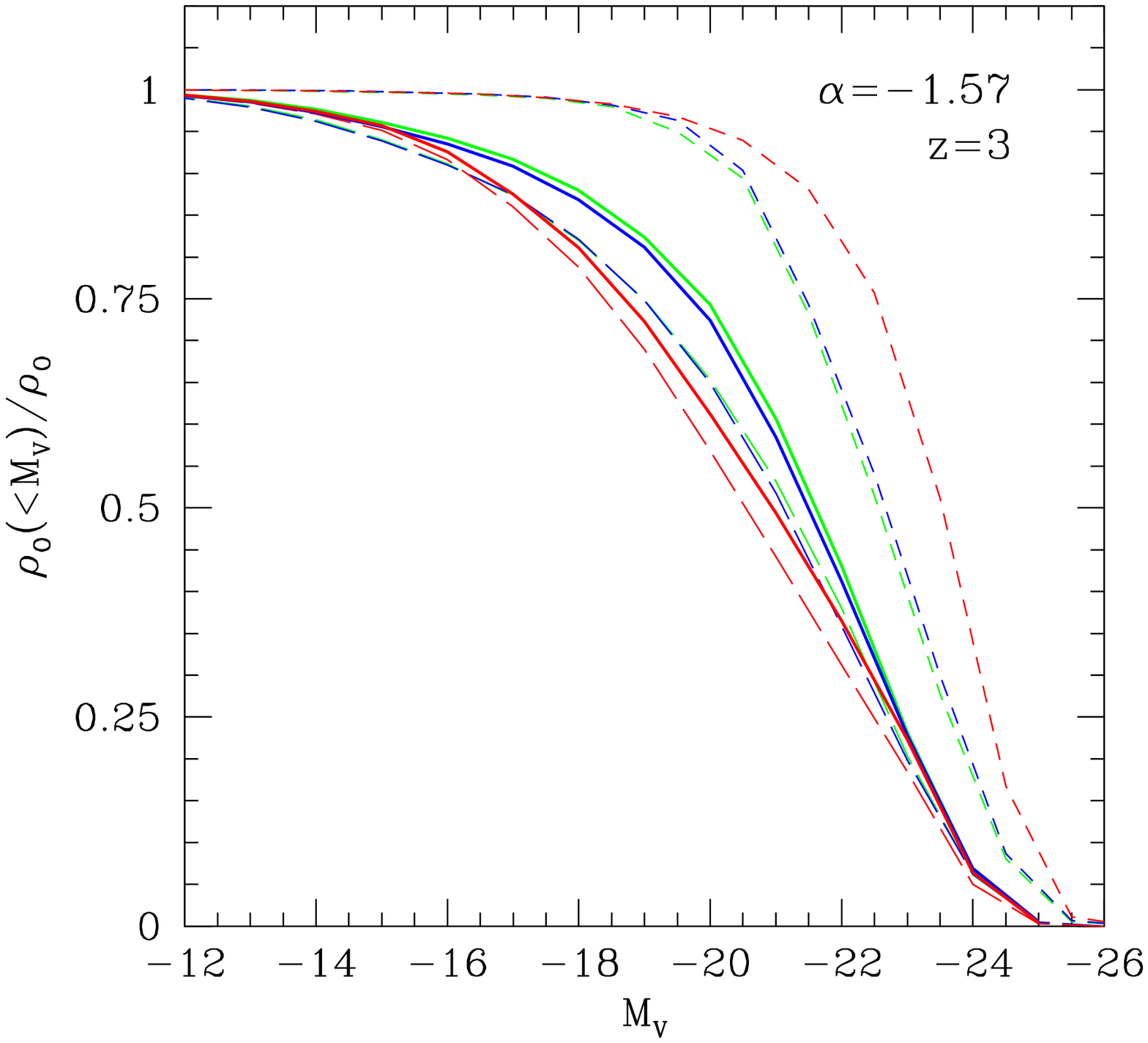,width=0.41\textwidth}

\caption{Fractional contribution to the cosmologically averaged gas-phase
oxygen density by galaxies brighter than $M_{\rm{V}}$, for the extinction
corrected modified Shapley \etal luminosity function (left) and the 
luminosity function of
faint end slope -1.57 (right). Green, blue and red curves correspond
to results for the K98, Salpeter and AY IMFs, respectively. Long-dashed
curves correspond to gas-phase oxygen, short-dashed to stellar oxygen,
and solid to the total.}
\label{fig:ocum}
\end{figure*}
In paper I the ``true'' $z\sim$3 galaxy luminosity function was determined 
by extinction correcting a modified version of the observationally
determined luminosity function (LF) of Shapley \etal (2001) of faint
end slope $\alpha=-1.85$. The original Shapley \etal (2001) LF, as well
as the extinction corrected, modified LF, are shown in Fig.~\ref{fig:LF}.
Also shown are the corresponding LFs obtained by fitting a LF
of faint end slope $\alpha=-1.57$ (see below) to the Shapley \etal (2001)
data. The faint end slopes of these $z\sim3$ LFs are steeper than what
is found for $z\sim0$ LFs. Results of semi-analytical galaxy formation
models (e.g., Lacey \etal 2005) indicate that this is a natural outcome
of hierarchical strucure formation scenarios.

Using the extinction corrected, modified Shapley \etal LF
in eq.\,4, and the $m_O(M_{\rm{V}}(L_V),\log(T))$ functions determined for the
three IMFs as described in the previous section, results in the 
distributions shown in Fig.~\ref{fig:cosmic}. For
the K98 IMF, the amount of oxygen in the $\log(T)\ge$4.5 phase is about 1.1
times the amount in the $\log(T)$=4 ``cold'' phase. For the S and AY IMFs,
the corresponding numbers are about 1.7 and 3.2, respectively. Hence, in
particular,
for the latter two IMFs, the majority of the gas-phase oxygen is in
the ``warm''/''hot'' rather than cold phase. Moreover, 
for the AY IMF, the hot to cold/warm gas-phase oxygen
mass ratio is 1.1, so the bulk of
the oxygen is actually in the hot $\log(T)$$\ga$5 phase,
rather than in the cold/warm
phase. These are some of the main results of this paper.

In Fig.~\ref{fig:cosmic} is also shown
the results of adopting a galaxy luminosity function with less steep
faint end slope than the Shapley \etal one. Following
paper I, results are presented for a luminosity function of faint end slope
$\alpha=-1.57$, as found for the rest-frame UV $z\sim$3 luminosity function 
by, e.g., Adelberger \& Steidel 2000). Although the results for
this LF are qualitatively similar to the ones described above, the
amounts of warm and hot phase oxygen decrease somewhat compared
to the cold phase ones. The warm/hot to cold 
gas-phase oxygen mass ratios are for this LF 0.8, 1.1 and 2.1 for
the K98, Salpeter and AY IMFs, respectively. Moreover, for the
AY IMF, the hot to cold/warm gas-phase oxygen mass ratio is unity.
     
An obvious question is now: for the different gas temperatures, which
galaxies contribute mainly to the cosmic gas-phase oxygen budget?
In Fig.~\ref{fig:cosmic_abunT_M_V} this is shown using the modified
Shapley \etal luminosity function, for the K98 and AY IMFs, 
respectively. It is seen that for the K98 IMF the main contribution to
the $\log(T)$=4.5 phase comes from small galaxies, $M_{\rm{V}}\sim-16$, and for
the AY IMF this is the case for the $\log(T)$=5 and 5.5 phases as well
(the reason for the large difference between the $M_{\rm{V}}$-distributions of 
$\log(T)$=4 and $\log(T)$=4.5, say, gas-phase metals is that, whereas
the $\log(T)$=4 gas is of fairly high density and associated with
all galaxies, $\log(T)$=4.5 gas is typically diffuse and associated
with halos of comparable virial temperature, i.e. fairly small galaxies).

Given this, and in relation to observational studies, it is clearly of 
interest to determine how large a fraction of the gas-phase oxygen is 
associated with galaxies to a certain limiting magnitude. In 
Fig.~\ref{fig:ocum} this is shown by the long-dashed curves 
(for the three IMFs considered) for the $\alpha=-1.85$ and $-1.57$ LFs.   
Typical Lyman Break galaxy metallicity determinations at $z\sim$3
only probe to about one magnitude below $L^*$, \ie $M_{\rm{V}}\sim-22$. 
For the $\alpha=-1.85$ LF, it is found that less than about 20\% of the 
cosmic gas-phase oxygen will be 
associated with galaxies brighter than this, irrespective of the
choice of IMF. For the $\alpha=-1.57$ LF the corresponding fraction
is about 30\%. Hence, most of the cosmic gas-phase oxygen is associated
with galaxies considerably fainter than $L^*$, and is also in this
sense ``missing''.

In calculating the results shown in Fig.~\ref{fig:ocum} it has been assumed
that the observational LFs maintain a constant faint end slope down to
$M_{\rm{V}}=-10.5$.
As can be seen from the figure, in particular for the
$\alpha=-1.85$ case this is somewhat critical to the shape of the curves shown.
At redshifts $z$$\sim$3 the LBGs luminosity function is only probed  down to a
few magnitudes below $L^*$ (but see below). Selecting galaxies using the Ly$\alpha$ line  it is
possible to probe 2-3 magnitudes fainter still (Fynbo \etal 2001, 2003; Gawiser
\etal 2006; Nilsson \etal 2007). This, however, only demonstrates the
existence of star-forming galaxies at these redshifts with M$_V\sim-17$ to
$-18$, but
the data are not good enough to put strong constraints on the shape of the
luminosity function. Jakobsson \etal (2005) find that the magnitudes of $z>2$
GRB selected galaxies are consistent with being drawn from the steep
luminosity function, but the statistics are still poor. We note, however, that
Bouwens \etal (2007) were actually able to probe the UV luminosity function 
at $z\sim4-6$ down to 4-5 magnitudes below $L^*_{UV}$ and find steep
faint end slopes, $\alpha\sim-1.7$.

Our calculations show, however, that if the slope of the
faint end LF flattens {\it significantly} at some brighter $M_{\rm{V}}$, then
the curves still fairly well represent the result, after appropriate
renormalisation at this limiting $M_{\rm{V}}$ (if the faint end
slope is assumed to be constant to even fainter than  $M_{\rm{V}}=-10.5$,
the results are not much changed compared to the $M_{\rm{V}}=-10.5$
results). 
Moreover, we note that neither
the Two-Degree Field Galaxy Redshift Survey (2dFGRS) $b_j$ local
luminosity function (Norberg \etal 2002), nor the Sloan Digital Sky
Survey (SDSS) U-band local luminosity function (Baldry \etal 2005)
show any flattening of the faint end slope to 5-6 magnitudes fainter
than $M_*$. Moreover, some galaxy cluster luminosity functions tend
to show a {\it steepening} faint end slope: Milne \etal (2007) find
a steepening of the slope in the Coma cluster at a $M_R$ range of
about $-14$ to $-9$, and Yamanoi \etal (2007) find a similar result for
the Hydra I cluster.

Given the significance of the results obtained in this section on the 
distribution of cosmic gas-phase metals, it is important to consider
the following two sources of potential model dependence. 

First, it
is possible that the results depend on the specific choice of
``base'' galaxies. We verified explicitly, however, that choices
of alternative sets of isolated sets of ``base'' galaxies, and subsequently
going through the procedure described in the previous section, lead to
very similar results as the ones described here. Hence, the results
presented are very robust to this.  

Second, it would be expected that the amount of (in particular) $\log(T)$=4
cold gas and metal mass available in a simulation at any given
time would depend on the star-formation efficiency adopted 
(Sec.~\ref{sect:simulations}). To this end, three
K98 galaxy simulations, of galaxies of $z$=3 $M_{\rm{V}}$=$-19.1$, $-21.5$, 
and $-22.3$
and surrounding regions, were run with twice the standard star-formation
efficiency. Comparing the results of these simulations to the
corresponding standard ones, it was found that the total mass of
gas-phase oxygen at $z$=3 was virtually unchanged, and that the mass
of $\log(T)$=4 oxygen was reduced by $\la$5-10\% compared to the standard
case. Hence, none of the main conclusions about $\log(T)$=4 metals, obtained in
this and the following sections, are affected by such a change. Moreover,
a doubling of the star-formation efficiency would lead to too small
$z$=0 disc galaxy gas fractions, and likely also too small disc galaxy angular
momenta (SGP03), likely too small $z$$\sim$3 DLA cross-sections
(Ellison \etal 2007, Sommer-Larsen \etal 2007), and star-formation
rates above the Kennicutt (1998) relation.
 
\section{\hbox{C\,{\sc iv}} properties of the low to intermediate density IGM}
\label{sect:CIV}
\begin{figure*}
\leavevmode
\psfig{file=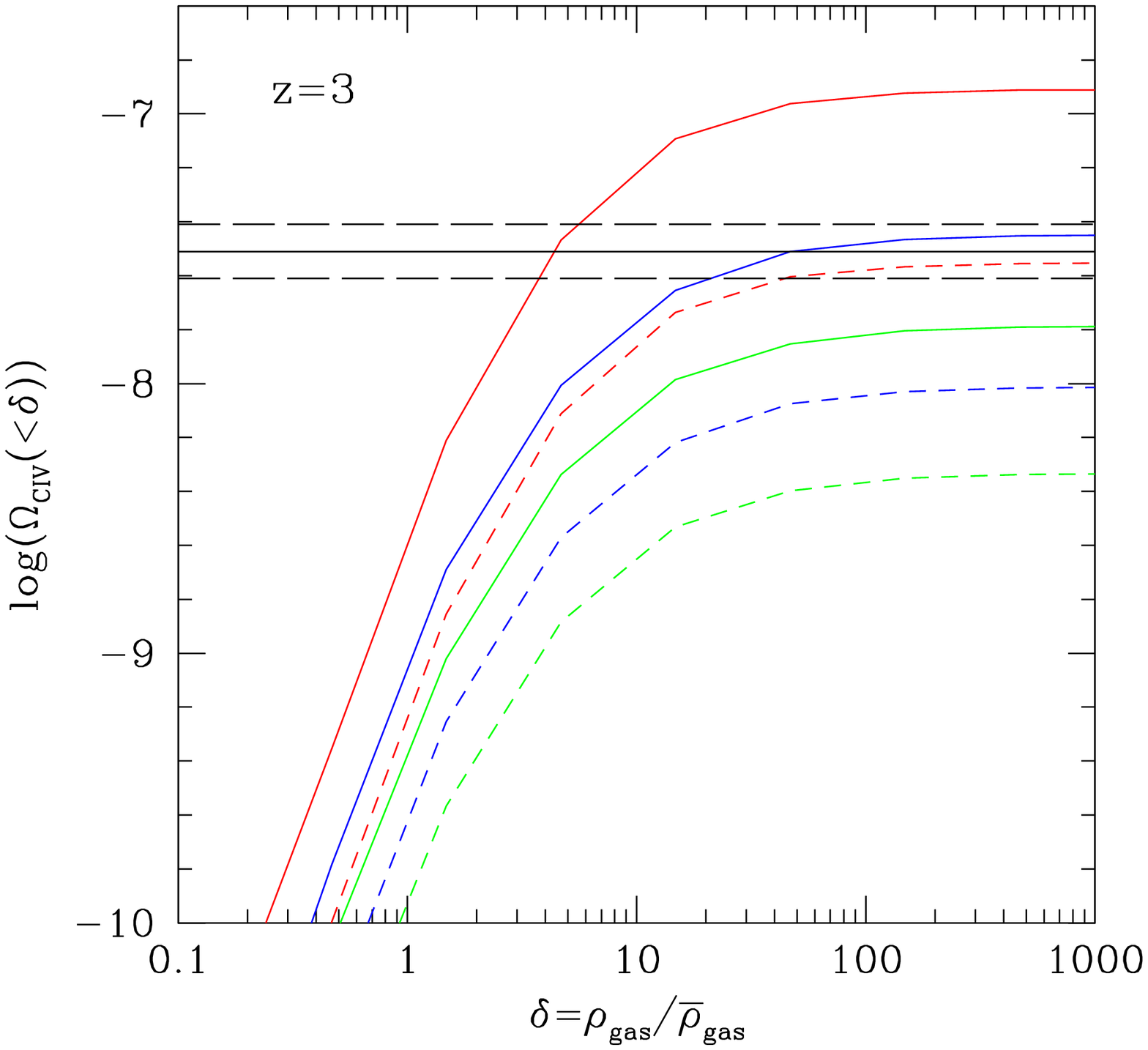,width=0.41\textwidth}
\psfig{file=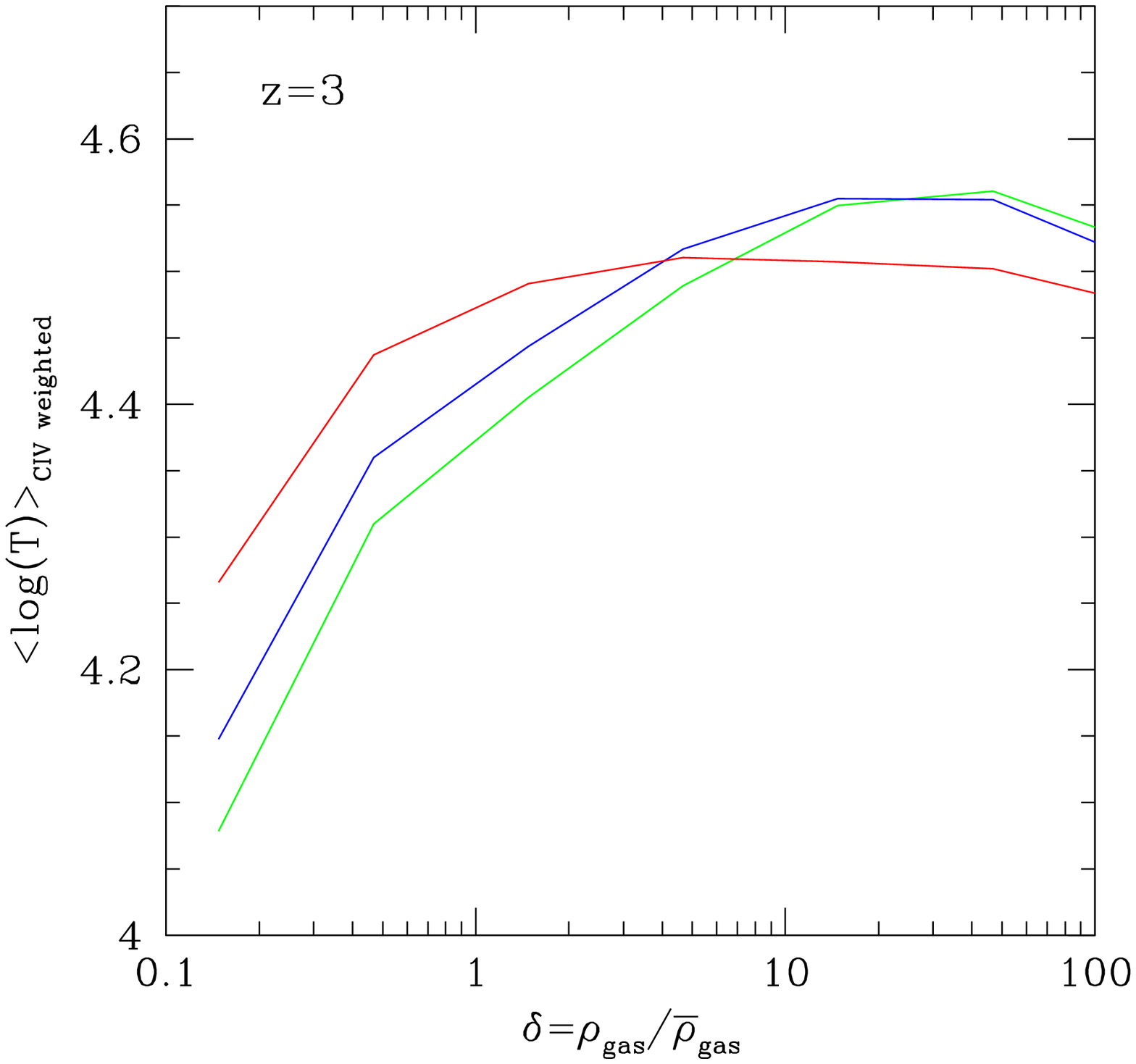,width=0.41\textwidth}

\caption{Cumulative cosmic \hbox{C\,{\sc iv}} density, in units of the critical,
$\Omega$(\hbox{C\,{\sc iv}}) as a function of the gas over-density 
$\delta=\rho_{\rm{gas}}/\bar{\rho}_{\rm{gas}}$ (left). Green, blue and red curves
correspond to the K98, Salpeter and AY IMFs. Solid curves corresponds
to upper limits, dashed to lower. Also shown, as a horizontal, black
solid line, is the median
of ten observational estimates at redshifts in the range 1.5-4.5 taken
from Songaila (2001), Boksenberg, Sargent \& Rauch (2003) and Songaila (2005).
Horizontal black dashed lines show 1-$\sigma$ deviations, where $\sigma$
is the variance of the observational estimates. The plot to the right
shows the median \hbox{C\,{\sc iv}} abundance weighted temperature of the IGM as
a function of gas over-density, for the three IMFs.}
\label{fig:CIV}
\end{figure*}
Interesting tests of the abundance and thermal properties of the low
to intermediate
density IGM are provided by OSO absorption line studies of the
Lyman metal forest, as probed by \hbox{C\,{\sc iv}}. 
Songaila (2001) first determined $\Omega$(\hbox{C\,{\sc iv}}) by 
integrating the
total column density of systems between $10^{12} \leq N$(\hbox{C\,{\sc
iv}})$ < 10^{15}$ cm$^{-2}$ --- see also Schaye \etal (2003). 
As shown by, e.g., OD06, this range of column densities trace moderate IGM
gas over-densities, $\delta=\rho_{gas}/\bar{\rho}_{gas}\sim 1-100$, or
$n_{\rm{H}}\sim10^{-5}-10^{-3}$ cm$^{-3}$, at $z$$\sim$3. To obtain a
corresponding estimate of $\Omega$(\hbox{C\,{\sc iv}}) based on our
simulations we adopted the following approach: 

To convert from total C (as given in our simulations) to \hbox{C\,{\sc iv}} abundances,
ionisation corrections are required. We base our conversion on ionisation 
corrections determined by OD06 using CLOUDY
(Ferland \etal 1998). OD06 assumed the Haardt \& Madau (2001) 
$z$$\sim$3 UV background field divided by a factor 1.6 at all redshifts, 
in order to
match the observed mean Ly$\alpha$ flux decrement, $D_{Ly\alpha}$. 
This may seem inconsistent with our use of the Haardt \& Madau (1996) UVB
in the cosmological simulations, but as shown by Croft \etal (1998),
because photo-ionisation is sub-dominant in gas dynamics such a 
correction yields virtually identical results as having done the 
simulations with the above (reduced) background. The resulting ionisation
corrections were kindly supplied to us by Ben Oppenheimer and Romeel 
Dav\'e, in the form of look-up tables.  

\begin{figure*}
\leavevmode
\psfig{file=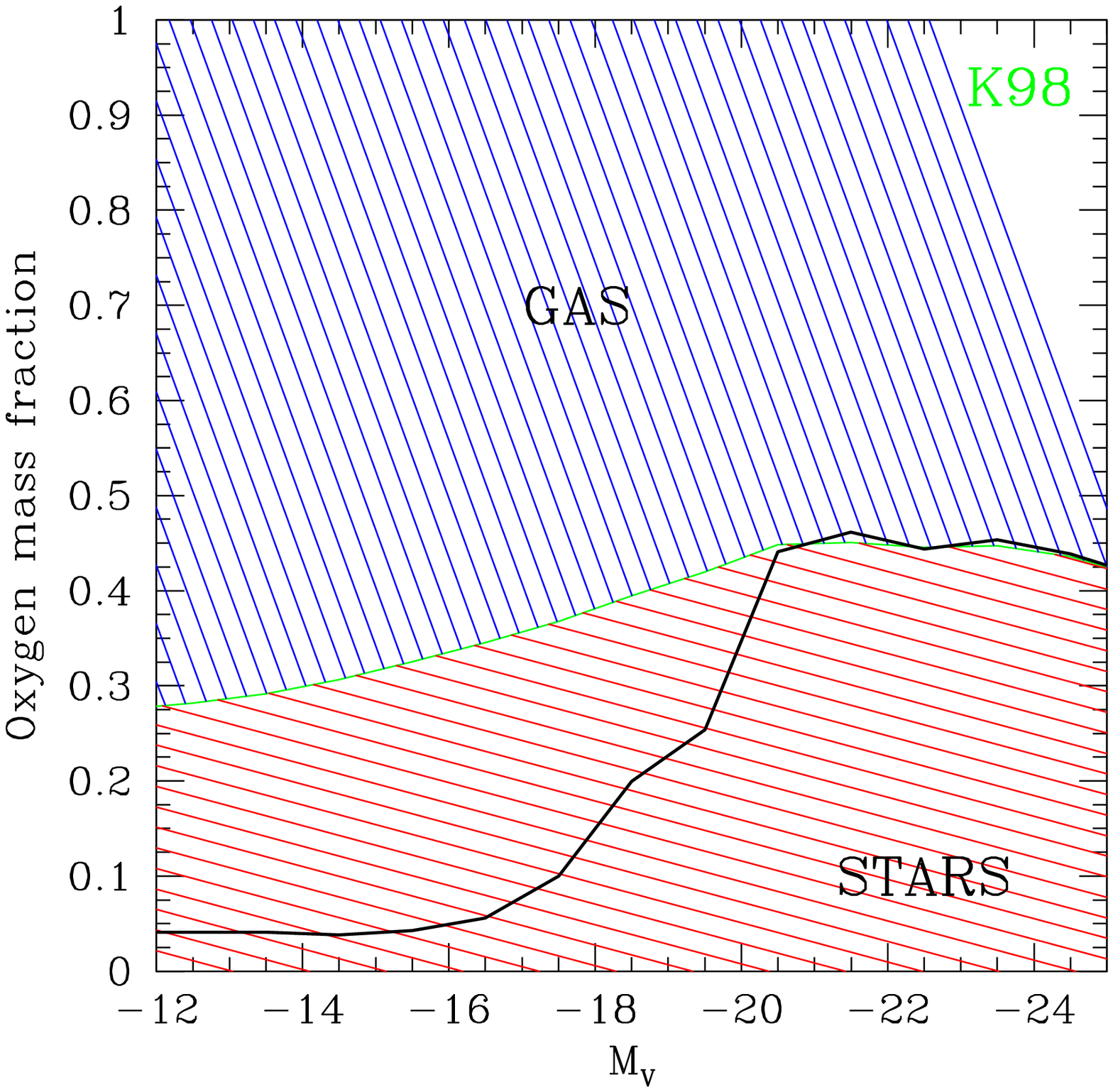,width=0.41\textwidth}
\psfig{file=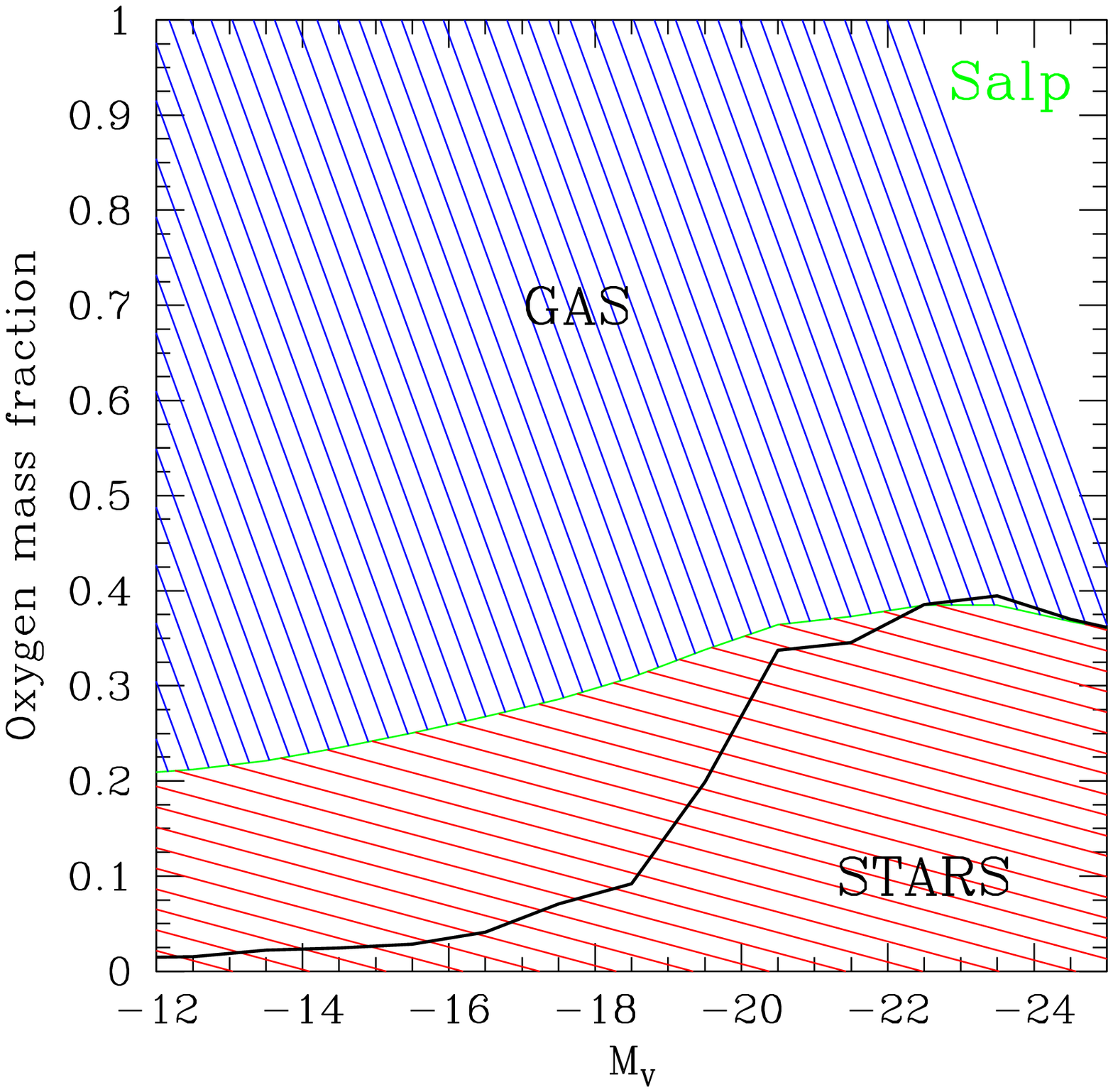,width=0.41\textwidth}

\leavevmode
\psfig{file=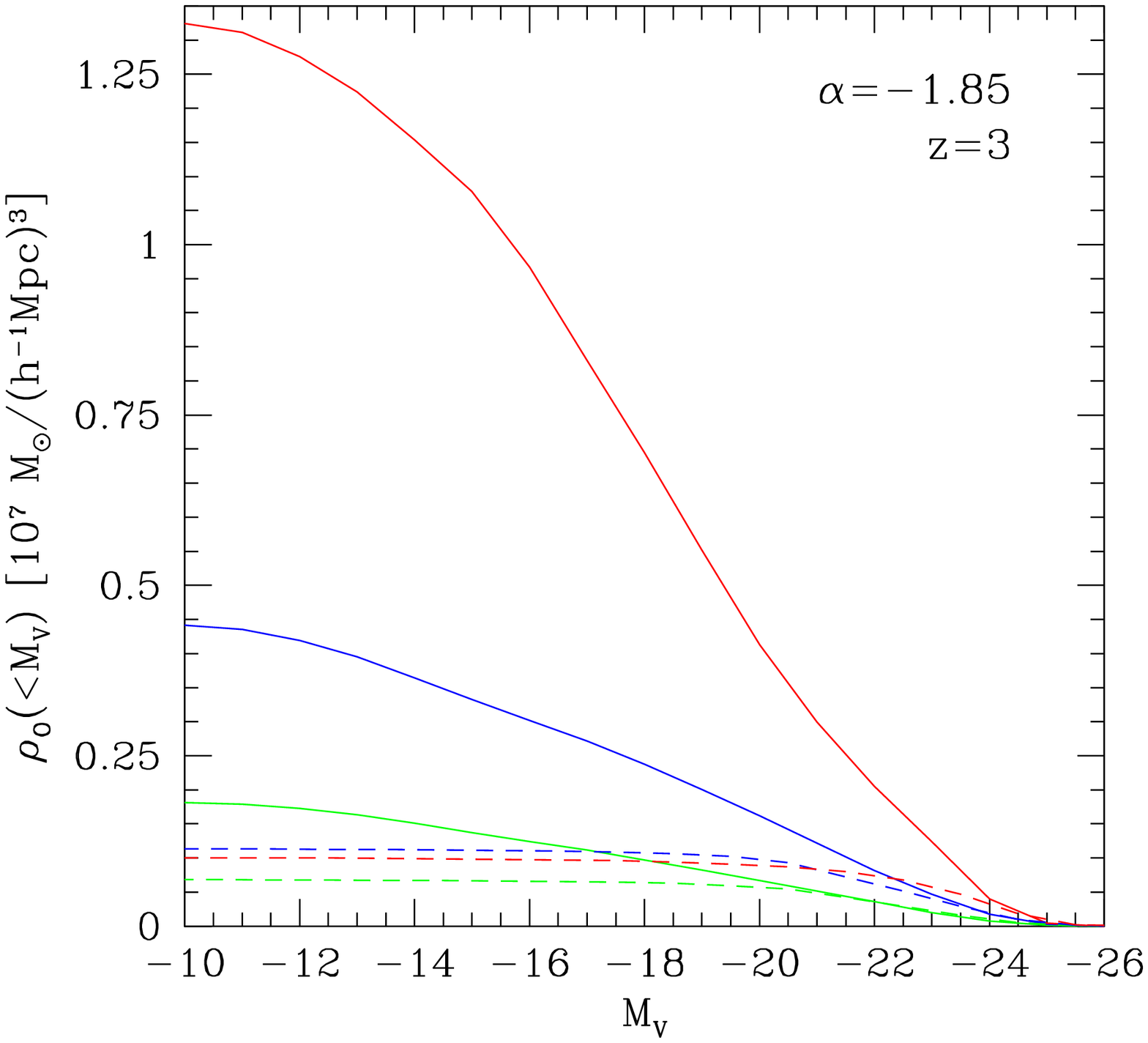,width=0.41\textwidth}
\psfig{file=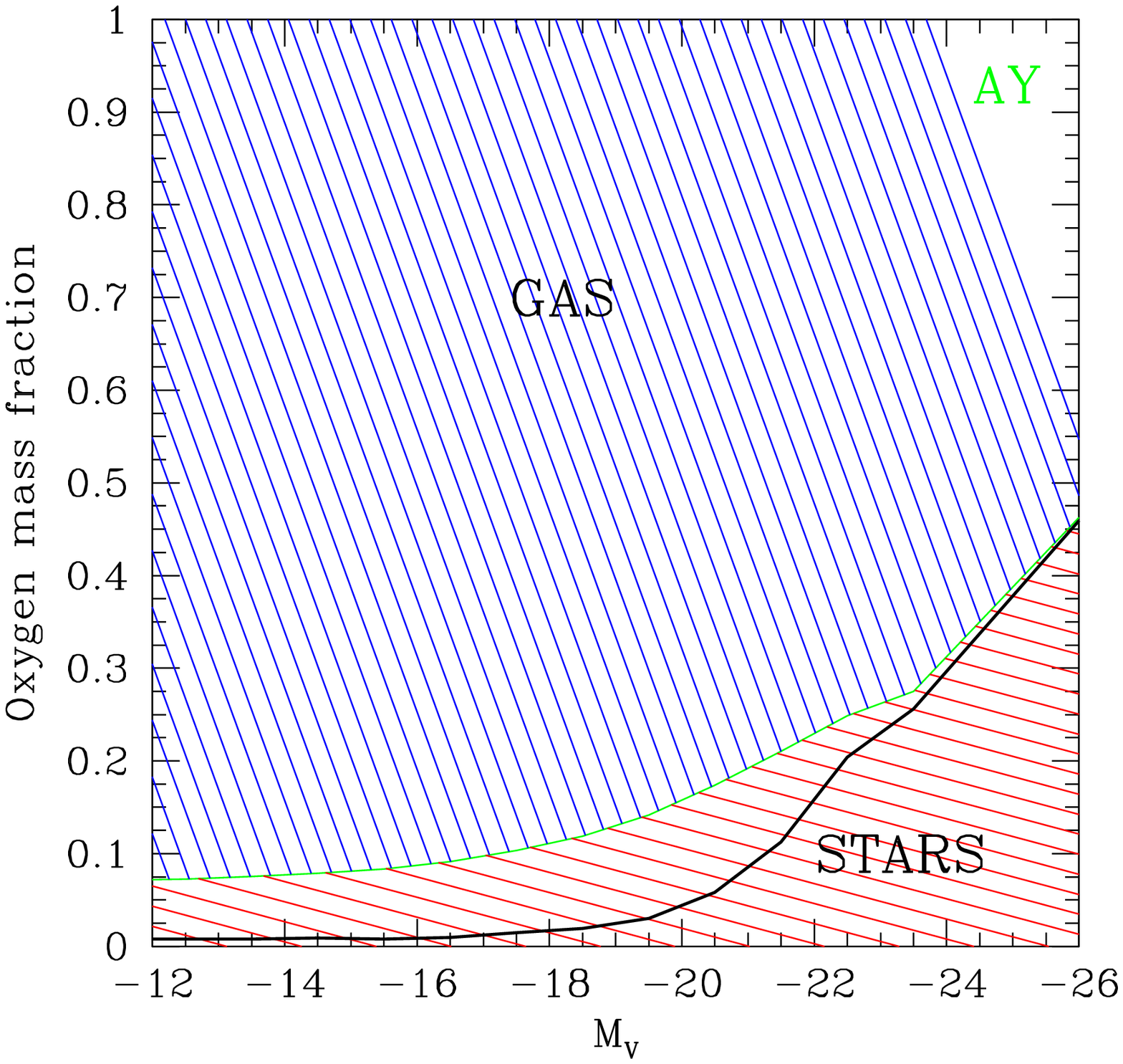,width=0.41\textwidth}
\caption{Cumulative fractional contributions of oxygen mass in stars and
gas, up to a given $M_{\rm{V}}$, are shown by the green solid lines,
separating the red and blue hatched regions (stellar/gas-phase
oxygen). All results are for the $\alpha=-1.85$ LF, and the plots shown
are for the K98 (top left), Salpeter (top right) and AY (lower right) IMFs.
Also shown in these plots, by black solid lines, is the fractional
partition between oxygen mass in stars and gas at any given $M_{\rm{V}}$.
Finally, in the lower left plot is shown the cumulative cosmic oxygen
density in gas+stars (solid lines) and stars only (dashed lines). In this
plot the IMFs are colour coded as follows: green: K98, blue: Salpeter,
red: AY. For comparison with the cosmological density parameter, see caption of 
Fig.~\ref{fig:cosmic}.}
\label{fig:gasstar}
\end{figure*}
In Fig.~\ref{fig:CIV} (left) is shown the cumulative 
$\Omega$(\hbox{C\,{\sc iv}}) as a function of $\delta$. For the two
simulated ``field'' galaxy regions the average ratio of the cumulative
\hbox{C\,{\sc iv}} mass to total (gas-phase) oxygen mass has been determined. This
has then been multiplied by $\Omega_{O,gas}$, determined on the
basis of the results obtained in the previous section. As is clear from
this section, the value of $\Omega_{O,gas}$ depends on the actual form
of galaxy luminosity function. To obtain lower and upper limits on
$\Omega_{O,gas}$, we evaluate it at $M_{\rm{V}}$=$-21$ and $-10$, respectively,
assuming the $\alpha=-1.85$ luminosity function.
$M_{\rm{V}}$=$-21$ corresponds approximately to the detection limit of
Shapley \etal (2001), whereas $M_{\rm{V}}$=$-10$ corresponds to assuming that
the faint end slope remains constant to this very faint magnitude.
As can be seen from Fig.~\ref{fig:ocum}, using the $\alpha=-1.57$
LF provides less conservative limits. In the figure is shown the
result of applying the lower and upper limits for the K98, Salpeter
and AY IMFs, respectively. Using other simulated regions, centered
on smaller galaxies, yields similar results. Using the proto-cluster
region to determine the average ratio of the cumulative
\hbox{C\,{\sc iv}} mass to total (gas-phase) oxygen mass results in lower values 
of $\Omega$(\hbox{C\,{\sc iv}}), due to the temperature dependence
of the C/\hbox{C\,{\sc iv}} ionisation correction combined with the higher average
IGM temperature of the proto-cluster region. However, proto-cluster
regions are very special sites, and, as can be seen from the results
of the previous section, the contribution to the cosmic metal production
associated with the very luminous galaxies typical of such regions,
$M_{\rm{V}}\sim-25$, is very small (in the proto-cluster regions containing
100-200 galaxies, almost half of the gas-phase metals are associated
with five galaxies of $M_{\rm{V}}\la-24$). Also shown in the figure is the
median value of 10 observational estimates in the redshift range
$z$=2-4, where the run of observed $\Omega$(\hbox{C\,{\sc iv}}) with redshift
is essentially flat (e.g., Songaila 2001, Schaye \etal 2003, Songaila 2005),
as well as
the observational variance. The observational estimates were taken from
Songaila (2001), Boksenberg, Sargent \& Rauch (2003) and Songaila (2005).

As can be seen, the chemical yield of the
K98 IMF appears too low to match the observational estimates, whereas
the Salpeter and AY IMFs can match the observations, and an IMF with
a yield in between these two appears optimal. We caution, however, that
our estimates of $\Omega$(\hbox{C\,{\sc iv}}) depend on the adopted
gas-phase oxygen to \hbox{C\,{\sc iv}} mass conversions. The optimal
would be to carry out simulations of $\ga$100 $h^{-1}$Mpc box size 
cosmological volumes at the (high) resolution of our galaxy formation
runs, but as mentioned previously, this is currently computationally
prohibitive. Hence, our result above on the cosmic IMF should be seen
as indicative only. The main point is that both the Salpeter and AY IMF
simulations can match the observational constraints on 
$\Omega$(\hbox{C\,{\sc iv}}).   

Observations of \hbox{C\,{\sc iv}} line widths for the above column density range can
be used to probe the thermal properties of the low-density IGM. Three
components contribute to the line widths: a) thermal broadening, b) spatial
broadening due to Hubble expansion across the physical extent of the absorber,
and c) turbulent broadening. The last component can be ignored, 
since the IGM at these over-densities is very quiescent (Rauch \etal
2005). 
Observational estimates of \hbox{C\,{\sc iv}} line widths by Boksenberg, Sargent \& 
Rauch (2003) indicate $b\simeq$10 km/s in the range $z\sim$1.5-4.5
for $10^{13} \leq N$(\hbox{C\,{\sc iv}})$ < 10^{14}$ cm$^{-2}$
absorbers. This can be used to derive upper limits to the 
\hbox{C\,{\sc iv}} abundance weighted IGM temperature. As the 
thermal component of the \hbox{C\,{\sc iv}} line width is given by 
$b_{th}=3.7 \sqrt{T/10^4}$, \hbox{C\,{\sc iv}} abundance weighted IGM temperatures of 
less than $\sim$5$\times$10$^4$ K are indicated. In Fig.~\ref{fig:CIV} 
(right) we show \hbox{C\,{\sc iv}} abundance weighted IGM temperatures a function 
of $\delta$ in the ``field'' galaxy regions for the three IMFs considered
(results for other regions are quite similar).
It is seen that models in
general satisfy the above IGM temperature criterion. Moreover, 
our predictions of the abundance weighted IGM temperatures agree well 
with the predictions of OD06 
for their best fitting ``momentum-driven'' wind model. 

Finally we note, that in the future it will be possible to probe
the metal enrichment of the IGM to even lower densities using 
\hbox{O\,{\sc vi}}--- see, e.g., Schaye \etal (2000).

\section{Combining the cosmic stellar and gas-phase oxygen distributions}
\label{sect:comb}
With the results on the cumulative cosmic gas-phase oxygen 
distributions presented in the previous section, and the results on the similar
distributions for the stellar oxygen from paper I, we can now assess the 
relative importance of
the two components. In Fig.~\ref{fig:ocum} results for the normalised
gas-phase oxygen 
distributions are shown by long-dashed curves, 
the stellar distributions are shown by the short-dashed curves, and the
combined distributions by the solid curves (for the three IMFs considered)
for the $\alpha=-1.85$ and -1.57 LFs (we note that stellar oxygen refers
to oxygen in galactic stars only, but since the amount of oxygen
in intergalactic stars is, in comparison, very small (paper I), we 
neglect this component in the following).

As can be seen, the combined
distributions are for all three IMFs dominated by the gas-phase oxygen,
and increasingly so going from the K98 to the AY IMF.
For the $\alpha=-1.85$ LF, it is found that less than about 25\% of the 
total cosmic oxygen is associated with galaxies brighter than $M_{\rm{V}}=-22$,
irrespective of the
choice of IMF. For the $\alpha=-1.57$ LF the corresponding fraction
is about 35\% for the AY IMF, and about 40\% for the K98 and Salpeter
IMFs.

Fig.~\ref{fig:gasstar} shows, for the $\alpha=-1.85$ luminosity function,
and the three IMFs, the cumulative partition between gas-phase oxygen and 
stellar oxygen as a function of $M_{\rm{V}}$. It is seen, that if the faint
end slope of -1.85 holds down to $M_{\rm{V}}\sim-12$, then the cumulative
gas-phase oxygen fractions are 72, 79 and 92\% for the K98, Salpeter
and AY IMFs, respectively, so, as stated above,
for all three IMFs, the amount gas-phase oxygen dominates over the amount
stellar oxygen. For the $\alpha=-1.57$ luminosity function these fractions
drop slightly to 64, 71 and 88\%.

As discussed in Sec.~\ref{sect:cosmic} the calculations assume a
constant faint end slope down to $M_{\rm{V}}$=$-10.5$. If the luminosity function
is assumed to display a significant flattening at a brighter magnitude
than this, then the figure can still be used to determine the 
gas-phase/stellar oxygen partition to such a limiting $M_{\rm{V}}$.
Assuming, for example, that the faint end slope is constant to at
least 6 magnitudes below $M^*$ (\ie $M_{\rm{V}}$$\sim-17$), 
which is the case for local luminosity
functions (Sec.~\ref{sect:cosmic}), results in lower limits on the
cumulative gas-phase oxygen fractions of 65, 72 and 90\% for the
$\alpha=-1.85$ luminosity function, \ie not much
different from the results quoted above.

\begin{figure}
\psfig{file=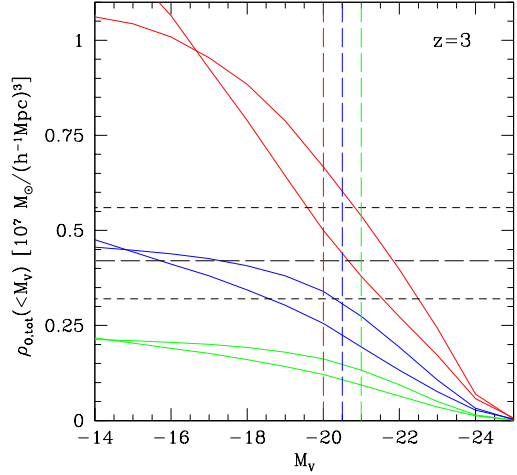,width=0.43\textwidth}
\caption{Contribution to the cosmologically averaged total
(gas-phase and stellar) oxygen density by galaxies brighter than $M_{\rm{V}}$, 
for the modified
Shapley \etal luminosity function, and the luminosity function of
faint end slope -1.57. Green, blue and red curves correspond
to results for the K98, Salpeter and AY IMFs, respectively.
(for comparison with the cosmological density parameter, see caption of 
Fig.~\ref{fig:cosmic}).
In the right side of the plot, the $\alpha=-1.57$ curves are
above the corresponding $\alpha=-1.85$ ones. Also shown, by horizontal
dashed lines, are the
constraints from the integrated cosmic oxygen production history derived
in Sec.~\ref{sect:UV}. Finally, the lower V-band luminosity limits,
at which the predictions for the various IMFs and LFs should be compared
to the integral constraints, are shown by green, blue and red vertical
dashed lines for the K98, Salpeter and AY IMFs, respectively --- see
text for details.}
\label{fig:cumOdens}
\end{figure}

We can now finally compare the integral constraints on the average
cosmic oxygen density at $z$=3, obtained in Sec.~\ref{sect:UV}, to
what is obtained by combining detailed high-resolution galaxy formation
simulations with the observed (corrected) $z$$\sim$3, V-band luminosity
function. In Fig.~\ref{fig:cumOdens} we show the cumulative oxygen
density versus $M_{\rm{V}}$ for the three IMFs and the two faint end luminosity
function slopes considered. Also shown are the constraints from the
``minimum'', ``median'' and ``maximum'' oxygen production rate density
models discussed in Sec.~\ref{sect:UV}. In order to enable a meaningful
comparison, the latter results should be compared to the former
evaluated at certain limiting magnitudes. These magnitudes should be chosen
to be brighter than or equal to about $M_{\rm{V}}\simeq-21.0$, $-20.5$ and 
$-20.0$
for the K98, Salpeter and AY IMFs, respectively, for the following reasons: 

Kennicutt (1998) finds a relation between UV luminosity and star formation
rate (SFR), viz.
\begin{equation}
SFR~(M_{\odot}\rm{yr}^{-1}) = 
1.4\times10^{-28}~\beta_{\rm{IMF}}~L_{\nu}(\rm{ergs}~\rm{s}^{-1}\rm{Hz}^{-1}),
\end{equation}
with $\beta_{\rm{IMF}}$=1.0 for the Salpeter IMF. Given that the yield
of the Kroupa (1998) IMF is smaller than for the Salpeter IMF, and vice 
versa for the Arimoto-Yoshii IMF, $\beta_{\rm{IMF}}$ will be larger and
smaller than unity for these two IMFs, respectively. Quantitatively we
find that $\beta_{\rm{IMF}}\simeq$1.7 and 0.4 for the K98 and AY IMFs,
respectively.

The absolute UV magnitude ($\lambda\sim$1500 {\AA}) characterising
the observed galaxy luminosity function at $z$$\sim$3 (\ie corresponding
to $L^*_{UV,z=3}$) is $M^*_{UV,z=3}$$\simeq-21.0$ (e.g., Steidel \etal
1999, after correction to the adopted cosmology, Sawicki \& Thomson 2006).
Following Sawicki \& Thomson (2006), and using eq.\,5 above, one can show
that this absolute magnitude corresponds to a (un-extincted) star formation
rate of about 15~$\beta_{\rm{IMF}}$ \Msol/yr. With a dust correction of
about a factor 5.5 at $z$$\sim$3 (Sec.~\ref{sect:UV}) the above 
observed $M^*_{UV,z=3}$ hence corresponds to a true (un-obscured)
SFR of about 82~$\beta_{\rm{IMF}}$ \Msol/yr.
We shall now consider two models for the $\lambda\sim$1500 {\AA}
extinction as a function of redshift for $z\ge$3. Model A assumes
a constant extinction factor of 5.5 at all $z\ge$3 (cf. Sec.~\ref{sect:UV}),
whereas model B is a low-extinction model assuming 
factors of 4.2, 3, 2 and 1.5 at $z$=3, 4, 5
and 6, respectively (cf. Bouwens \etal 2006). For model A, a galaxy
of {\it observed} $L_{UV}=0.1\times$$L^*_{UV,z=3}$ will be characterised
by a SFR$\sim$8~$\beta_{\rm{IMF}}$ \Msol/yr. From our large sample of 
galaxy models we
find that galaxies with such star formation rates at {\it any} $z\ge$3
will have an (un-extincted) $M_{\rm{V}}$ at $z$=3 brighter than about $-22.0$, 
$-21.5$
and $-20.5$ for the K98, Salpeter and AY IMFs, respectively.
For model B, the above $M^*_{UV,z=3}$ corresponds to a true SFR of
about 63~$\beta_{\rm{IMF}}$ \Msol/yr, and
galaxies of observed $L_{UV}=0.1\times$$L^*_{UV,z=3}$  
will have SFRs of about 6.3, 4.5, 3.0 and 2.3 \Msol/yr at $z$=3, 4, 5
and 6, respectively. We find that only galaxies of (un-extincted) $M_{\rm{V}}$
at $z$=3 brighter than about $-21.5$, $-21.0$ and $-20.5$ for the above three
IMFs, respectively, will satisfy this. Moreover, even assuming (very
conservatively)
zero extinction at $z\ga$6, corresponding to a ``limiting'' (un-extincted) 
SFR of about 1.5 \Msol/yr, all galaxies of such $z\ga$6 SFRs
will be be brighter than about $M_{\rm{V}}$=$-21.5$, $-21.0$ and $-20.5$ at $z$=3 for
the three IMFs, respectively. Assuming 
lower luminosity limits of $M_{\rm{V}}$=$-21.0$, $-20.5$ and $-20.0$ for the three
IMFs is hence very conservative --- these 
limits are indicated in Fig.~\ref{fig:cumOdens} by vertical dashed lines. 

From Fig.~\ref{fig:cumOdens} it follows that galaxy models based on the
K98 IMF can not meet the constraint set by the observed UV luminosity
density history --- the oxygen (and general metal) yield of this IMF is
simply not sufficiently large. The same is the case for the Salpeter IMF, 
though the $\alpha=-1.57$ model comes close to matching the lower bound of
0.32$\times 10^7$ \Msol/($h^{-1}$Mpc)$^3$ (see also below). 
On the other hand, models based on the Arimoto-Yoshii IMF, match the
integral constraint well. 

Summarising, one of the main results obtained in
this paper is that galaxy formation models based on the Kroupa (1998) IMF
(and any other IMF of similar chemical yield) are strongly excluded by
the cosmic enrichment history constraint. Other arguments, why the average
cosmic IMF must have a larger chemical yield than what is typical for a 
solar neighborhood one, have been
given by, e.g., Portinari \etal (2004), D'Antona \& Caloi (2004), 
Serjeant \& Harrison (2005), Lucatello \etal (2005), Loewenstein (2006), 
Prantzos \& Charbonnel 
(2006) and Weidner \& Kroupa (2006), but see also Elmegreen (2006).  

Perhaps even more interestingly,
models based on the Salpeter IMF, the arguably most extensively used model 
of the
stellar IMF, are also excluded, though the $\alpha=-1.57$ models only
marginally so. Given that the Salpeter IMF is routinely used in
translating UV luminosities into star formation rates for high redshift
galaxies, this is also an important result. Although it would be 
inappropriate to assign a high statistical significance to this result,
given all the uncertainties involved, it is obvious from 
Fig.~\ref{fig:cumOdens} that the models based on the more top-heavy
Arimoto-Yoshii IMF provide a much better match to the median value
for the integral constraint. This is in particular the case, since the
above lower absolute luminosity limits of $M_{\rm{V}}$=$-21.0$, $-20.5$ and 
$-20.0$ 
for the three IMFs, respectively, are likely to be very conservative.

\section{Missing metals and comparison to other works}
\label{sect:mm}
In Sec.~\ref{sect:cosmic} it was shown that for the $\alpha=-1.85$ $z$$\sim$3 
luminosity function, less than about 1/4 of the cosmic oxygen is associated
with Lyman Break galaxies sufficiently bright, $M_{\rm{V}}$$\ga-22$, for direct 
abundance determination, using oxygen lines emitted from HII regions around
young stars in the galaxies (e.g., Pettini \etal 2001). For the $\alpha=-1.57$
luminosity function this fraction increases slightly, to about 35\%.
Furthermore,
in Sec.~\ref{sect:comb} it was shown that for both luminosity functions 
the major part of the cosmic oxygen is in
the gas-phase, rather than in stars. In particular, for the Salpeter and
Arimoto-Yoshii IMFs, which emerge from the previous section as the more
plausible, the gas-phase oxygen fraction exceeds 70\%. 

FSB05 found a factor of about five discrepancy between
the amount of metal in the stars of Lyman Break galaxies and what is
predicted from the integrated UV luminosity density history. They
assumed typical Lyman Break galaxy stellar masses of 
2~$\times 10^{10}$ \Msol, corresponding to $M_{\rm{V}}$$\sim-22.5$, \ie close
to $M^*_{V,z=3}$. 

In Fig.~\ref{fig:gasstar} is shown, for the
$\alpha=-1.85$ luminosity function, the cumulative stellar cosmic oxygen 
density. For the K98 IMF, the ratio between the stellar oxygen density to
$M_{\rm{V}}=-21$ and what is obtained from the median model 
(Sec.~\ref{sect:comb}) is 0.10. For the
Salpeter and AY IMFs, evaluated at $M_{\rm{V}}=-20.5$ and $-20$, respectively,
the corresponding ratios are 0.17 and 0.20. For the $\alpha=-1.57$
luminosity function the corresponding ratios are 0.14, 0.24 and 0.29,
for the K98, Salpeter and AY IMFs, respectively. Given that the
AY models over-predict the metallicities of Lyman Break galaxies
of $M_{\rm{V}}\sim-22$ to $-23$ by about 0.2 dex relative to observations (paper I),
the above ratios for the AY models should be reduced to 0.13 and 0.18 for
the $\alpha=-1.85$ and -1.57 luminosity functions, respectively. 

The above results are obtained, however, by including stellar oxygen 
mass all the
way down to $M_{\rm{V}}$$\sim-20.5$. If one only includes stellar oxygen mass
to $M_{\rm{V}}$$\sim-22.5$, all the above fractions are reduced by about a
factor of two. Hence the discrepancy discussed above, denoted by
FSB05 and others as ``the missing metals problem'', is actually
about {\it twice} larger than originally found by FSB05.

DO06 predicted that at $z$$\sim$3, about
50\% of the gas-phase metals should reside in the diffuse IGM 
(gas outside the virial radii of galaxy halos, and
of $T<3\times$10$^4$ K). We can not compare the results
found in the previous sections directly to theirs, since different gas-phase 
criteria have been used. However, our $\log(T)$=4.5 gas criterion is
similar to theirs for diffuse IGM, though we also include $\log(T)$=4.5 metals
inside of galaxy virial radii in our estimate (see also below; note
also that we find almost all the $\log(T)$=4 phase metals to reside in gas of 
fairly high density, $n_{\rm{H}}$$\ga$0.1 cm$^{-3}$, so there is essentially no 
overlap between this phase and DO06's diffuse phase). 

For the $\alpha=-1.85$ LF we find $\log(T)$=4.5 to total gas phase
metal fractions of 34, 31 and 26\% for the K98, Salpeter and AY IMFs,
respectively. For the $\alpha=-1.57$ LF, the corresponding fractions are
21, 20 and 18\%. If we try to mimick the criterion of DO06 better, by
selecting all metals in gas of $T$$<$3$\times$10$^4$ K and 
$\delta=\rho_{gas}/\bar{\rho}_{gas}$$\le$100, then for the field
galaxy regions we find ``diffuse IGM'' oxygen fractions of 4, 4 and 6\%
for the K98, Salpeter and AY IMFs, respectively. If we use the
proto-cluster regions, which are the only simulations at our
disposal with a resolution comparable to the (fairly modest) numerical
resolution of DO06's 32 $h^{-1}$Mpc box size simulation, the corresponding
fractions drop to $\sim$1\% for all IMFs. However, we stress that
the temperature of the proto-cluster IGM is larger than that of the
average IGM, but in any case significantly lower ``diffuse IGM''
metal fractions, than predicted by DO06, are indicated.  

\begin{figure*}
\leavevmode
\psfig{file=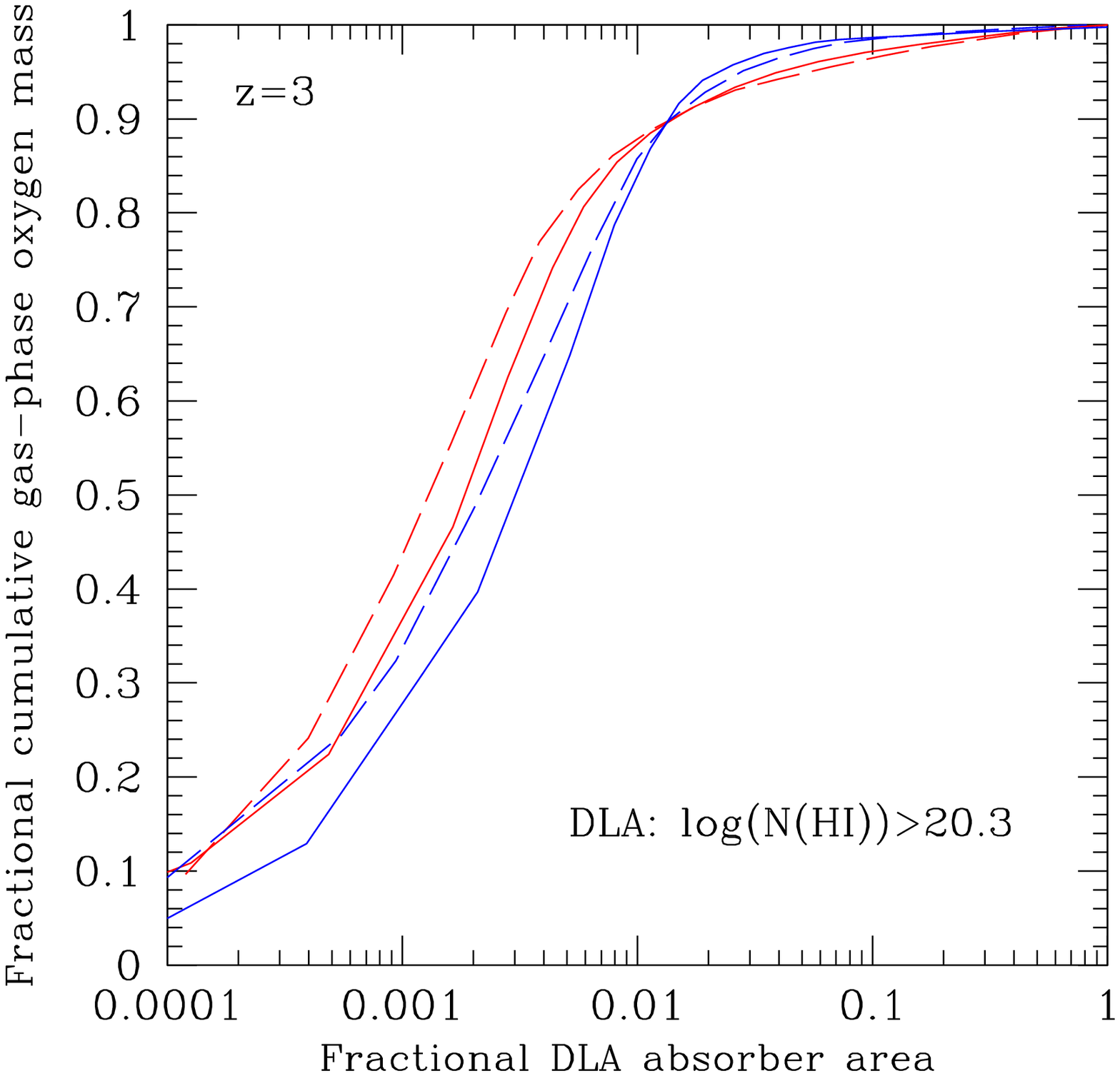,width=0.33\textwidth}
\psfig{file=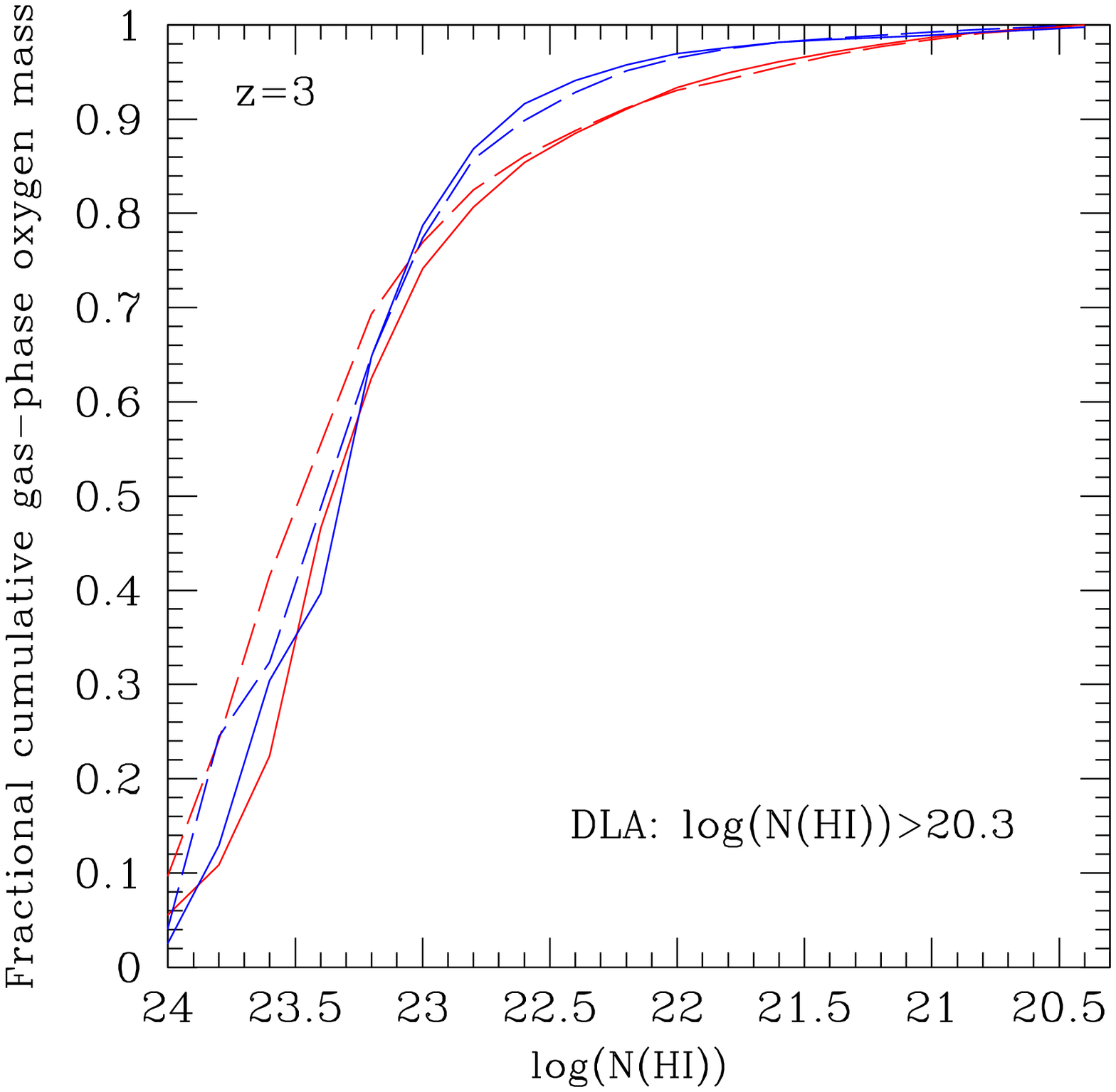,width=0.33\textwidth}
\psfig{file=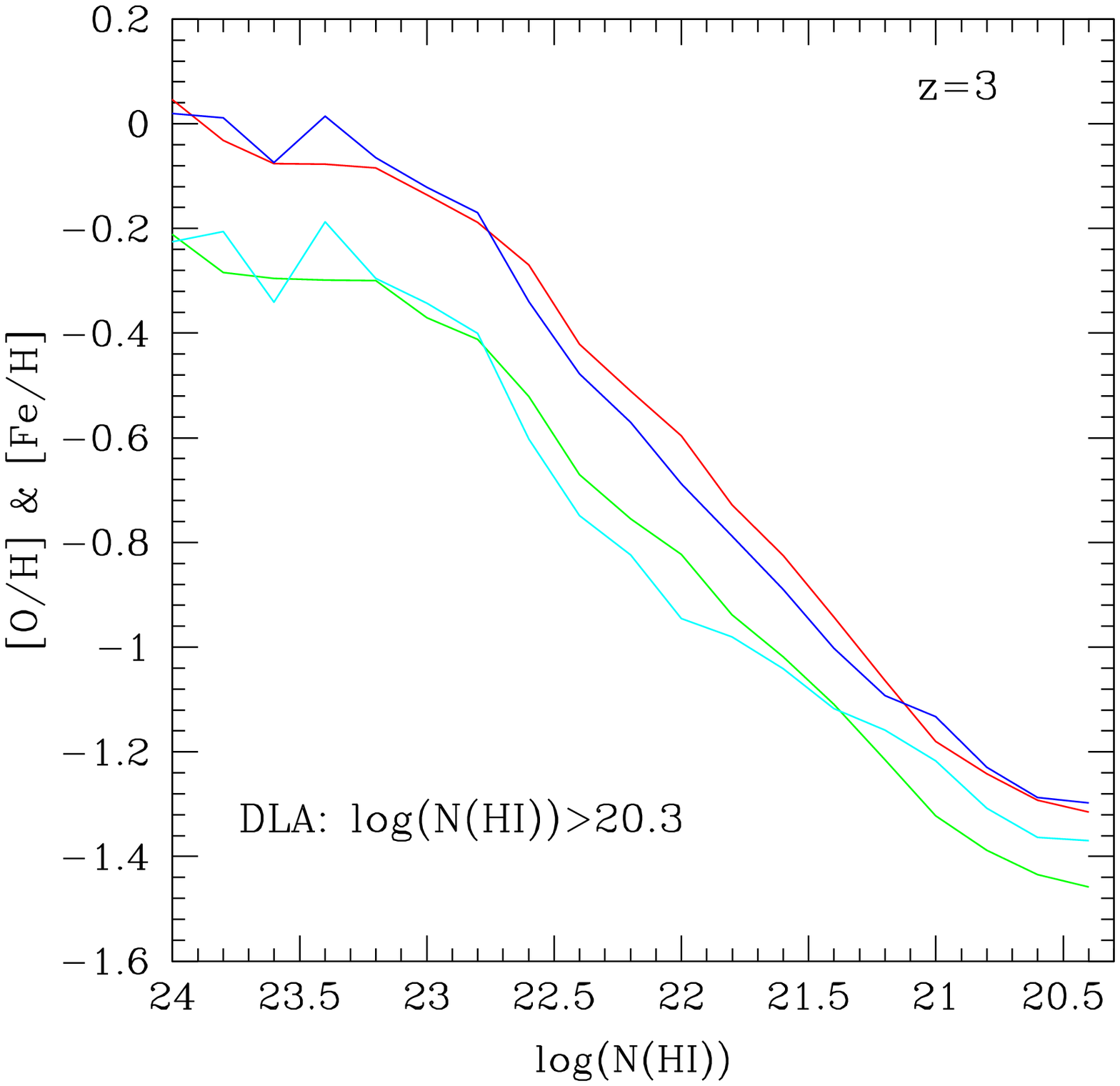,width=0.33\textwidth}

\caption{{\it Left}: Fractional cumulative oxygen mass in the absorbers above 
log(N(HI))=20.3 versus
the cumulative absorber area, starting at the highest HI surface
densities. Red and blue curves correspond to the proto-galaxies of $z$=0 
$V_c$=245 km/s and 180 km/s, respectively. ``Face-on'' projections
are shown by solid and ``edge-on'' by dashed curves. {\it Center}:
Same, but versus HI column density. {\it Right}: Oxygen and iron
abundance versus HI column density; for clarity only ``face-on''
projections are shown. Red and blue curves show [O/H], green and
cyan [Fe/H].}
\label{fig:DLA}
\end{figure*}
DO06 used the Springel \& Hernquist (2003) ``sub-grid'' approach to
model star-formation. They predict that about 20\% of the gas-phase metals 
reside in star-forming gas. In our models, which invoke explicit
two-phase modeling of the ISM, gas of $T$$\sim$$10^4$ K and 
$n_H$$\ga$0.1 cm$^{-3}$ is potentially star-forming. Hence we compare
our results for the $\log(T)$=4 gas-phase metal content to the above of DO06
(as noted above, almost all $\log(T)$=4 metals reside in gas of 
$n_H$$\ga$0.1 cm$^{-3}$). 
For the $\alpha=-1.85$ LF we find $\log(T)$=4 to total gas phase
metal fractions of 47, 37 and 26\% for the K98, Salpeter and AY IMFs,
respectively. For the $\alpha=-1.57$ LF, the corresponding fractions are
57, 47 and 32\%. Hence we find somewhat larger metal fractions in
star-forming gas, than do DO06. A more important difference is, however,
that DO06 at $z$$\sim$3 find no metal containing gas at densities
$n_H$$\ga$0.01 cm$^{-3}$. This result is strongly at variance with
our results, and it would seem difficult for such models to match
the large amount of DLA cross section observationally detected at
such redshifts, although further analysis obviously is required to 
clarify this. 

Finally, we can compare the $z$=3 gas phase metal temperature distribution
predicted by DO06 to our results. Qualitatively, our predictions for
the K98 and Salpeter IMFs agree with that of DO06, in yielding temperature
distributions, which peak at $T$$\sim$$10^4$ K and decrease towards larger
temperatures. However, for the AY IMF the results are very different.
Moreover, quantitatively our results for the K98 and Salpeter IMFs disagree
significantly with those of DO06 at the ``high-$T$'' end: DO06 find that
$\la$5\% of the gas-phase metals reside in gas of $T$$>$$10^{4.75}$ K.
In comparison, we find for the $\alpha=-1.85$ LF that the corresponding 
fractions
are 20, 32 and 51\% for the K98, Salpeter and AY IMFs, respectively, and
for the $\alpha=-1.57$ LF, 21, 32 and 50\%. Hence for any of the IMFs
considered, we predict significantly larger fractions of ``high-$T$''
gas-phase metals, than do DO06.  

\subsection{Metals in the cold gas-phase and DLA absorbers}
\label{sect:DLA}
The $\log(T)$=4 phase is found to be the most prominent metal containing 
gas-phase for the K98 and Salpeter IMFs, and also to be
quite prominent for the AY IMF. As the metals in this phase typically reside
in gas of densities $n_H$$\ga$0.1 cm$^{-3}$, one would expect
a significant fraction of these metals to be situated at column densities 
typical of DLAs, viz. N(HI)$\ge$10$^{20.3}$ 
cm$^{-2}$. However, in general DLAs have very low abundances - significantly 
less
than Galactic stars with similar ages (Pettini \etal 1990, 1994; Lu \etal 
1996; Kulkarni \& Fall 2002; Prochaska \etal 2003; Akerman \etal 2005; 
Zwaan \etal 2005; Erni \etal 2006).
Expressed in terms of the global metal content of the LBG stars,
we predict similar to three times as large amounts of metals in the $\log(T)$=4
phase (depending on the IMF), whereas observations of DLAs only indicate
a fraction of about 10-20\% (e.g., FSB05, Prochaska \etal 2006).  
It is obviously of importance to understand the reason for this
apparent discrepancy. Although we will defer a thorough discussion
of DLAs to a forthcoming paper (Sommer-Larsen \etal 2007), 
we briefly address the above issue in the following.

Ellison \etal (2007) analysed two of the very high resolution simulations,
described in the following section, in relation to DLA properties.
The two simulations represent the formation and evolution of two
disc galaxies, of $z$=0 characteristic circular velocities 
$V_c$=245 and 180 km/s. The galaxies were selected from a larger sample
to represent two different disc formation evolutionary paths: for the
$V_c$=180 km/s galaxy the disc starts growing steadily already by 
$z$$\sim$2.5, whereas for the other galaxy disc growth is merger
induced, and the disc grows strongly from $z$$\sim$1 to 0 (see also
Robertson \etal 2004). 
Ellison \etal determined the DLA/sub-DLA characteristics of these two 
proto-galaxies at $z$=3.6, 3.0 and 2.3, with focus on determining
neutral column density distributions, and the probability of detecting
coincident 100 kpc-scale DLA/sub-DLA absorption in individual
galaxies at such redshifts --- we refer the reader to Ellison \etal for
more detail, as well as images of the objects. 

Here we build on the analysis by Ellison \etal of these
two (proto-)galaxies. In Fig.~\ref{fig:DLA} (left) we show the
normalised cumulative oxygen mass in the absorber above log(N(HI))=20.3 versus
the cumulative absorber area, starting at the highest HI surface
densities, log(N(HI))$\sim$24, and going down to log(N(HI))=20.3.
For each galaxy results for ``face-on'' and one ``edge-on'' projections
are shown. As can be seen, about 90\% of the total oxygen mass of the
absorber resides in 1\% of the total absorber area for both projections.
From Fig.~\ref{fig:DLA} (center) it is moreover seen that this oxygen
mass is associated with column densities log(N(HI))$\ga$22.5. DLAs of
such high column densities have never been detected in QSO spectra, which 
of course is not surprising given the above findings. Moreover, at such high
densities, formation of molecular hydrogen is likely to take place (Schaye
2001b; Zwaan \& Prochaska 2006). Although the process of
H$_2$ formation is not included in the hydro/gravity simulations, it is
clear that the properties of the high density gas can be significantly
affected by H$_2$ formation (e.g., Pelupessy \etal 2006). 
Greve \& Sommer-Larsen (2007) showed that the effects of H$_2$ formation
can be approximately determined post-process on the basis of the
cosmological galaxy formation simulations --- this will be one of the
main topics of a forthcoming paper on DLAs (Sommer-Larsen \etal 2007).
H$_2$ molecules have been detected in DLAs, but only at relatively low
fractions of less than a tenth relative to atomic hydrogen (Ledoux \etal 2003).
Moreover, H$_2$ is preferentially detected in the highest metallicity DLAs (Petitjean \etal
2006). At this point it is sufficient to note that a) only a very small
fraction of QSO sight-lines will probe the high-metallicity regions of DLAs,
and b) due to H$_2$ formation these regions may possibly be difficult to probe
in neutral hydrogen (Zwaan \& Prochaska 2006). We also note from
Fig.~\ref{fig:DLA} (right) that the metal abundances of the high column density
regions are significantly larger than what is typically observed in DLAs,
whereas at log(N(HI))$\la$22 they are not (e.g., Pettini \etal 2003). Finally,
we note that the high column density regions will likely be characterised by
significant dust contents, which could further bias against selecting QSO
sight-lines passing through such regions. Evidence for dust in DLAs
has been reported (Pei, Fall \& Bechtold 1991; Vladilo \etal 2006). This
may naturally explain why DLA column densities of log(N(HI))$\ga$22 are very
rare (Vladilo \& P\'eroux 2005, Vladilo \etal 2006). However, studies of
radio selected DLAs (free from dust-bias) have found similar column density and
metallicity distributions as for optically selected DLA samples  (Ellison \etal
2001, 2004; Ellison, Hall \& Lira 2005; Akerman \etal 2005; Jorgenson \etal
2006). Also, Murphy \& Liske (2004) find no excess absorption towards QSOs with
DLAs (but this study is based on an optically selected DLA sample). It remains
to be clarified to which extent dust bias is important for DLA column density
and metallicity studies, especially at log(N(HI))$\ga$22.  
In particular, a larger sample of radio selected DLAs is needed 
(Jorgenson \etal 2006, Ellison 2007, private communication).

We note that DLAs at $z\approx2$ with near solar metallicity have
been found (Ledoux \etal 2002). Also, the Sloan Digital Sky Survey has been
used to strongly increase the sample of DLAs (Prochaska \etal 2005) and among
these 5\% have very strong metal-line absorption (Herbert-Fort \etal 2006).
Some of these 5\% are metal rich DLAs, although some could be relatively 
metal poor DLAs with very large hydrogen column densities. About 60\%
of the metal-strong DLAs have detected molecular hydrogen (Milutinovic \etal
in preparation).

Interestingly, DLAs with log(N(HI))$\ga$22 have been detected in the spectra of
the optical afterglows of Gamma-Ray Bursts (Watson \etal 2006; Jakobsson \etal
2006; Prochaska \etal 2007).  The selection function for GRB sight-lines is
completely different than the cross-section selection of QSO DLAs. GRB
sight-lines are strongly correlated with the blue light of their host galaxies
(Bloom, Kulkarni \& Djorgovski 2002; Fruchter \etal 2006) most likely tracing
the location of the most massive stars (M$\ga$20M$_{\odot}$,  Larsson \etal
2007). Significant dust obscuration is observed for at least some of the
log(N(HI))$\ga$22 GRB DLAs (Watson \etal 2006). Furthermore, it has been found
that the metallicities inferred for DLAs in GRB hosts are systematically higher
than for QSO DLAs and that GRBs hence may offer a new probe of early cosmic
chemical evolution complementary to the QSO DLAs and LBGs (Fynbo \etal 2006;
Savaglio 2006; Prochaska \etal 2007).

In conclusion, the low metallicities inferred from QSO DLAs do not 
exclude that significant amounts of metals can reside in 
the $\log(T)=4$ phase. The reason for this is most likely that the total 
cross-section for this component is only of order a few percent of the 
total DLA cross-section (see also Johansson \& Efstathiou 2006) 
and that the current DLA samlpes are too small 
to uncover them (see also Zwaan \& Prochaska 2006), but dust bias, 
as well as H$_2$ formation, will probably also be of importance.

\section{Numerical resolution}
\label{sect:numres}
\begin{figure}
\psfig{file=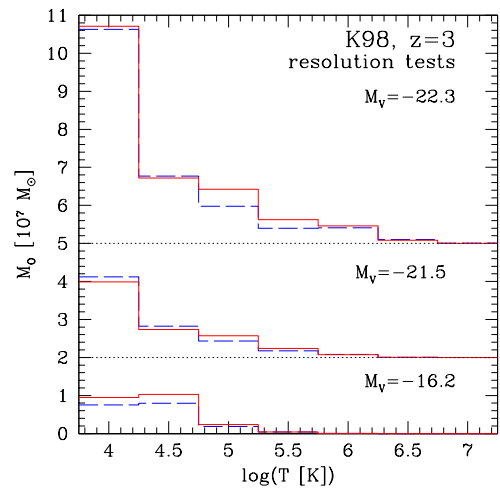,width=0.43\textwidth}
\caption{Gas-phase oxygen temperature distributions for three 
proto-galactic regions simulated at both very high (blue dashed
histograms) and normal (red solid histograms) numerical resolution.
The three proto-galaxies become at $z$=0 disc galaxies of characteristic
circular velocities $V_c$=245, 180 and 66 km/s, respectively. 
The oxygen masses for the smaller galaxy have been multiplied by
a factor 50 for clarity. The slight offset between the results for the
smaller galaxy is due to high resolution proto-galactic region being
slightly smaller than the corresponding normal resolution one.}
\label{fig:numres}
\end{figure}
In SGP03 and Sommer-Larsen (2006) it has been shown
that the results of cosmological galaxy formation and evolution using
our code are, in general, robust to changes of the numerical resolution.
However, in this paper we present gas-phase metal temperature distributions
and other results, which have not been presented before. It is hence
clearly important to demonstrate that these results are resolution robust 
as well. To this end we carried out three very high resolution
K98 IMF simulations, of eight times higher mass and two times higher force
resolution compared to ``standard''. We simulated
three (proto-)disc galaxies, which at $z$=0 have 
characteristic circular velocities of 245, 180 and 66 km/s. The galaxy
forming regions were represented by a total of about 2.2, 1.2 and 1.3 million
particles, compared to the about 290\,000, 150\,000 and 210\,000 particles used 
at ``normal'' resolution (for computational reasons the region
simulated at very high resolution for the last galaxy was slightly smaller
than the corresponding ``normal'' resolution one).
Gas and star particle masses were 
$m_{\rm{gas}}$=$m_*$=9.1$\times$10$^4$ and 2.5$\times$10$^3$
$h^{-1}$M$_{\odot}$ for the first two galaxies and the last one, 
respectively. The corresponding dark matter particle masses were
$m_{\rm{DM}}$=5.2$\times$10$^5$ and 1.4$\times$10$^4$ $h^{-1}$M$_{\odot}$.
Moreover, gravitational (spline) 
softening lengths of $\epsilon_{\rm{gas}}$=$\epsilon_*$=190 and 
$\epsilon_{\rm{DM}}$=340 $h^{-1}$pc, respectively, were adopted for the
first two galaxies, and 57 and 100 $h^{-1}$pc for the latter.
Other results for the first two runs have already been presented
in Sommer-Larsen (2006), Razoumov \& Sommer-Larsen (2006), Greve \&
Sommer-Larsen (2007), Laursen \& Sommer-Larsen (2007) and Ellison \etal
(2007).

In Fig.~\ref{fig:numres} is shown the gas-phase oxygen temperature
distributions for the three very high resolution runs, together with
the results of the normal resolution ones --- note that the oxygen
masses for the smaller galaxy have been multiplied by a factor 50 for
clarity. It is seen that the general agreement is very good; the
slight disagreement for the small galaxy is due to the fact that the
region simulated at very high resolution was slightly smaller than
the corresponding normal resolution one.

Due to computational limitations we did not perform similar resolution
tests for the two other IMFs considered, but we have no reason to
believe that simulations based on these IMFs would perform less
well in resolution tests. We conclude
that the results obtained in this paper are robust to resolution changes.
\section{Summary and conclusions}
\label{sect:conclusions}
The global temperature distribution of the cosmic gas-phase 
oxygen at $z$$\sim$3 has been determined by combining high resolution 
cosmological simulations of individual proto-galactic, as well as larger, 
regions with extinction-corrected, observationally based, V-band 
(rest-frame) galaxy luminosity functions (LFs) of faint end slopes 
$\alpha=-1.85$ and $-1.57$.
The simulations have been performed with three 
different stellar initial mass functions (IMFs), a Kroupa (K98),
a Salpeter (S) and an Arimoto-Yoshii (AY), spanning a range
of a factor of five in chemical yield and specific SNII energy feedback.
Gas-phase oxygen is binned according to $T$ as log$(T)\la4.0$ (``cold''),
log$(T)\sim4.5$ (``warm''), and log$(T)\sim$5.0, 5.5, 6.0, 6.5, 7.0 
(``hot'' phases). Below we summarize results for the $\alpha=-1.85$ LF,
but results for the $\alpha=-1.57$ LF are similar.

Oxygen is found to be distributed over all $T$ phases, in particular for
the (``top-heavy'') AY IMF. 
But, at variance with previous works, it is found that,
for the K98 and Salpeter IMFs, the most important phase is the cold one,
which contains 47 and 37\% of all gas-phase oxygen, mainly in gas at fairly
high density, $n_{\rm{H}}$$\ga$0.1 cm$^{-3}$, and potentially star-forming.
Moreover, the cold phase alone contains 
1.3, 1.5 and 3.2 times the mass of oxygen in galactic stars for the three 
IMFs. The
implications of this in relation to observational damped Lyman-$\alpha$ 
absorber (DLA) studies are
discussed on the basis of very-high resolution simulations of two
(proto-)disc galaxies, with emphasis on oxygen and iron abundances.
It is concluded, that the reason why current DLA surveys only detect
a cold ISM metal fraction of about 20\% relative to the metal mass
in galactic stars is that the total 
cross-section for the high-metallicity component is only of order a 
few percent of the total DLA cross-section, and that the current DLA samlpes 
are too small to uncover it. Moreover, in addition, dust bias, as well as 
H$_2$ formation, will likely also be of importance.

In relation to ``missing metals'' it is found that the ratio of
gas-phase to stellar oxygen mass is 2.7, 3.9 and 13, and
the ratio of warm+hot to cold gas-phase oxygen mass is 1.1, 1.7 
and 3.2 for the three IMFs. 
For the AY IMF, the hot phases actually contain more oxygen than the 
cold+warm. 

In conclusion, a significant fraction of
the cosmic oxygen may be difficult or impossible to detect.
In addition, it is found that less than about 20-30\% of the cosmic 
oxygen will be associated with galaxies 
brighter than $M_{\rm{V}}\sim-22$, \ie the faintest galaxy luminosities 
probed by current LBG metallicity determinations (about one mag. below $L^*$). 
Hence, 70-80\% of the cosmic oxygen is also in this sense ``missing''.

From the LBG based, $\lambda$$\sim$1500 {\AA} UV luminosity
density history at $z$$\ge$3, we obtain an essentially IMF independent
constraint on the mean cosmic oxygen density at $z$=3. We compare this
to what is obtained from our models, for the three different IMFs.
We find that the (solar neighbourhood type) K98 IMF is strongly excluded,
as the chemical yield is simply too small, the Salpeter is marginally
excluded, and the AY matches the constraint well. The optimal IMF
would have a yield intermediate between the S and AY. The K98 IMF can 
only match the data if the $\lambda$$\sim$1500 {\AA} 
extinction corrections have been overestimated by factor of $\sim$4,
which seems highly unlikely, cf. Reddy \& Steidel (2004).

Using carbon abundances, and C to \hbox{C\,{\sc iv}} 
ionisation corrections, we estimate
 $\Omega$(\hbox{C\,{\sc iv}}) at moderate IGM
gas over-densities, $\delta=\rho_{\rm{gas}}/\bar{\rho}_{\rm{gas}}\sim 1-100$ for
the three IMFs, and compare to observational results. As above, we find 
that the yield of the K98 IMF is too small to match the data, whereas
models based on the Salpeter and AY IMFs can match the data, with the
optimal IMF in between the two. Moreover, we show that for all IMFs, 
\hbox{C\,{\sc iv}} abundance weighted
IGM temperatures are moderate, $T$$\la$4$\times$10$^4$ K, consistent 
with observational constraints on \hbox{C\,{\sc iv}} line widths.   
\section*{Acknowledgments}
We are very grateful to Ben Oppenheimer and Romeel Dav\'e for supplying
us with the C/\hbox{C\,{\sc iv}} ionisation correction look-up tables used in 
this work. We have benefited from discussions with Romeel Dav\'e, 
Sara Ellison, Peter Johansson, Cedric Lacey, Peter Laursen, Ben Oppenheimer, 
Laura Portinari and Alex Razoumov. The comments by the anonymous referee
considerably helped in improving the presentation of the paper. The TreeSPH
simulations were performed on the SGI Itanium II facility provided by
DCSC. The Dark Cosmology Centre is funded by the DNRF.
This research was supported by the DFG cluster of excellence ``Origin and
structure of the Universe''. 

\end{document}